\documentclass[
prd,
nofootinbib,
floatfix,superscriptaddress,
showpacs ,amssymb, aps]{revtex4}

\usepackage{comment}
\usepackage{graphicx} 
\usepackage{subfigure}
\usepackage{amsmath,amssymb}
\usepackage{bm}
\usepackage{hyperref}
\usepackage{natbib}
\usepackage{wasysym}

\begin{document}

\title{Cross-correlation search for continuous gravitational waves from \\a compact object in  SNR 1987A in LIGO Science Run 5}

\author{L. Sun}
\email{lings2@student.unimelb.edu.au} 
\author{A. Melatos} 
\email{amelatos@unimelb.edu.au}
\affiliation{School of Physics, University of Melbourne, Parkville, VIC 3010, Australia}
\author{P. D. Lasky}
\affiliation{Monash Centre for Astrophysics, School of Physics and Astronomy, Monash University, VIC 3800, Australia}
\affiliation{School of Physics, University of Melbourne, Parkville, VIC 3010, Australia}
\author{C. T. Y. Chung} 
\author{N. S. Darman} 
\affiliation{School of Physics, University of Melbourne, Parkville, VIC 3010, Australia}

		\begin{abstract}
			 We present the results of a cross-correlation search for gravitational waves from SNR 1987A using the second year of LIGO Science Run 5 data. The frequency band 75--450\,Hz is searched. No evidence of gravitational waves is found. A 90\% confidence upper limit of $h_0 \leq 3.8\times10^{-25}$ is placed on the gravitational wave strain at the most sensitive frequency near 150\,Hz. This corresponds to an ellipticity of $\epsilon \leq 8.2\times10^{-4}$ and improves on previously published strain upper limits by a factor $\approx4$. We perform a comprehensive suite of validations of the search algorithm and identify several computational savings which marginally sacrifice sensitivity in order to streamline the parameter space being searched. We estimate detection thresholds and sensitivities through Monte-Carlo simulations.
		\end{abstract}
		
		
		
		
		\maketitle
		
\section{Introduction}
\label{sec:codedescription}
Neutron stars in young supernova remnants are excellent targets for ground-based gravitational wave interferometers such as the Laser Interferometer Gravitational Wave Observatory (LIGO). Young neutron stars may be more promising targets than their older counterparts for three reasons. First, searches for periodic gravitational waves from younger neutron stars can more easily reach or probe below the indirect upper limits inferred from the spin-down rate $\dot{\nu}$ (where $\dot{\nu}$ is measured) or the age (where $\dot{\nu}$ is unknown). Indirect wave strain upper limits are proportional to $\dot{\nu}$ or inversely proportional to the age and hence are larger for younger neutron stars \cite{Abbott2007S2,Knispel2008, Riles2013}; see also Eq. (\ref{eq:h0upperlim}) in Section \ref{sec:xieffect}. Second, less time has passed in young objects for their crusts and interiors to settle down and erase historical nonaxisymmetries frozen-in at birth. Third, young objects spin down rapidly, driving crust-superfluid differential rotation which can excite nonaxisymmetric flows in the high-Reynolds-number interior \cite{Peralta2006,Melatos2010,Melatos2012}. For a review of gravitational wave generation mechanisms in neutron stars, see Ref. \cite{Abbott2007S2,Lasky2015}. On the other hand, even if young neutron stars emit gravitational waves strongly, their rapid spin down is a disadvantage for detection, because the phase of the gravitational wave signal evolves rapidly. A prohibitively large set of matched filters is needed for a coherent search, if a radio ephemeris is unavailable \cite{chung11}. Hence less sensitive semi-coherent search strategies are favoured \cite{brady00,Krishnan2004,Abbott2008,dhurandhar08,Riles2013,Wette2015}. 

LIGO achieved its design sensitivity over a wide band during its fifth and sixth science runs (S5 \cite{lscinstrument09} and S6 \cite{2012arXiv1203.2674T}, respectively). Data from S5 and S6 have been analysed in several searches for continuous wave sources in supernova remnants targeting specific, known sources like the Crab pulsar \cite{crab08, crab09, knownpul09}, Cassiopeia A \cite{wette08,abadie2010A}, other young pulsars with radio or X-ray ephemerides \cite{knownpul09, knownpul2014}, and young supernova remnants \cite{Aasi2015}. Broadband, all-sky searches have also been carried out for unknown sources, some of which may turn out post-discovery to reside in supernova remnants \cite{allsky09, lsceah09, Aasi2014}. Although no detections resulted from these searches, upper limits have been placed on parameters of astrophysical interest, e.g. the maximum ellipticity and internal magnetic field strength of the Crab pulsar \cite{crab08, crab09} and the amplitude of r-mode oscillations in Cassiopeia A \cite{abadie2010A}.

In this paper, we report on the search for periodic gravitational waves from a possible neutron star in one of the youngest and closest known supernova remnants, SNR 1987A. The remnant was produced by a Type II core collapse supernova which occurred in February 1987 in the Large Magellanic Cloud (right ascension $\alpha$ = 5h 35m 28.03s, declination $\delta$ = $-$69$^\circ$ 16$^\prime$ 11.79$^{\prime\prime}$, distance $d = 51.4$ kpc); see reviews by \citet{panagia08} and \citet{immler20yrs07}. The gravitational wave search relies on the semi-coherent cross-correlation algorithm \cite{dhurandhar08}, which has also been used in searches for gravitational waves from the low-mass X-ray binary Sco X-1 \cite{Whelan2015,Messenger2015}. Although the noise power spectral density of the LIGO S5 run is higher than that of the first Advanced LIGO observation run (O1), there are strong reasons to look for gravitational waves from SNR 1987A in the earlier data set. For example, the S5 run is considerably longer than O1, and the expected gravitational-wave amplitude during S5 is larger than during O1, given that the neutron star has aged significantly in the intervening ten years, which amount to 35 per cent of the object's age.

The structure of the paper is as follows. In Section II, we discuss the evidence for a neutron star in SNR 1987A and briefly review the results of previous gravitational wave searches. Section III summarizes the theory and implementation of the cross-correlation algorithm and the associated astrophysical phase model. Section IV reports on the verification tests performed on synthetic data containing pure noise and injected signals and evaluates the sensitivity penalty exacted when averaging over source orientation and polarization in order to reduce computational cost. In Section V, we calculate the sensitivity of the search as a function of the frequency and spin-down rate. Section VI presents the results obtained from running the search on LIGO S5 data and interprets the results astrophysically.

\section{A neutron star in SNR 1987A?}
\label{sec:NS_in_1987a}
\subsection{Indirect evidence for formation}
No neutron star has yet been detected electromagnetically in SNR 1987A, either reproducibly as a pulsar or as a nonpulsating central compact object \cite{Gotthelf2008}. Nevertheless, strong theoretical evidence exists for the existence of a neutron star in SNR 1987A from detailed studies of the progenitor \cite[e.g.][]{Arnett1989,Podsiadlowski1992}, and the coincident worldwide detection of core collapse neutrinos from the supernova event \cite{aglietta87, hirata87, bionta87, bahcall87}. Although no pulsar detection has been confirmed, numerous searches have placed upper limits on the flux and luminosity at radio \cite[$<115$ $\mu$Jy at 1390 MHz; ][]{immler20yrs07}, optical/near-UV \cite[$<8 \times 10^{33}$\,ergs s$^{-1}$; ][]{graves05}, and soft X-ray \cite[$<2.3 \times 10^{34}$ erg s$^{-1}$;][]{burrows00} wavelengths. \citet{middleditch00} reported finding an optical pulsar in SNR 1987A with a frequency of 467.5\,Hz, modulated sinusoidally with a $\sim$ 1-ks period, consistent with precession given an ellipticity of $\epsilon \sim 10^{-6}$. The pulsations disappeared after 1996 \cite{middleditch00} and were never confirmed independently.

One possible reason why a pulsar has not yet been detected is that its magnetic field is too weak. The weak-field theory is supported by some theoretical models, in which the field grows after the neutron star is formed over $\sim 10^3$ yr, e.g. due to thermomagnetic effects \cite{blandford88, reisenegger03,Pons2009}. In a related scenario, the magnetic field of a millisecond pulsar intensifies (linearly or exponentially) from $\sim 10^{10}$\,G  at birth to $\sim 10^{12}$\,G after $\sim$ 0.3--0.7\,kyr, before the pulsar spins down significantly \cite{michel94}. On the other hand, the neutron star may be born with a strong magnetic field, which is amplified during the first few seconds of its life by dynamo action \cite[e.g.][]{duncan92, bonanno05}. Population synthesis calculations combined with measurements of the known spin periods of isolated radio pulsars imply a distribution of birth magnetic field strengths $B_0$ satisfying $\log B_0 =12.65 \pm 0.55$ ($1\,\sigma$ range) \cite{hartman97, arzoumanian02, faucher06}. Several birth scenarios for the pulsar in SNR 1987A were considered by \citet{ogelman04} in this context, who concluded that the maximum magnetic dipole moment is $1.1 \times 10^{26}$\,G cm$^3$, $2.5 \times 10^{28}$\,G cm$^3$, and $2.5 \times 10^{30}$\,G cm$^3$ for birth periods of 2\,ms, 30\,ms, and 0.3\,s respectively. The dynamo model also accommodates a magnetar in SNR 1987A, with magnetic dipole moment exceeding $2.4 \times 10^{34}$\,G cm$^3$, regardless of the initial spin period \cite{ogelman04}.

Estimates of the birth spin of the putative pulsar in SNR 1987A are more uncertain. Simulations of the bounce and post-bounce phases of core collapse produce proto-neutron star spin periods between 4.7 ms and 140 ms, proportional to the progenitor's spin period \cite{ott06}. Some population synthesis studies, which infer the radio pulsar velocity distribution from large-scale 0.4\,GHz pulsar surveys, favour shorter millisecond birth spin periods \cite{arzoumanian02}, while others argue the opposite ($300\pm150$\,ms; $1\,\sigma$ range) \cite{faucher06}. Faint, non-pulsed X-ray emission from SNR 1987A was first observed four months after the supernova and decreased steadily in 1989 \cite{dotani87, inoue91}, leading to the suggestion that a neutron star could be powering a plerion, which is partially obscured by a fragmented supernova envelope. A model of the plerion's X-ray spectrum, with a magnetic field of $10^{12}$\,G and an expansion rate of $5 \times 10^8$\,cm s$^{-1}$, fits the X-ray data for a pulsar spin period of 18\,ms.

Despite their indirect nature, the above studies broadly justify a search for gravitational waves from SNR 1987A at frequencies from $\sim 50$\,Hz to $450$\,Hz (i.e. twice the spin frequency), bracketing the most sensitive portion of the LIGO band. The range of frequency derivative searched in this paper, namely from $10^{-13}\,{\rm Hz\,s}^{-1}$ to $10^{-6}\,{\rm Hz\,s}^{-1}$, is consistent with a magnetic field between $10^9\,{\rm G}$ and $10^{12}\,{\rm G}$ at the present epoch and hence with the $B_0$ values above. It is also consistent with maximum ellipticity in the range $10^{-5}\lesssim \epsilon \lesssim 10^{-4}$, if the spin down is gravitational wave dominated.

\subsection{Indirect gravitational radiation limits}
\label{sec:xieffect}
Neither $\nu$ nor $\dot{\nu}$ are known for the putative neutron star in SNR 1987A, so one is unable to infer an indirect spin-down upper limit on the characteristic wave strain $h_0$ by assuming that all the observed spin-down luminosity $4\pi^2I\nu\dot{\nu}$ (where $I$ is the stellar moment of inertia) goes into gravitational radiation \cite{wette08,Riles2013}. However, by a similar energy conservation argument, one can place an upper limit on $h_0$ in terms of the object's age, $T_\text{age}$ \cite{abadie2010A,chung11,Riles2013}, viz.

\begin{equation}
\label{eq:h0upperlim}
h_0\leq \frac{1}{D}\left(\frac{5GI\left|\xi\right|}{2c^3T_\text{age}}\right)^{1/2},
\end{equation}
with
\begin{equation}
	\label{eq:xi}
	\xi = \frac{1}{n-1}\left[1-\left(\frac{\nu_b}{\nu}\right)^{1-n}\right],
\end{equation}
where $G$ is Newton's gravitational constant, $c$ is the speed of light, $D$ is the distance to the source, $n$ is the braking index defined via $\dot{\nu} \propto \nu^n$ (assumed constant here for simplicity), $\nu_b$ is the spin frequency at birth, and $\left|\xi\right|^{-1}T_\text{age}=-\nu/\dot{\nu}$ is proportional to the characteristic electromagnetic spin-down time-scale  \cite{wette08}.

\begin{figure}
	\centering
	\scalebox{0.41}{\includegraphics{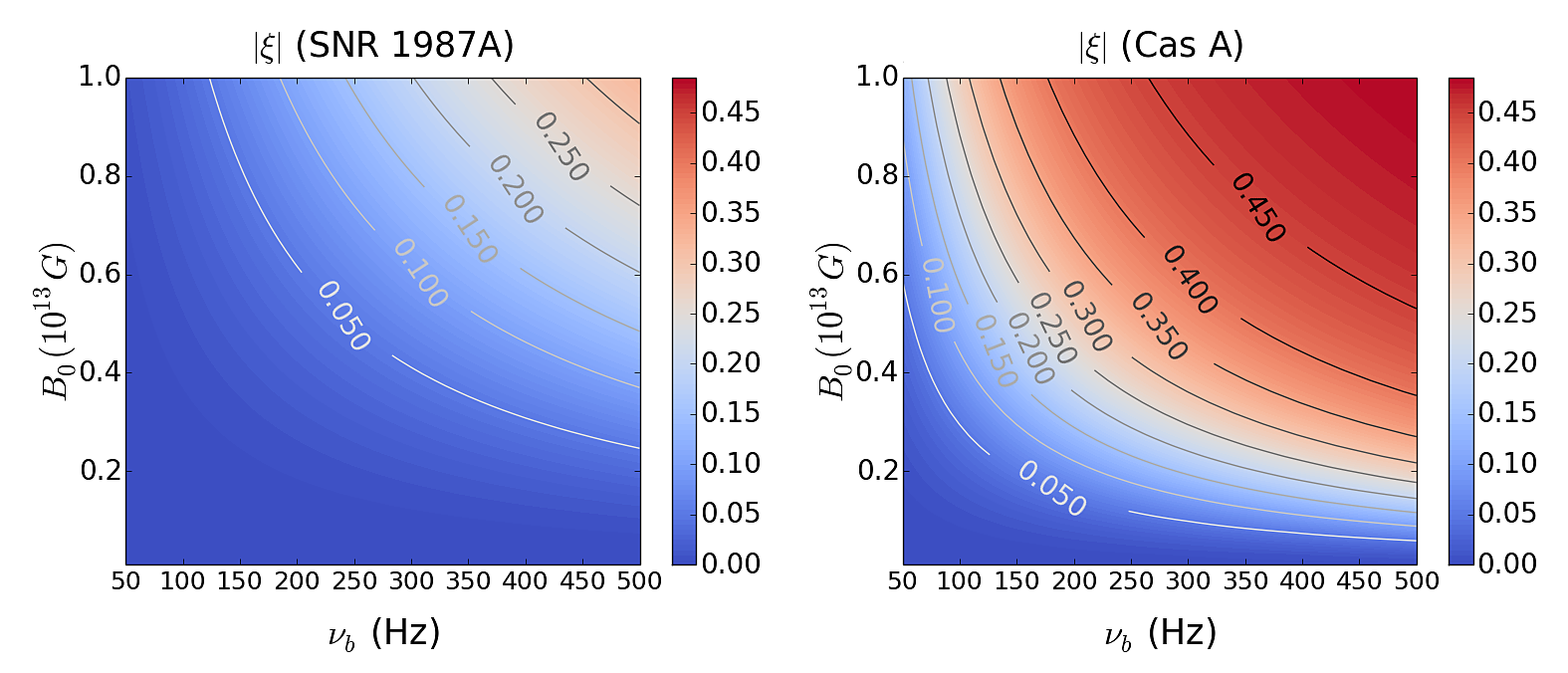}}
	\caption[Contours of $\left|\xi\right|$ for SNR 1987A (left) and Cas A(right)]{Contours of age factor $\left|\xi\right|$ in equation (\ref{eq:xi}) for SNR 1987A (left) and Cas A(right) as a function of the spin frequency at birth $\nu_b$ and dipole magnetic field $B_0$ with $n=3$. }
	\label{fig:xi}
\end{figure}

The factor $\left|\xi\right|$ in equation (\ref{eq:h0upperlim}) is normally neglected when quoting indirect limits under the assumption $\nu\ll\nu_b$ \cite{abadie2010A,chung11}. This assumption is reasonable for objects like Cas A but not for SNR 1987A, where $T_\text{age}$ is much less than $-\nu/\dot{\nu}$ for many reasonable choices of birth spin and magnetic field \cite{palomba05}. Figure \ref{fig:xi} illustrates the point. It displays contours of $\left|\xi\right|$ as a function of $\nu_b$ and dipole magnetic field $B_0$, assuming purely electromagnetic spin down ($\dot{\nu} \propto B_0^2 \nu^n$, $n=3$) for simplicity. The spin-down model is described in more detail in Section \ref{sec:phasemodel}. The left panel contours (SNR 1987A; $T_\text{age}=19$\,yr) satisfy $\left|\xi \right|\ll 1$ except in the top right corner of the plot (e.g. $\nu_b \gtrsim 350\,\text{Hz}$, $B_0 \gtrsim 7 \times 10^{12}$\,\text{G} for $\left|\xi\right| \gtrsim 0.25$). By contrast the right panel contours (Cas A;  $T_\text{age}\simeq 333$\,yr) satisfy $\left|\xi\right| \gtrsim 0.25$ over more of the plot, as befits an older object with $\nu \ll \nu_b$.

The indirect upper limit on $h_0$ is inversely proportional to $T_\text{age}$. Hence it is harder to reach observationally for older neutron stars. Younger objects like the putative neutron star in SNR 1987A generally have a higher limit on $h_0$, although not as high as one would expect assuming $\dot{\nu} \approx \nu T_\text{age}^{-1}$ in view of the ``$\xi$ effect" discussed above. The fact that young neutron stars with $\nu \approx \nu_b$ spin down slower than $\sim \nu T_\text{age}^{-1}$ aids detection by reducing dramatically the number of matched filters required to track the phase evolution. The latter advantage is further discussed in Section \ref{sec:brakingindex}.

\subsection{Previous gravitational wave searches}
 The likely existence of a young neutron star in SNR 1987A makes it a good target for gravitational wave searches \cite{piran88, nakamura89}.
 A coherent matched filtering search was carried out in 2003 with the TAMA 300 detector, searching $1.2 \times 10^3$ hours of data from its first science run over a 1-Hz band centered on 934.9\,Hz, assuming a spin-down range of (2--3)$\times 10^{-10}$\,Hz s$^{-1}$. The search yielded an upper limit on the wave strain of $h_0 \leq 5 \times 10^{-23}$ \cite{soida03}. An earlier matched filtering search was conducted using $10^2$ hours of data taken in 1989 by the Garching prototype laser interferometer. The latter search was carried out over 4-Hz bands near 2\,kHz and 4\,kHz, did not include any spin-down parameters, and yielded a strain upper limit of $h_0 \leq 9 \times 10^{-21}$ \cite{niebauer93}.
 
 The most sensitive gravitational wave search to date for SNR 1987A was conducted with the radiometer pipeline using LIGO S5 data \cite{Eric2011}. This search yielded an upper limit on the wave strain of $h_0 \approx 1.57 \times 10^{-24}$ (90\% confidence level) in the most sensitive frequency range near 160\,Hz. It is noted that the radiometer analysis always assumes a circularly polarised signal, so that in a case of random polarisation like the one discussed in this paper, the equivalent radiometer strain upper limit needs to be converted to a more conservative value, by multiplying a sky position dependent factor of 2.248 \cite{MessengerNote}. The above upper limit $h_0 \approx 1.57 \times 10^{-24}$ has already been converted from the original value stated in Ref. \cite{Eric2011}.

A coherent search for SNR 1987A based on the optically derived \citet{middleditch00} spin parameters requires 30 days of integration time and at least $10^{19}$ search templates covering just the frequency and its first derivative \cite{santostasi03}. Of course, the optical detection has not been confirmed independently, so one may have $\nu\ll467.5$\,Hz in reality, reducing $\dot{\nu} \propto {\nu}^3$ and hence the number of templates. Nevertheless, as a rule, young objects do spin down rapidly, and five or six higher-order frequency derivatives must be searched typically in order to accurately track the gravitational wave phase. In order to sidestep this problem, \citet{chung11} proposed an astrophysically motivated phase model which describes the spin down in terms of the ellipticity, magnetic field, and electromagnetic braking index of the source instead of its frequency derivatives. The model is most useful if the braking index varies slowly during the observation, in a sense to be defined precisely in Section \ref{sec:brakingindex}. At the time of writing, it is unclear on astrophysical grounds whether a slowly varying braking index is favoured or disfavoured by theoretical arguments (e.g. \cite{melatos97,Contopoulos2006}) and the sparse observational data available \cite{Livingstone2007}.

\section{Search pipeline}
\label{sec:pipeline}
\subsection{Cross-correlation algorithm}
\label{sec:xcorralg}
The theoretical basis of the cross-correlation algorithm was described in detail by \citet{dhurandhar08}. Here we summarize briefly the key results that are necessary for the algorithm's implementation. 

The algorithm operates on interferometer data in the form of short Fourier transforms (SFTs) \cite{Riles2013}, usually of 30\,min duration. It outputs a cross-correlation detection statistic called the $\rho$ statistic. SFTs are multiplied pair-wise according to some criterion (e.g. time lag or interferometer combination) to form a raw cross-correlation variable
\begin{equation}
\label{eq:yalpha}
\mathcal{Y}_{IJ} = \frac{\tilde{x}^*_{k_I, I} \tilde{x}_{k_J,J}}{ (\Delta T)^2},
\end{equation}
where $I$ and $J$ index the pair of SFTs $\tilde{x}_{k_I, I}$ and $\tilde{x}_{k_J,J}$, $k_I$ and $k_J$ are the indices of the frequency bins of the two SFTs, and $\Delta T$ denotes the length of the SFTs. The gravitational wave signal is assumed to be concentrated in a single frequency bin in each SFT, i.e. $\Delta T \ll \nu/\dot{\nu}$ due to sidereal motion and pulsar spin down.

The frequency range spanned by the two SFTs is the same, but the signal does not appear in the same frequency bin in $\tilde{x}_I$ and $\tilde{x}_J$. The specific frequency bins with indices $k_I$ and $k_J$ multiplied in (\ref{eq:yalpha}) are related by the time lag between the pair and between interferometers, as well as spin-down and Doppler effects.  For an isolated source, the instantaneous signal frequency at time $t$ is given by
\begin{equation}
\label{eq:freq}
\nu(t) = \hat{\nu}(t) \left( 1 + \frac{\mathbf{v}\cdot\mathbf{n}}{c} \right),
\end{equation}
where $\hat{\nu}(t)$ is the instantaneous signal frequency in the rest frame of the source, $\mathbf{v}$ is the detector velocity relative to the source, and $\mathbf{n}$ is the unit vector pointing from the detector to the source. The instantaneous signal frequencies in SFTs $I$ and $J$, $\nu(T_I)$ and $\nu(T_J)$, are calculated at the times corresponding to the midpoints of the SFTs, $T_I$ and $T_J$. The frequency bin $k_J$ is therefore shifted from $k_I$ by an amount $\lfloor\Delta T[\nu(T_J) - \nu(T_I)]\rfloor$, where $\lfloor\ldots\rfloor$ denotes the largest integer smaller than $(\ldots)$ \cite{dhurandhar08}. For convenience, we drop the subscripts $k_I$ and $k_J$ henceforth. 

The $\rho$ statistic comprises a weighted sum of $\mathcal{Y}_{IJ}$ over all pairs $(I,\,J)$. The relative weights of the pairs in the $\rho$ statistic are controlled by the polarization amplitudes and phase of the signal and the interferometer antenna pattern. These variables are packaged within the signal cross-correlation function $\tilde{\mathcal{G}}_{IJ}$, defined as
\begin{equation}
\label{eqngij} 
\tilde{\mathcal{G}}_{IJ} = \frac{1}{4} \exp(-i\Delta\Phi_{IJ})\exp\{-i \pi \Delta T[\nu(T_I) - \nu(T_J)]\} \left[F_{I+} F_{J+} \mathcal{A}_+^2 + F_{I\times} F_{J\times} \mathcal{A}^2_{\times} - i(F_{I+} F_{J\times} - F_{I\times}F_{J+}) \mathcal{A}_+ \mathcal{A}_{\times}\right].
\end{equation}
In (\ref{eqngij}) we define $\Delta \Phi_{IJ} = \Phi_I(T_I) - \Phi_J(T_J)$, where $\Phi_I(T_I)$ is the signal phase at time $T_I$. The terms in the second square brackets in (\ref{eqngij}) depend on the polarization angle $\psi$, and the inclination angle $\iota$ between $\mathbf{n}$ and the rotation axis of the pulsar, according to
\begin{eqnarray}
\label{eq:aplus} \mathcal{A}_+ &=& \frac{1}{2}(1 +  \cos^2 \iota) ,\\
\label{eq:across} \mathcal{A}_\times &=& \cos \iota ,\\
\label{eq:fplus} F_+(t; \mathbf{n}, \psi) &=& a(t; \mathbf{n}) \cos 2\psi + b(t; \mathbf{n}) \sin 2\psi,\\
\label{eq:ftimes} F_\times(t; \mathbf{n}, \psi) &=& b(t; \mathbf{n}) \cos 2\psi - a(t; \mathbf{n}) \sin 2\psi.
\end{eqnarray}
Here $a(t;\mathbf{n})$ and $b(t; \mathbf{n})$ are the detector response functions for a given sky position, defined in equations (12) and (13) of \citet{jara98}. A geometrical definition is also given in \citet{prixwhelan07}. The gravitational wave strain tensor is
\begin{equation}
\label{eq:wave_strain}
\overleftrightarrow{h}(t) = h_0 [\mathcal{A}_+ \cos \Phi(t) \overleftrightarrow{e}_+ + \mathcal{A}_\times \sin \Phi(t) \overleftrightarrow{e}_\times],
\end{equation} 
where $h_0$ is the characteristic gravitational wave strain, and $\overleftrightarrow{e}_{+, \times}$ are the basis tensors for the plus (+) and cross ($\times$) polarizations in the transverse-traceless gauge.\footnote{We alert the reader to an error in equation (3.10) of \citet{dhurandhar08}, which omits the factor of $\exp\{-i \pi \Delta T [\nu(T_I) - \nu(T_J)]\}$ arising from the choice of time origin of the Fourier transforms. Also see \citet{Whelan2015}.}

With the above definitions, the $\rho$ statistic is given by the weighted sum
\begin{equation}
\label{eq:rho}
\rho = \Sigma_{IJ} (u_{IJ} \mathcal{Y}_{IJ} + u^*_{IJ} \mathcal{Y}^*_{IJ}),
\end{equation}
where the weights are defined by
\begin{equation}
\label{eq:ualpha}
u_{IJ} = \tilde{\mathcal{G}}_{IJ}^*/\sigma_{IJ}^2,
\end{equation}
and
\begin{equation}
\label{eq:sigmasquare}
\sigma^2_{IJ} = S_n^{(I)}(\nu_I) S_n^{(J)}(\nu_J)/ (4 \Delta T^2), 
\end{equation}
is the variance of $\mathcal{Y}_{IJ}$ in the absence of a signal, where $S_n^{(I)}(\nu_I)$ is the power spectral density of SFT $I$ at frequency $\nu_I=\nu(T_I)$. For each frequency and sky position that is searched, we obtain \textit{one} real value of $\rho$, which is a sum of the Fourier power from all the pairs. Ignoring self-correlations (i.e. no SFT is paired with itself), the mean of $\rho$ is predicted to satisfy 
\begin{equation}
\label{eq:meanrho}
\mu_\rho = h_0^2 \sum_{IJ} \lvert \tilde{\mathcal{G}}_{IJ} \rvert^2/ \sigma_{IJ}^2.
\end{equation}
 In the limit of zero signal, the variance of $\rho$ is 
 \begin{equation}
 \label{eq:variancerho}
 \sigma^2_\rho = 2 \sum_{IJ} \lvert \tilde{\mathcal{G}}_{IJ}\rvert^2 /\sigma^2_{IJ}. 
 \end{equation}
 In the presence of a strong signal, and if self-correlations are included, $\mu_\rho$ and $\sigma^2_\rho$ scale as $h_0^2$ \cite{dhurandhar08}. The number of pairs is limited by computational resources. Summing over all possible pairs, which is normally prohibitive computationally, returns the same result as a fully coherent search.

In principle, one should search over the unknowns $\cos \iota$ and $\psi$ when computing $\rho$ through (\ref{eq:rho}). However, this adds to the already sizeable computational burden occasioned by searching over pulsar spin parameters (see Section \ref{sec:phasemodel}), when the number of SFT pairs is large. Accordingly, it is customary to average over $\cos \iota$ and $\psi$ when computing $\tilde{\mathcal{G}}_{IJ}$, assuming uniform priors on both variables. The result is
\begin{equation}
\label{eq:galphave} 
\langle \tilde{\mathcal{G}}_{IJ}\rangle_{\cos \iota, \psi} = \frac{1}{10} \exp(-i \Delta \Phi_\text{IJ}) \exp\{-i \pi \Delta T[\nu(T_I) - \nu(T_J)]\} (a_I a_J + b_I b_J),
\end{equation}
with $a_I = a (T_I; \mathbf{n})$ and $b_I = b (T_I; \mathbf{n})$. One then computes the detection statistic $\rho_\text{av}$ by replacing $\tilde{\mathcal{G}}_{IJ}^*$ by $\langle \tilde{\mathcal{G}}_{IJ}^*\rangle$ in (\ref{eq:ualpha}). Similarly, the mean and variance of $\rho_\text{av}$ can be computed by replacing $\tilde{\mathcal{G}}_{IJ}$ by $\langle \tilde{\mathcal{G}}_{IJ}\rangle$ in (\ref{eq:meanrho}) and (\ref{eq:variancerho}). Note, importantly, that $\rho_\text{av}$ is not equal to $\langle \rho \rangle_{\cos\iota, \psi}$, because $\mathcal{Y}_{IJ}$ depends implicitly on $\cos \iota$ and $\psi$ if $h_0\not =0$. Once a first-pass search with $\rho_\text{av}$ is complete, a follow-up search on any promising candidates can be performed, which searches explicitly over $\cos \iota$ and $\psi$ to achieve maximum sensitivity. Tests in Section \ref{sec:averaging} illustrate that the detection statistic resulting from (\ref{eq:galphave}) is approximately 10$-$15\% lower (up to 47 \% lower in rare cases) than if the exact $\cos \iota$ and $\psi$ values are used.

\subsection{Astrophysical phase model}
\label{sec:phasemodel}
The cross-correlation algorithm in Section \ref{sec:xcorralg} must be accompanied by a parameterized model for the phase and frequency of the signal as functions of time, in terms of which we express the factors $\Delta \Phi_{IJ} $ and $\nu(T_I) - \nu(T_J)$ in equations (\ref{eqngij}) and (\ref{eq:galphave}). A target like SNR 1987A raises special challenges in this regard. It is young and spins down rapidly, accumulating phase by an amount $k$ proportional to $\nu^{(k)}T_\text{obs}^{k+1} \approx \nu T_\text{obs}(\xi T_\text{obs}/T_\text{age})$ from the $k^\text{th}$ term of the Taylor expansion of the phase evolution after an observation time $T_\text{obs}$. One therefore needs approximately $N_\text{total} \propto \xi^{10} T_\text{obs}^{10}T_\text{lag}^5$ templates to keep the overall phase error below $\pi/4$ with SFT time separation $T_\text{lag}=1$\,hr \cite{chung11} by tracking terms up to and including $\nu^{(4)}$. As noted in Section \ref{sec:xieffect}, $N_\text{total}$ is suppressed strongly by the factor $\xi^{10}$, with $\xi^{10} \ll 1$ for SNR 1987A.

In the special but astrophysically plausible situation where the braking index $n = \nu\ddot{\nu}/\dot{\nu}^2$ changes slowly with time, one can take advantage of an alternative model for the gravitational wave phase introduced in \citet{chung11}, stated in terms of astrophysical parameters (i.e. the magnetic field strength and the neutron star ellipticity) instead of spin frequency derivatives. The model tracks the phase by assuming that the spin-down torque is the direct sum of gravitational-wave and electromagnetic components to a good approximation, with
\begin{eqnarray}
\label{eq:spindownmodel_full}
\dot{\nu} &=& -\frac{32 \pi^4 G \epsilon^2 I \nu^5}{5 c^5} - \frac{2 \pi^3 R_{\star}^6 B^2 \nu^{n_\text{em}}}{3 \mu_0 I c^3} \left(\frac{\pi R_{\star}}{c}\right)^{n_\text{em}-3}\\
\label{eq:numodel} &=& - Q_1' \nu^5 - Q_2' \nu^{n_\text{em}}.
\end{eqnarray}
In equation (\ref{eq:numodel}), $R_{\star}$ is the neutron star radius, $B$ is the polar magnetic field, and $n_\text{em}$ is the electromagnetic braking index. If the electromagnetic torque is proportional to a power of $\nu$, then $\nu$ must enter the torque in the combination $R_\star \nu/c$, (i.e. the ratio of $R_\star$ to the characteristic lever arm, the light cylinder distance, $c/2 \pi \nu$) on dimensional grounds. Hence, in terms of an arbitrary reference frequency, $\nu_{\rm{ref}}$, we write $\dot{\nu} = - Q_1 \left(\nu/\nu_{\rm{ref}}\right)^5 - Q_2 \left(\nu/\nu_{\rm{ref}}\right)^{n_\text{em}}$, with $Q_1 = Q_1' \nu_{\rm{ref}}^5$ and $Q_2 = Q_2' \nu_{\rm{ref}}^{n_\text{em}}$. Throughout this paper, we set $\nu_\text{ref} = 1$\,Hz without loss of generality.

The spin-down model (\ref{eq:numodel}) tracks the phase in terms of four parameters: $\nu_0, Q_1', Q_2'$ and $n_\text{em}$. Theoretically one expects $n_\text{em}=3$ for a magnetic dipole in vacuo \cite{Contopoulos2006}, but observations of radio pulsars find $1.8\leq n_\text{em}<3.0$ \cite{Livingstone2007}, and there exist several theoretical mechanisms consistent with $n_\text{em}<3$ \cite{melatos97, palomba05, Arons2007}. If $n_\text{em}$ is truly constant, then phase tracking requires $N_\text{total} \propto \xi^{6} T_\text{obs}^3 T_\text{lag}^3$ templates \cite{chung11}, and the search problem simplifies considerably. However, observations returning $n_\text{em}\neq 3$ raise the spectre of $n_\text{em}$ evolving as the star spins down; indeed the extended dipole braking model predicts $n_\text{em} \to 3$ as $t \to \infty$ \cite{melatos97}. If $n_\text{em}$ evolves too rapidly, it negates the advantage of (\ref{eq:numodel}) relative to a Taylor expansion $\{\nu, \dot{\nu}, \ddot{\nu}, \cdots\}$. This issue is quantified in Section \ref{sec:brakingindex}, and the parameter range where (\ref{eq:numodel}) remains useful is determined.

\subsection{Numerical algorithm}
\label{sec:numericalalgrithm}

The cross-correlation algorithm is implemented as part of the LIGO data analysis software suite (LAL\footnote{https://www.lsc-group.phys.uwm.edu/dawsg/projects/lal.html} and LALApps\footnote{https://www.lsc-group.phys.uwm.edu/dawsg/projects/lalapps.html}) in a general-purpose form. Firstly, the user can choose to search over the gravitational wave frequency at the start of the observation, denoted by $\nu_0$, and up to two frequency derivatives  in the Taylor expansion of the phase model ($\nu_0, \dot{\nu}, \ddot{\nu}$), or use the astrophysical model described in \citet{chung11} to search over $\nu_0, Q_1, Q_2$ and $n_\text{em}$. Secondly, one can choose to target a single sky coordinate, or search within a grid of sky coordinates. Thirdly, the user can choose to run the search using a particular value for the inclination and polarization angles ($\iota$ and $\psi$) or average over these variables. Finally, the user can decide to pair up Short Fourier Transforms (SFTs) from only the same interferometer, different interferometers, or a combination.

\begin{figure}
\centering
\scalebox{0.7}{\includegraphics{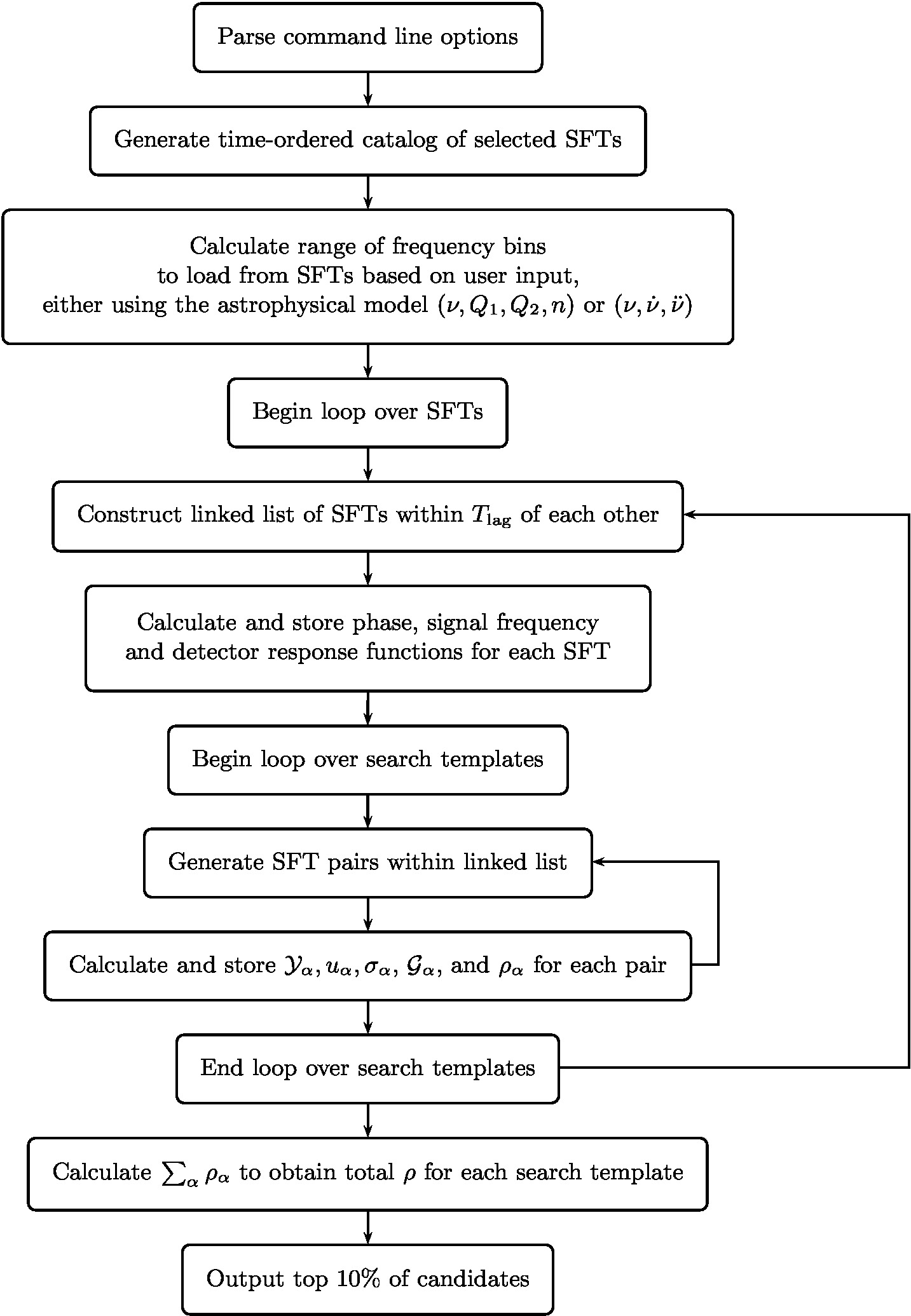}}
\caption[Flowchart summarising the algorithm used in LALApps.]{Flowchart summarising the cross-correlation algorithm used in the LALApps utility. $\alpha$ labels the SFT index pair ($I,J$).}
\label{fig:searchflow}
\end{figure}

The flow chart in Figure \ref{fig:searchflow} summarises the numerical algorithm. Firstly, the command-line options are parsed, and the relevant SFTs are located and read into a time-ordered catalogue. Only the frequency bins corresponding to the user-specified search frequency range are extracted from the SFTs. Each SFT is paired with another which satisfies the user-specified selection criteria, e.g. the maximum time lag $T_\text{lag}$. For each unique SFT pair, the code loops over each search template and calculates the corresponding normalised cross-correlation statistic, $\rho$, for each template.

An important issue encountered in the implementation is the heavy use of central processing unit (CPU) virtual memory. When pairing up an entire year's worth of SFTs ($\sim 10^5$ SFTs $\sim$ 1\,Terabyte), it is not feasible to load and store all the SFTs in virtual memory while looping over the search templates. Instead, we construct a time-ordered linked list, which contains only SFTs within a sliding window of length $T_\text{lag}$, i.e. a first-in-first-out queue. The signal phase, frequency, and detector response functions are calculated at each $T_I$. Then SFT pairs are constructed within the sliding window, and we loop over the search templates. For each pair, we calculate and store the quantities $\mathcal{Y}_\alpha, u_\alpha, \sigma_\alpha, \mathcal{G}_\alpha$, and $\rho_\alpha = \sum_\alpha (u_\alpha\mathcal{Y}_\alpha + u^*_\alpha \mathcal{Y}^*_\alpha)$ which are defined in \cite{dhurandhar08} and Section \ref{sec:xcorralg}; to simplify notation, we use the subscript $\alpha$ to denote the index pair ($I,J$) \cite{dhurandhar08}. As the window slides forward, we delete the SFT at the head of the linked list, add the next SFTs to its tail (as long as it satisfies the user-specified multiplication condition), and repeat the process. Once the loop over all possible pairs is finished, the final value of the detection statistic $\rho = \sum_\alpha \rho_\alpha$ for a particular search template is calculated. Finally, we output $\rho$ (normalised by its standard deviation) along with the relevant search parameters used. Typically, we search up to $\sim 10^9$ templates and filter the output so that e.g. only the highest 10\% of $\rho$ values are saved.

\section{Algorithm Verification}
\label{sec:testing}

\subsection{Distribution of $\rho/\sigma_\rho$ when searching over pure noise}
\label{sec:noise}
A basic consistency check is to run the search on simulated noise with no injected signal. The detector noise time series $n(t)$ is typically assumed to be Gaussian with zero mean. In this situation, according to equation (\ref{eq:yalpha}), $\mathcal{Y}_\alpha$ is related to the noise power spectra in the SFTs centred at $T_I$ and $T_J$. Hence $\rho_\alpha$ is a product of two independent Gaussian variables with zero mean. Its probability density function (PDF) is a modified Bessel function of the second kind of order zero, with zero mean and finite variance \cite{dhurandhar08}. Applying the central limit theorem, the sum of a large number of such zero-mean variables tends to a Gaussian random variable \cite{feller57}, as the number of SFT pairs, $N_\text{pairs}$, increases. 

The mean $\mu_\rho$ and variance $\sigma_\rho^2$ of $\rho$ in the low-signal limit are given by \cite{dhurandhar08}
\begin{eqnarray}
\label{eq:murho} \mu_\rho &=& h_0^2 \sum_\alpha(u_\alpha \mathcal{G}_\alpha + u_\alpha^* \mathcal{G}_\alpha^*),\\
\label{eq:sigmarho} \sigma_\rho^2 &=& 2 \sum_\alpha \lvert u_\alpha \rvert^2 \sigma_\alpha^2,
\end{eqnarray}
where $h_0$ vanishes for pure noise, and $u_\alpha, \mathcal{G}_\alpha$, and $\sigma_\alpha$ are defined in equations (\ref{eq:ualpha}), (\ref{eqngij}) and (\ref{eq:sigmasquare}) respectively. We note that these equations exclude self-correlations (i.e. pairing an SFT with itself), and equation (\ref{eq:sigmarho}) assumes $h_0 \ll |n(t)|$. We discuss how to generalise beyond the small-signal limit in Section \ref{sec:rhosignal}.
The code outputs the normalised cross-correlation statistic, $\rho/\sigma_\rho$, whose PDF should have zero mean and unit variance for pure noise.
We emphasize that the mean $\mu$ and variance $\sigma^2$ of the PDF of $\rho/\sigma_\rho$ should not be confused with the mean and variance of the pre-normalised $\rho$ distribution, given by (\ref{eq:murho}) and (\ref{eq:sigmarho}).

The simulated Gaussian noise is generated using the standard LALApps utility. This utility creates SFTs for user-specified values of signal strength $h_0$, single-sided power spectral density $[S_n (\nu)]^{1/2}$, SFT length $T_\text{SFT}$, total observation time $T_\text{obs}$, and signal parameters $\nu(t), \alpha, \delta, \cos \iota,$ and $\psi$. In order to vary $N_\text{pairs}$ for testing purposes, we generate separate sets of 30-minute SFTs for five different values of $T_\text{obs}$ ranging from 1\,hour (2 SFTs per interferometer) to 1\,year (17532 SFTs per interferometer) with zero signal strength ($h_0 = 0$) and random signal parameters. The standard analytic approximation of the single-sided power spectral density as a function of the signal frequency is \cite{damour01b}
\begin{equation}
\label{eq:psd} S_n (\nu)^{1/2} \approx \alpha_0 \left[ \alpha_1 \left(\frac{ \nu}{150\,\text{Hz}}\right)^{-56} + \alpha_2 \left(\frac{\nu}{150\,\text{Hz}}\right)^{-4.52} + \alpha_3 \left(\frac{\nu}{150\,\text{Hz}}\right)^2 + \alpha_4 \right]^{1/2},
\end{equation}
with $\alpha_0 = (9 \times 10^{-46})^{1/2}$\,Hz$^{-1/2}$, $\alpha_1 = 4.49$, $\alpha_2 = 0.16$, $\alpha_3 = 0.32$, and $\alpha_4 = 0.52$.
In real LIGO data, variable phenomena like seismic noise make $S_n (\nu)$ time-dependent on time-scales of hours to days. Simulated noise does not suffer from this problem. SFTs are simulated for only two interferometers (H1 and L1), and the SFTs for each interferometer span identical times.

\begin{figure}
\centering
\scalebox{0.18}{\includegraphics{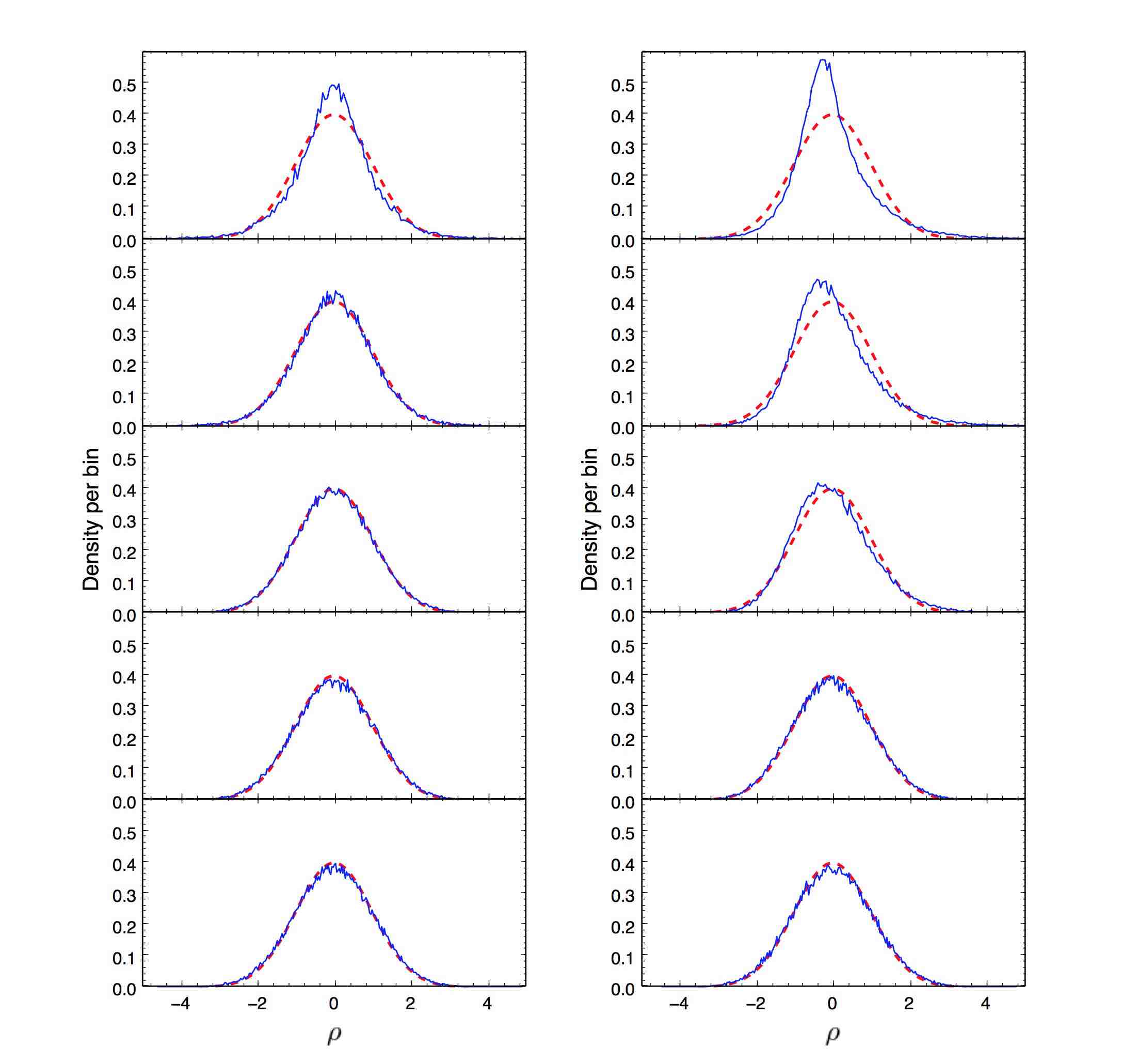}}
\caption[PDFs of $\rho$ for $T_\text{lag}$ = 0\,s and $T_\text{lag} = 3600$\,s for $T_\text{obs} =$ 1 hour, 5 hours, 1 day, 1 month, and 1 year.]{PDFs of $\rho/\sigma_\rho$ (solid curves) for $T_\text{lag}$ = 0\,s (left) and $T_\text{lag} = 3600$\,s (right) for $T_\text{obs} =$ 1\,hr, 5\,hr, 1\,day, 1\,month, and 1\,yr (top to bottom). For reference, the dashed curve shows a Gaussian with zero mean and unit variance.}
\label{fig:NSS}
\end{figure}

For each set of SFTs, we run the search using a frequency band of $10^{-2}$\,Hz, and a frequency resolution of $10^{-4}$\,Hz, corresponding to 100 search templates for each pair ($T_\text{lag}, T_\text{obs}$). We consider five values of $T_\text{obs}$ (1\,hr $\leq T_\text{obs} \leq$ 1\,yr) and two values of $T_\text{lag}$ for each $T_\text{obs}$, viz. 0\,s and 3600\,s. Setting $T_\text{lag} = 0$ correlates only SFTs from different interferometers. This ensures that all pairs, and the resulting $\rho_\alpha$ values, are completely independent. For $T_\text{lag} = 3600$\,s, each SFT is paired with three others if we include data from two interferometers, and five others if we include data from three interferometers. In this case, the same SFT contributes to more than one $\rho_{\alpha}$. As a result, the $\rho_{\alpha}$ values are not statistically independent. \citet{Coyne2016} discussed the correction to $\rho$ for dependent $\rho_{\alpha}$, finding that $\rho$ is distributed as a $\chi^2$ distribution with two degrees of freedom instead of a Gaussian distribution (for more details, see \cite{Coyne2016}). This correction is crucial to an intermediate-duration search ($T_{\rm obs} \lesssim 10^4\,{\rm s}$). In this paper, we carry out a long-duration search ($T_{\rm obs}=1$\,yr). To test the above effect, we compare the PDFs of $\rho$ for $T_{\rm lag}=0$ ($\rho_\alpha$ independent) and $T_{\rm lag}=3600$\,s ($\rho_\alpha$ dependent) for $T_{\rm obs}$ in the range $1\,{\rm hr} \leq T_\text{obs} \leq 1\,$yr. The full results are presented below in Table~\ref{tab:noisestatistics} and Figure \ref{fig:NSS}. In brief, they confirm that the correction in Ref.~\cite{Coyne2016} is appreciable for $T_{\rm obs}\lesssim1$\,day but negligible for $T_{\rm obs}=1$\,yr. The experiment is repeated 1000 times for each pair ($T_\text{lag}, T_\text{obs}$) with 100 templates, and the statistics of the resulting $10^5$ $\rho/\sigma_{\rho}$ values are compiled. 

The mean $\mu$ and standard deviation $\sigma$ of the $\rho/\sigma_\rho$ PDFs are presented in Table \ref{tab:noisestatistics}. The values of $\mu$ lie within the 95\% confidence limits\footnote{The 95\% confidence limits for $\mu$ are $\pm 1.96 \sigma N_\text{trial}^{-1/2} \approx \pm 0.0064$, where $N_\text{trial}$ is the number of trials ($10^5$).} and deviate from zero by at most 0.0053. The values of $\sigma$ however, are systematically $\sim 4$\% larger than unity and appear to increase with $T_\text{obs}$. The reason for this discrepancy is unclear. We keep this issue in mind as the analysis proceeds. Discrepancies at the $\lesssim$5\% level are not expected to impact the search results significantly.

\begin{table}
\centering
\setlength{\tabcolsep}{8pt}
\begin{tabular}{lllcc}
\hline
\hline
$T_\text{obs}$ & $N_\text{pairs}$ & $T_\text{lag}$ (s) & $\mu$& $\sigma$ \\
\hline
1 hour & 2 & 0  & $-$0.003393 & 1.035527\\
 & 6 & 3600 & 0.001966 & 1.033288\\
5 hours & 10  & 0 & 0.005309 & 1.037305\\
  & 46 & 3600 & $-$0.002081 &  1.044258\\
1 day & 48 &  0 & 0.000937  & 1.040644\\
 & 236 & 3600 & 0.002111 &  1.044826\\
1 month & 1440 & 0 & 0.000613 & 1.042374\\
 & 7196 & 3600 & 0.000095 & 1.039875\\
1 year &  17532 & 0 & $-$0.004183 & 1.040686\\
 & 87656 & 3600 & 0.003961 & 1.047173\\
\hline
\hline
\end{tabular}
\caption[Mean $\mu$ and standard deviation $\sigma$ of $10^5$ values of $\rho/\sigma_\rho$ for a search over simulated Gaussian noise using observation times $T_\text{obs} =$ 1 hour, 5 hours, 1 day, 1 month, and 1 year.]{Mean $\mu$ and standard deviation $\sigma$ of $10^5$ values of the normalized cross-correlation statistic $\rho/\sigma_\rho$ for a search over simulated Gaussian noise for observation times satisfying 1\,hr $\leq T_\text{obs} \leq$ 1\,yr. For each value of $T_\text{obs}$, the number of SFT pairs $N_\text{pairs}$ is listed, along with the maximum SFT pair separation, $T_\text{lag}$.}
\label{tab:noisestatistics}
\end{table}

Figure \ref{fig:NSS} displays PDFs of $\rho/\sigma_\rho$ (solid curves) for the trials listed in Table \ref{tab:noisestatistics}. From top to bottom, the panels show $\rho/\sigma_\rho$ for $T_\text{obs}$ running from 1\,hr to 1\,yr for $T_\text{lag} = 0$\,s (left panel) and $T_\text{lag} = 3600$\,s (right panel). By way of comparison, Gaussian PDFs with zero mean and unit standard deviation are overplotted as dashed curves in each panel. The PDFs for $T_\text{obs}=1$\,hr are clearly non-Gaussian. For $T_\text{lag} = 0$\,s (top row, left panel), the distribution is symmetric about zero, but more sharply peaked than a Gaussian. For $T_\text{lag} = 3600$\,s (top row, right panel), the distribution peaks more sharply than a Gaussian and is significantly skewed. As $T_\text{obs}$ increases, the PDFs for both $T_\text{lag}$ values approach a Gaussian. For $T_\text{obs}=1$\,month (fourth row in Figures \ref{fig:NSS}), the difference is nearly imperceptible by eye.

\begin{figure}
\centering
\scalebox{0.14}{\includegraphics{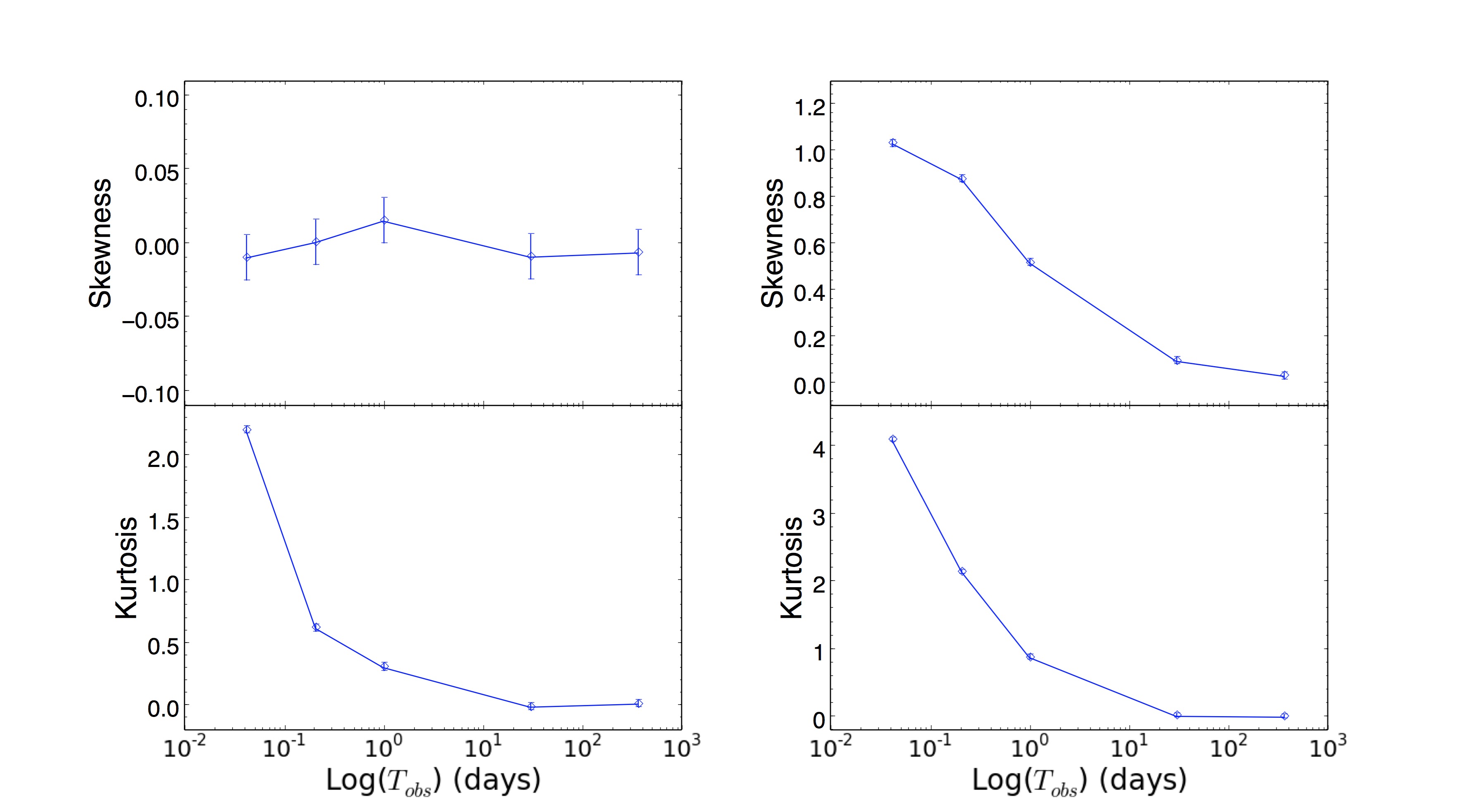}}
\caption[Skewness and kurtosis excess of $\rho/\sigma_\rho$ as a function of total SFT duration, for $T_\text{lag} = 3600$\,s and SFTs containing pure Gaussian noise.]{Skewness, $\gamma_1$ (top panels), and kurtosis excess, $\gamma_2 - 3$ (bottom panels), of $\rho/\sigma_\rho$ as functions of $T_\text{obs}$, for $T_\text{lag}$ = 0\,s (left) and $T_\text{lag} = 3600$\,s (right) when SFTs contain only Gaussian noise. The error bars (small vertical lines overplotted on the points) have peak-to-peak amplitudes of twice the standard error of skewness (top panels) and twice the standard error of kurtosis (bottom panels) (see text for definition).}
\label{fig:skewness}
\end{figure}

We quantify the Gaussianity of the PDFs in Figure \ref{fig:NSS} by plotting their skewness and kurtosis excess in Figure \ref{fig:skewness} as functions of $T_\text{obs}$. The skewness of a distribution, which measures its reflection asymmetry, is defined as $\gamma_1 = \mu_3/\mu_2^{3/2}$, where $\mu_2$ and $\mu_3$ are the second and third central moments. For a Gaussian, $\gamma_1$ is zero. The kurtosis measures the peakiness and is defined as $\gamma_2 = \mu_4/\mu_2^2$, where $\mu_4$ is the fourth central moment. For a Gaussian, one has $\gamma_2 = 3$. The kurtosis excess, $g_2 = \gamma_2 - 3$, therefore equals zero for a Gaussian. Figure \ref{fig:skewness} displays $\gamma_1$ and $g_2$ as functions of $\log(T_\text{obs})$ for $T_\text{lag} = 0$\,s (left) and $T_\text{lag} = 3600$\,s (right). Error bars of size $\pm 2 s_s$ and $\pm 2 s_k$ are overplotted, where $2 s_s = 2 \sqrt{6/N_\text{trial}} = 0.0155 $ is twice the standard error of skewness, $2 s_k = 2 \sqrt{24/N_\text{trial}} = 0.031$ is twice the standard error of kurtosis, and $N_\text{trial} = 10^5$ is the total number of trials \cite{tabachnick96}. For $T_\text{lag} = 3600$\,s, the skewness and kurtosis decrease from $\gamma_1 = 1.03, g_2 = 4.09$ for $T_\text{obs}=1$\,hr to $\gamma_1 = 0.0277, g_2 = 0.0006$ for $T_\text{obs}=1$\,yr. For $T_\text{lag} = 0$\,s, when there is no overlap between SFT pairs and all $\rho_\alpha$ values are independent, the kurtosis also decreases as $T_\text{obs}$ increases, from $g_2 = 2.204$ for $T_\text{obs}=1$\,hr to $g_2 = 0.014$ for $T_\text{obs}=1$\,yr. However, the skewness remains roughly centred at zero, fluctuating between $\gamma_1 = -0.010$ (for $T_\text{obs}=1$\,hr) and $\gamma_1 = 0.015$ (for $T_\text{obs}=1$\,day), which is within the standard errors. The shape of the $\rho/\sigma_\rho$ PDF is therefore significantly affected by $T_\text{obs}$ and, to a lesser extent, $T_\text{lag}$. However, for $T_\text{obs} \geq 1$\,yr, the PDFs for $T_\text{lag}=0$\,s and $T_\text{lag}=3600$\,s agree with theoretical predications to an accuracy of better than 95\% in $\mu$, $\sigma$, $\gamma_1$ and $g_2$. The above results show that for intermediate-duration searches with $T_{\rm obs} \lesssim 10^4\,{\rm s}$ (i.e. $T_\text{obs} =1$\,hr, 5\,hr in our test), the skewness and kurtosis deviate significantly from the expected values in a Gaussian distribution, and hence the correction to $\rho$ discussed in Ref.~\cite{Coyne2016} is required. However, in a long-duration search with $T_\text{obs} = 1$\,yr, the above moments of $\rho$ match those of a Gaussian distribution to an accuracy above 95\%, so the correction is negligible for the search in this paper.

\subsection{Distribution of $\rho/\sigma_\rho$ as a function of signal strength}
\label{sec:rhosignal}
The introduction of a gravitational wave signal changes the distribution of $\rho$ and $\rho/\sigma_\rho$. Most notably, the mean and variance increase with the signal strength. In Appendix A of \citet{dhurandhar08}, the statistics of the $\rho$ distribution are recalculated, including self-correlations and $\mathcal{O}(h_0^2)$ terms which are left out in the main body of their analysis. In the absence of self-correlations, which we neglect in this paper, one obtains
\begin{eqnarray}
\label{eq:generalmu} \mu_\rho &=& h_0^2 \sum_{IJ} (u_{IJ} \tilde{\mathcal{G}}_{IJ} + u_{IJ}^* \tilde{\mathcal{G}}_{IJ}^*) ,\\
\label{eq:generalsigma} 
\sigma_\rho^2 &=& 2 \sum_{IJ} \lvert u_{IJ} \rvert^2 \sigma^2_{IJ} + \frac{h_0^2}{\Delta T} \sum_{I \neq J} \lvert u_{IJ} \rvert^2 \left[\tilde{\mathcal{G}}_{II} S_n^{(J)} + \tilde{\mathcal{G}}_{JJ} S_n^{(I)}\right]+ \mathcal{O}(h_0^4),
\end{eqnarray}
for the mean and variance of $\rho$ respectively. For the normalised statistic $\rho/\sigma_\rho$, whose noise-only PDF is a Gaussian with zero mean and unit variance, the mean and variance in the presence of a signal are given by
\begin{eqnarray}
\label{eq:generalnormmu}  \mu &=& 2^{1/2} h_0^2 \sum_\alpha  \lvert u_\alpha \rvert,\\
\label{eq:generalnormsigma} \sigma^2 &=& \sum_{I \neq J} \lvert u_{IJ} \rvert^2 \left[ \frac{S_n^I S_n^J}{4 \Delta T^2} + \frac{h_0^2}{2 \Delta T} \left(\tilde{\mathcal{G}}_{II} S_n^J + \tilde{\mathcal{G}}_{JJ} S_n^I\right)\right] + \mathcal{O}(h_0^4),
\end{eqnarray}
respectively, where $S_n^I$ is the single-sided power spectral density squared for the SFT centred at $t=T_I$. Self-correlation terms are not included in (\ref{eq:generalnormmu}) and (\ref{eq:generalnormsigma}). Again, we stress that $\mu$ and $\sigma$ in (\ref{eq:generalnormmu}) and (\ref{eq:generalnormsigma}) are not derived from (\ref{eq:generalmu}) and (\ref{eq:generalsigma}) simply by dividing the latter equations by $\sigma_\rho$ and $\sigma_\rho^2$ respectively. The PDF of $\rho/\sigma_\rho$ is not truly Gaussian for pure noise, if dependent pairs are included \cite{Coyne2016}. However, the results in Section \ref{sec:noise} demonstrate that the impact is negligible (i.e. the moments match those of a Gaussian distribution to an accuracy above 95\%) for $T_\text{obs} = 1$\,yr, so we do not correct for this effect in this paper, as discussed in Section \ref{sec:noise}.

\begin{figure}
\centering
\scalebox{0.7}{\includegraphics{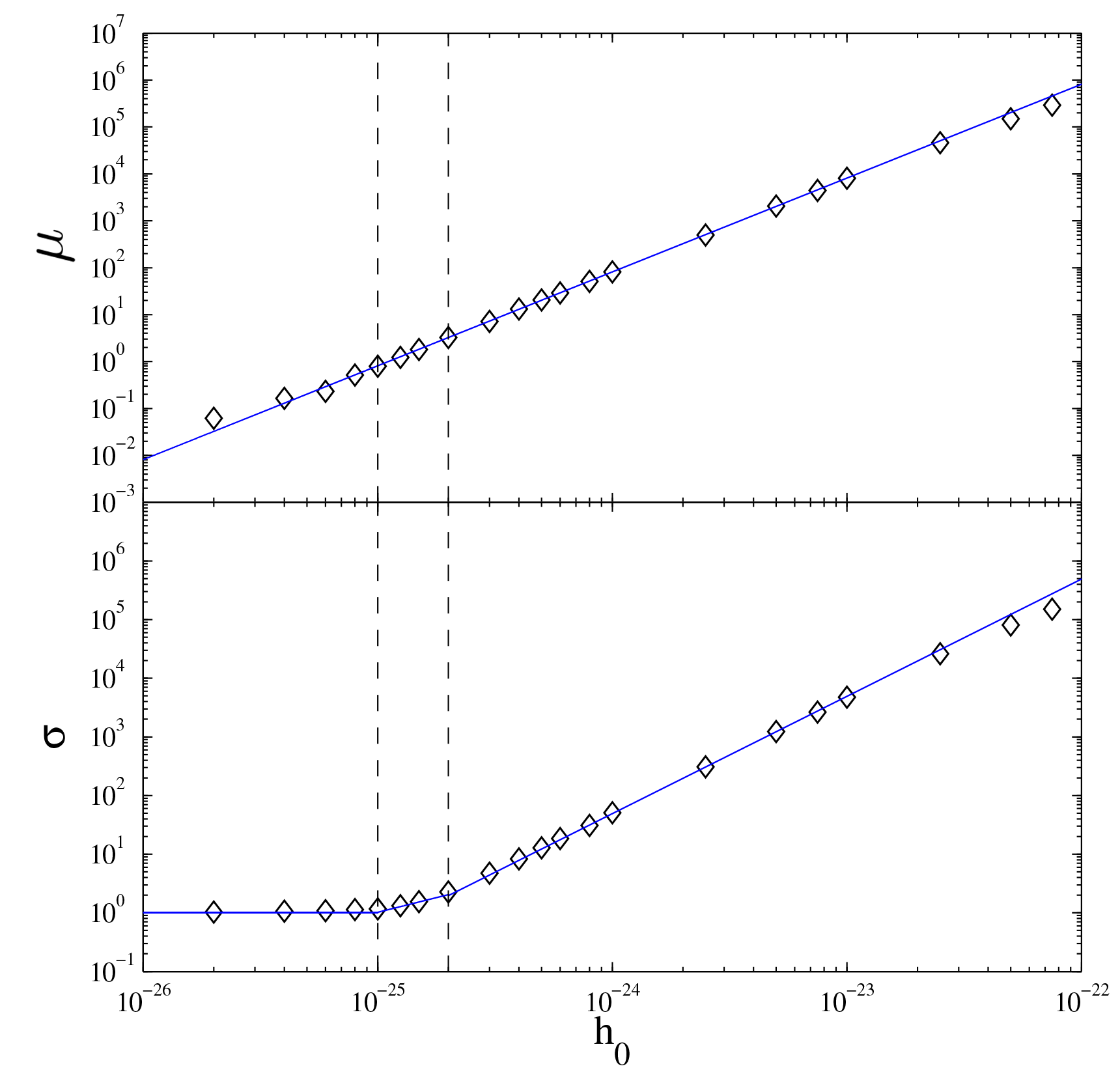}}
\caption[The mean and standard deviation of the normalized detection statistic $\rho/\sigma_\rho$ as a function of injected gravitational wave strain $h_0$.]{The mean $\mu$ (top panel) and standard deviation $\sigma$ (bottom panel) of the normalized cross-correlation statistic $\rho/\sigma_\rho$ as a function of injected gravitational wave strain $h_0$. The injected signals have arbitrary parameters $\alpha, \delta, \cos \iota,$ and $\psi$, and a fixed frequency of 150.1\,Hz with zero spin-down. Each point comes from $10^3$ $\rho/\sigma_\rho$ values for given $h_0$. The solid curves match the approximation predicted by equations (\ref{eq:generalnormmu}) and (\ref{eq:generalnormsigma}), and the vertical dashed lines mark the three $h_0$ regimes ($h_0 \ll |n(t)|$, $h_0 < |n(t)|$ and $h_0 \gtrsim |n(t)|$).}
\label{fig:rhovsh0150hz}
\end{figure}

We test (\ref{eq:generalnormmu}) and (\ref{eq:generalnormsigma}) against numerical results by injecting signals into simulated Gaussian noise with wave strains ranging between $1 \times 10^{-26} \leq h_0 \leq 7.5 \times 10^{-23}$ at 150.1\,Hz and zero spin-down. The $h_0$ range covers the regimes $h_0 \ll |n(t)|$, $h_0 < |n(t)|$ and $h_0 \gtrsim |n(t)|$. Again, we use the LALApps utility to generate $10^3$ SFTs for each $h_0$ value, with arbitrary signal parameters $\alpha, \delta, \cos \iota,$ and $\psi$. We take $T_\text{obs}=1$\,yr, $T_\text{lag} = 3600$\,s, and search over a 0.01\,Hz band centred on the signal frequency with a frequency resolution of $10^{-4}$\, Hz. We only search the chosen $\alpha, \delta, \cos \iota$, and $\psi$ values of the injected signal. From the theory, $\rho/\sigma_\rho$ is maximized at the injected frequency value, which is 150.1\,Hz in this case, and this maximum value appears to be dominant if the signal is strong enough. For verification purposes, we extract the $10^3$ $\rho/\sigma_\rho$ values at 150.1\, Hz for each $h_0$ value tested. The mean and standard deviation of the $10^3$ $\rho/\sigma_\rho$ values are calculated and shown in Figure \ref{fig:rhovsh0150hz}.

The top panel of Figure \ref{fig:rhovsh0150hz} plots $\mu$ as a function of $h_0$. For $h_0\leq 8 \times 10^{-26}$, $\mu$ gets very close to zero, which is as expected in the low signal limit. Above $h_0 = 8 \times 10^{-26}$, $\mu$ increases from $\approx 0.8$ at $h_0 = 1 \times 10^{-25}$ to $\approx 3 \times 10^5$ at $7.5 \times 10^{-23}$, growing $\propto h_0^2$ as expected from equation (\ref{eq:generalnormmu}).

The bottom panel of Figure \ref{fig:rhovsh0150hz} shows $\sigma$ as a function of $h_0$. One can distinguish three regimes. For $h_0 \leq 1 \times 10^{-25}$, i.e. $h_0 \ll |n(t)|$, the signal is too small to be detectable, giving the same, unit standard deviation as the results obtained in Section \ref{sec:noise} for pure noise. In the intermediate regime $1 \times 10^{-25} \leq h_0 \leq 2 \times 10^{-25}$ between the vertical dashed lines, where the signal is small but still detectable, $\sigma$ grows approximately $\propto h_0$ as predicted by (\ref{eq:generalnormsigma}), and the $\mathcal{O}(h_0^4)$ terms are negligible. As $h_0$ increases further, $\sigma$ tends towards the scaling $\sigma \propto h_0^2$. Above $h_0 \approx 10^{-24}$, the $\mathcal{O}(h_0^4)$ terms in (\ref{eq:generalnormsigma}) are dominate. In practice, realistic astrophysical signals are unlikely to fall in the third regime. For an optimistic yet realistic signal strength satisfying $10^{-25} \lesssim h_0 \lesssim 10^{-24}$, $\mu$ and $\sigma$ scale approximately as $h_0^2$, as predicted by (\ref{eq:generalnormmu}) and (\ref{eq:generalnormsigma}).

\subsection{Averaging over $\cos \iota$ and $\psi$}
\label{sec:averaging}

The inclination and polarization angles, $\iota$ and $\psi$, which modulate the amplitude of a gravitational wave signal, are not known for the compact object in SNR 1987A, and should strictly be included in the set of search parameters. To economize computationally however, it is often preferable to average over $\cos \iota$ and $\psi$ in a first-pass search. If a suitable candidate is identified, follow-up searches can include these parameters, once the number of templates is narrowed down. In this subsection, we quantify the loss of sensitivity occasioned by the averaging process.

When averaging over $\cos \iota$ and $\psi$, the unaveraged signal cross-correlation function $\tilde{\mathcal{G}}_{IJ}$ in equation (\ref{eqngij}) is replaced by the averaged version \cite{jara98}
\begin{equation}
\label{eqn:avg}
\langle \tilde{\mathcal{G}}_{IJ} \rangle_{\psi, \cos \iota} = \frac{1}{4 \pi} \int_0^{2\pi} d\psi \int_{-1}^{1} d(\cos \iota) \, \tilde{\mathcal{G}}_{IJ}.
\end{equation}
The result is given by equation (\ref{eq:galphave}), applying the averaged versions of equations (\ref{eq:aplus})--(\ref{eq:ftimes}) \cite{dhurandhar08}.

Let $\cos\iota_\text{real}$ and $\psi_\text{real}$ denote the $\cos\iota$ and $\psi$ values of an injected signal, and let $\cos\iota_\text{test}$ and $\psi_\text{test}$ be the associated search variables in a mock search. We create 400 injections on a uniformly spaced $20\times20$ grid of  $\cos\iota_\text{real}$ and $\psi_\text{real}$ values, using the LALApps utility as in Section \ref{sec:noise}. Signals are injected into 1 year of 30-min SFTs (from H1 and L1) with $h_0/\sqrt{S_n(\nu)} = 9.635$\,Hz$^{1/2}$ at frequency 991.413\,Hz. Sky coordinates ($\alpha, \delta$) are chosen arbitrarily to be $(1.16357, -0.0439203)$. For each injection, we run two mock searches using (1) $\tilde{\mathcal{G}}_{IJ}$, the exact signal parameters ($\alpha, \delta$), and a $10\times10$ grid of $\cos\iota_\text{test}$ and $\psi_\text{test}$ values, and (2) $\langle \tilde{\mathcal{G}}_{IJ}\rangle_{\cos \iota, \psi}$ and the exact signal parameters ($\alpha, \delta$). Both searches analyse the same SFTs across a frequency band of full width 0.003\,Hz centred on the injected signal frequency, with a resolution of $10^{-5}$\,Hz and $T_\text{lag}=3600$\,s. We extract the maximum normalised statistics $\rho/\sigma_\rho$ at the injected frequency. Among the 100 values of $\rho/\sigma_\rho$ from the first set of searches, we denote the maximum, mean, and minimum values with ${(\rho/\sigma_\rho)}_\text{max}$, ${(\rho/\sigma_\rho)}_\text{mean}$, and ${(\rho/\sigma_\rho)}_\text{min}$ respectively. From the second search, ${(\rho/\sigma_\rho)}_\text{avg}$ denotes the single normalized statistic returned by using the averaged cross-correlation function $\langle \tilde{\mathcal{G}}_{IJ} \rangle_{\psi, \cos \iota}$. We emphasize that ${(\rho/\sigma_\rho)}_\text{mean}$ and ${(\rho/\sigma_\rho)}_\text{avg}$ are different quantities; the former involves $\tilde{\mathcal{G}}_{IJ}$, while the latter involves $\langle \tilde{\mathcal{G}}_{IJ} \rangle_{\psi, \cos \iota}$.

\begin{figure}
	\centering
	\scalebox{0.11}{\includegraphics{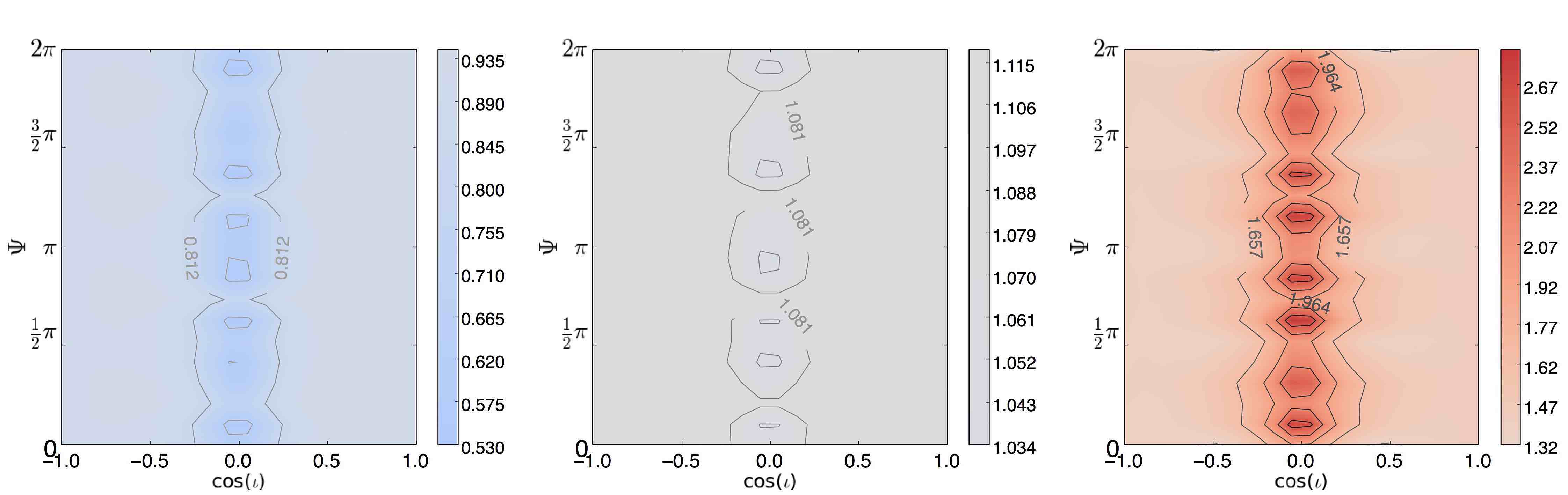}}
	\caption[The comparison of general cross-correlation function and averaged cross-correlation function]{Ratio of the normalised detection statistic ${(\rho/\sigma_\rho)}_\text{avg}$ computed with the averaged cross-correlation function $\langle \tilde{\mathcal{G}}_{IJ}\rangle_{\cos \iota, \psi}$ divided by the maximum, mean and minimum values of the normalised detection statistic using the unaveraged $\tilde{\mathcal{G}}_{IJ}$ as a function of the injection angles ($\cos\iota_\text{real}$, $\psi_\text{real}$) (left, middle, right panels respectively). The averaged statistic $\langle\tilde{\mathcal{G}}_{IJ}\rangle_{\cos \iota, \psi}$ performs nearly as well as $\tilde{\mathcal{G}}_{IJ}$ for a fraction of the computational cost, sacrificing $\lesssim 50\%$ sensitivity in the (rare) worst cases (left panel) and $\lesssim 10\%$ sensitivity typically (middle panel). It yields better sensitivity in the best cases (right panel).}
	\label{fig:avgcompare}
\end{figure}

Figure \ref{fig:avgcompare} compares ${(\rho/\sigma_\rho)}_\text{avg}$ to ${(\rho/\sigma_\rho)}_\text{max}$ (left panel), ${(\rho/\sigma_\rho)}_\text{mean}$ (middle panel), and ${(\rho/\sigma_\rho)}_\text{min}$ (right panel). The relevant ratios are plotted as contours on the plane spanned by $\cos\iota_\text{real}$ and $\psi_\text{real}$. ${(\rho/\sigma_\rho)}_\text{avg}$ is plotted as the numerator in order to make the comparison straightforward. The left panel corresponds to trials where ($\cos\iota_\text{test}$, $\psi_\text{test}$) happens to be close to ($\cos\iota_\text{real}$, $\psi_\text{real}$), in which ${(\rho/\sigma_\rho)}_\text{avg}$ is expected to be smaller than ${(\rho/\sigma_\rho)}_\text{max}$. We find $0.534 \leq (\rho/\sigma_\rho)_\text{avg}/(\rho/\sigma_\rho)_\text{max} \leq 0.943$ for the 400 injections. These results from the worst case for using $\langle \tilde{\mathcal{G}}_{IJ}\rangle_{\cos \iota, \psi}$, yet the loss in sensitivity is tolerable. The middle panel plots the ratio of ${(\rho/\sigma_\rho)}_\text{avg}$ to the mean value of $\rho/\sigma_\rho$ among the 400 injections. The ratio fluctuates slightly between 1.035 to 1.118; using $\langle \tilde{\mathcal{G}}_{IJ}\rangle_{\cos \iota, \psi}$ typically sacrifices $\lesssim 10\%$ sensitivity and can even improve it slightly for certain ($\cos\iota_\text{test}$, $\psi_\text{test}$) combinations. The right panel compares ${(\rho/\sigma_\rho)}_\text{avg}$ with ${(\rho/\sigma_\rho)}_\text{min}$, when ($\cos\iota_\text{test}$, $\psi_\text{test}$) is far from ($\cos\iota_\text{real}$, $\psi_\text{real}$). The ratio ranges between 1.33 and 2.802; i.e. there is a significant advantage in using $\langle\tilde{\mathcal{G}}_{IJ}\rangle_{\cos \iota, \psi}$. In every panel, the results appear to depend more on $\cos\iota$ than $\psi$, but, near $\cos\iota \approx 0.0$ where the signal is weakest, the variation with $\psi$ is more apparent. This matches expectations: $\tilde{\mathcal{G}}_{IJ}$ depends more strongly on $\cos\iota$ via $\mathcal{A}_+$ and $\mathcal{A}_\times$ in equations (\ref{eq:aplus}) and (\ref{eq:across}) than on $\psi$ via $F_+$ and $F_\times$ in equations (\ref{eq:fplus}) and (\ref{eq:ftimes}). When the inclination angle approaches 90 degrees (i.e. $\cos\iota \approx 0.0$), the gravitational wave strain in equation (\ref{eq:wave_strain}) is smaller than for smaller inclination angles, and hence the sensitivity sacrificed by averaging $\tilde{\mathcal{G}}_{IJ}$ is more obvious when there is a weaker signal. In every panel, for $|\cos\iota| \lesssim 0.25$, the contours make periodic patterns along the vertical axis caused by variation of $\psi$ with period approximately equal to $\pi$, as expected from the periodic functions $F_+$ and $F_\times$ in equations (\ref{eq:fplus}) and (\ref{eq:ftimes}).

In summary, $\langle\tilde{\mathcal{G}}_{IJ}\rangle_{\cos \iota, \psi}$ performs nearly as well as $\tilde{\mathcal{G}}_{IJ}$ for a fraction of the computational cost, sacrificing $\lesssim 50\%$ sensitivity in the (rare) worst cases and $\lesssim 10\%$ sensitivity typically.

\subsection{Astrophysical spin-down parameters}
\label{sec:spindown}
As described in Section \ref{sec:numericalalgrithm}, one can choose to search over the Taylor coefficients ($\nu_0, \dot{\nu}, \ddot{\nu}$) or the parameters ($\nu_0, Q_1, Q_2, n_\text{em}$) that define the astrophysical spin-down model described in Section \ref{sec:phasemodel}. The latter approach performs better when $n_\text{em}$ is constant to a good approximation over the observation time. In this subsection we quantify the relative performance of the two approaches and how that the relevant ``observation time" is $T_\text{lag}$ rather than $T_\text{obs}$, because the cross-correlation algorithm is semi-coherent. The computational cost of the search is analysed in \citet{chung11}.

\subsubsection{Astrophysical model versus Taylor expansion}
\label{sec:astro_vs_taylor}
We begin by running a single search for an injected signal that is spinning down using both the astrophysical model and Taylor expansion. We inject a signal into 30-min SFTs (from H1 and L1) for the 1-yr observation period with the parameters listed in Table \ref{tab:sig-params}, which lie in the typical ranges discussed in Section \ref{sec:NS_in_1987a}. Note that the utility LALApps was not written to accommodate a general spin-down model in the form (\ref{eq:spindownmodel_full}) for generating synthetic data, so we input the frequency and its first three derivatives instead, as calculated from (\ref{eq:spindownmodel_full}).

\begin{table}
	\centering
	\setlength{\tabcolsep}{8pt}
	\begin{tabular}{lll}
		\hline
		\hline
		Injection parameter & Value & Units \\
		\hline
		$h_0/\sqrt{S_n(\nu)}$& 3.33 & Hz$^{1/2}$ \\
		${\nu_0}_\text{signal}$ & 150.1 & Hz \\
		$\dot{\nu}_{0_{\rm signal}}$ & $-2.67 \times 10^{-8}$ & Hz\,s$^{-1}$\\
		$\ddot{\nu}_{0_{\rm signal}}$& $2.37 \times 10^{-17}$ & Hz\,s$^{-2}$ \\
		$\dddot{\nu}_{0_{\rm signal}}$& $-3.80 \times 10^{-26}$ & Hz\,s$^{-3}$\\
		\hline
		\hline
	\end{tabular}
	\caption[signal parameters]{Injection parameters used to create the synthetic data analysed in Section \ref{sec:astro_vs_taylor}. The frequency derivatives correspond to spin-down parameters ${Q_1}_\text{signal}=3.5\times10^{-19}\,\text{Hz\,s}^{-1}$ and ${Q_2}_\text{signal}=1\times10^{-17}\,\text{Hz\,s}^{-1}$ (i.e. $\epsilon=4.52 \times 10^{-4}$ and $B=4.05 \times 10^{11}$\,G) according to equations (\ref{eq:spindownmodel_full}) and (\ref{eq:numodel}).}
	\label{tab:sig-params}
\end{table}

Two searches are carried out with this mock data set for $T_\text{lag}=3600$\,s. The first search uses the astrophysical model. The second uses the Taylor expansion. The search parameter ranges encompass the injected signal and are quoted in Table \ref{tab:search-params}. For now we take $n_\text{em}=3$ to be constant. The evolution of $n_\text{em}$ is discussed in Section \ref{sec:brakingindex}.

\begin{table}
	\centering
	\setlength{\tabcolsep}{8pt}
	\begin{tabular}{lllll}
		\hline
		\hline
		&Search parameter & Range width & Resolution & Units \\
		\hline
		Astrophysical model &$\nu_0$ & 0.5 & 0.005 & Hz \\
		&$Q_1$& $5\times10^{-19}$ & $0.05\times10^{-19}$ &Hz\,s$^{-1}$ \\
		&$Q_2$& $2\times10^{-17}$ & $0.02\times10^{-17}$ &Hz\,s$^{-1}$ \\
		\hline
		Taylor series &$\nu_0$ & 0.5 & 0.005 & Hz \\
		&$\dot{\nu}_0$& $2\times10^{-8}$ & $0.01\times10^{-8}$ &Hz\,s$^{-1}$ \\
		& $\ddot{\nu}_0$& $2\times10^{-17}$ & $0.01\times10^{-17}$ &Hz\,s$^{-2}$ \\
		\hline
		\hline
	\end{tabular}
	\caption[search parameters]{Search parameter ranges using the astrophysical model and the Taylor series. The ranges are centred on the injected signals. We take $n_\text{em}=3$ to be constant and discuss the evolution of $n_\text{em}$ in Section \ref{sec:brakingindex}. The range width column defines the domain of the search parameter assuming that it is centred on the injection.}
	\label{tab:search-params}
\end{table}

\begin{figure}
	\centering
	\scalebox{0.13}{\includegraphics{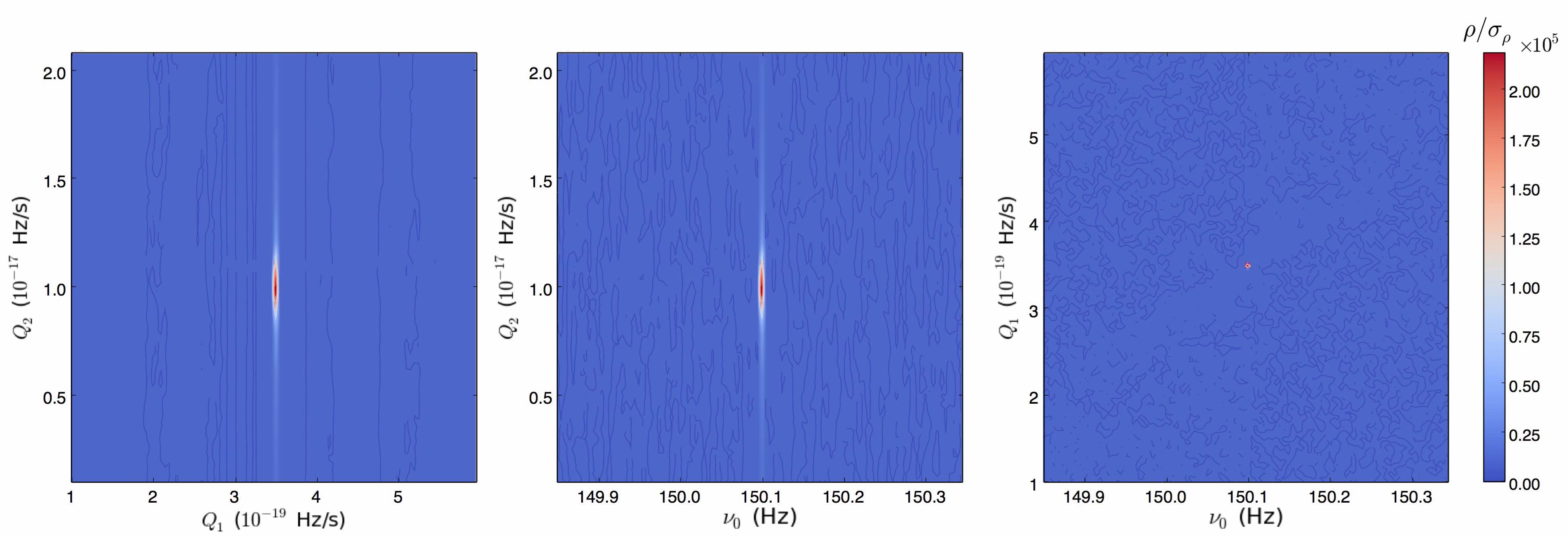}}
	\caption[Contour plots displaying the normalised detection statistic $\rho/\sigma_\rho$ as functions of every two parameters among ($\nu_0, Q_1, Q2$) being searched.]{Normalised detection statistic $\rho/\sigma_\rho$ as a function of trial parameter value pairs from the set $\{\nu_0, Q_1, Q_2\}$. The injected values are ${\nu_0}_\text{signal}=150.1$\,Hz, ${Q_1}_\text{signal}=3.5\times10^{-19}\,\text{Hz\,s}^{-1}$, ${Q_2}_\text{signal}=1\times10^{-17}\,\text{Hz\,s}^{-1}$.}
	\label{fig:oneTrialQ1Q2}
\end{figure}

\begin{figure}
	\centering
	\scalebox{0.13}{\includegraphics{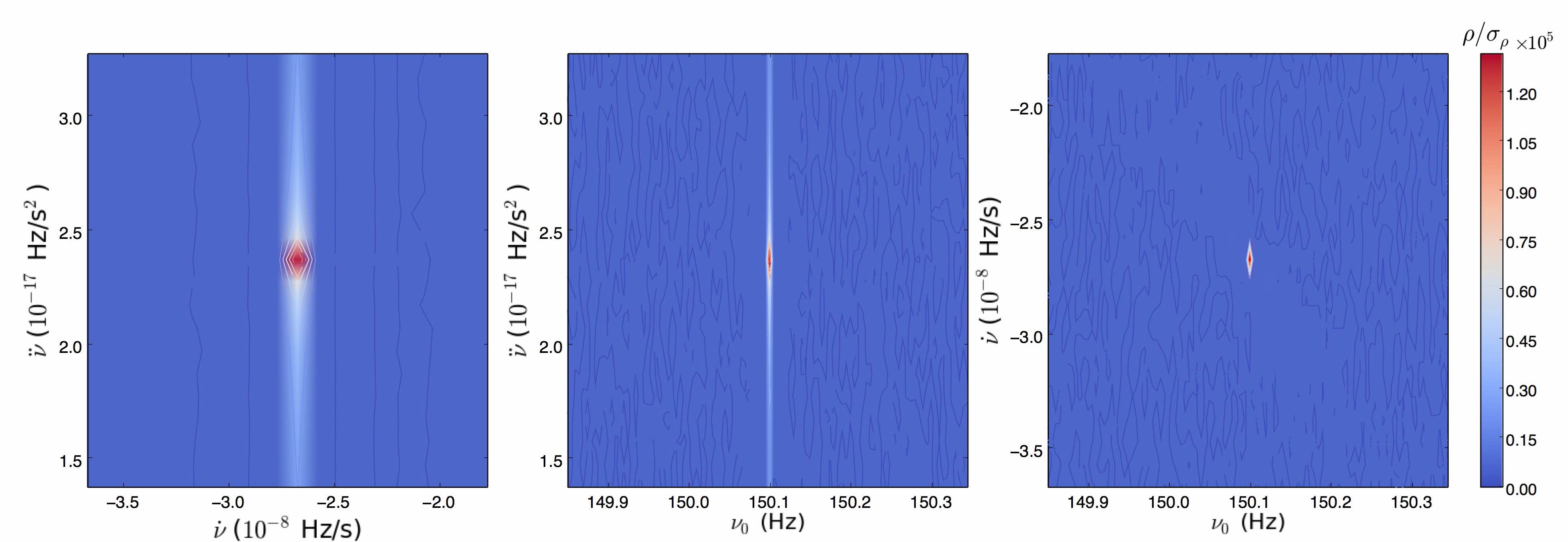}}
	\caption[Contour plots displaying the normalised detection statistic $\rho/\sigma_\rho$ as functions of every two parameters among ($\nu_0, \dot{\nu}, \ddot{\nu}$) being searched.]{Normalised detection statistic $\rho/\sigma_\rho$ as a function of trial parameter value pairs from the set $\{\nu_0, \dot{\nu}, \ddot{\nu}\}$. The injected values are ${\nu_0}_\text{signal}=150.1$\,Hz, $\dot{\nu}_{0_{\rm signal}}=-2.67 \times 10^{-8}\,\text{Hz\,s}^{-1}$, $\ddot{\nu}_{0_{\rm signal}}=2.37 \times 10^{-17}\,\text{Hz\,s}^{-2}$, $\dddot{\nu}_{0_{\rm signal}}=-3.80 \times 10^{-26}\,\text{Hz\,s}^{-3}$.}
	\label{fig:oneTrialfdot}
\end{figure}

Figure \ref{fig:oneTrialQ1Q2} presents the normalised detection statistic $\rho/\sigma_\rho$ from the first search as a function of parameter pairs from the set $\{\nu_0, Q_1, Q_2\}$ in three separate contour plots. Similarly, Figure \ref{fig:oneTrialfdot} presents contours of $\rho/\sigma_\rho$ as a function of parameter pairs from the set $\{\nu_0, \dot{\nu}, \ddot{\nu}\}$. The statistic peaks when the trial parameter values are closest to the injected values (${\nu_0}_\text{signal}, {Q_1}_\text{signal}, {Q_2}_\text{signal}$) or (${\nu_0}_\text{signal}, \dot{\nu}_\text{signal},\ddot{\nu}_\text{signal}$) as expected. The first search generates a higher maximum ($\rho/\sigma_\rho\approx 2 \times 10^5$) than the second ($\rho/\sigma_\rho \approx 1.3 \times 10^5$), because only the first and second frequency derivatives are searched in the second test, whereas the first search tracks the phase exactly. The superiority of searching astrophysical parameters becomes more dominant when we inject a signal with faster spin-down rate.

\subsubsection{Including or excluding $Q_1$ and $Q_2$}
\label{sec:include_exclude_q1q2}
One may doubt whether the astrophysical spin-down model is correct and how the search can benefit from including spin-down parameters. We now test how much sensitivity is sacrificed by searching over $\nu_0$ and neglecting spin down, as compared to searching a combination of ($\nu_0, Q_1, Q_2$) according to the astrophysical spin-down model (\ref{eq:spindownmodel_full}). Again, we assume $n_\text{em}$ is constant for simplicity; cf. Section \ref{sec:brakingindex}. We inject signals with a range of wave strains $h_0$ but identical $\nu_0=150.1$\,Hz. For a specific wave strain, a grid of $15 \times 17$ values of ${Q_1}_\text{signal}$ and ${Q_2}_\text{signal}$ are chosen within the ranges listed in Table \ref{tab:sig-range-Q1Q2}. The signal parameters are astrophysically relevant in line with the discussion in Section \ref{sec:NS_in_1987a} and affordable from the perspective of computing cost. Each signal, which is spinning down, is injected into 30-min SFTs (from H1 and L1) for a whole year. Two sets of searches, excluding and including $Q_1$ and $Q_2$ in the search parameters, are run over the parameter ranges in Table \ref{tab:search-params-Q1Q2} with $T_\text{lag}=3600$\,s, targeted at the same injections. We analyse only the largest $\rho/\sigma_\rho$ value returned. 

\begin{table}
	\centering
	\setlength{\tabcolsep}{8pt}
	\begin{tabular}{ll}
		\hline
		\hline
		Injection parameter & Astrophysical parameter \\
		\hline
		$1 \times 10^{-22} \leq {Q_1}_\text{signal} \leq 1.64 \times 10^{-18}\,\text{Hz\,s}^{-1}$ & $7.65 \times 10^{-6} \leq \epsilon \leq 9.79\times 10^{-4}$  \\
		$1 \times 10^{-21} \leq {Q_2}_\text{signal} \leq 1 \times 10^{-13}\,\text{Hz\,s}^{-1}$ & $4.05 \times 10^9$\,G $\leq B \leq 4.05 \times 10^{13}$\,G\\
		\hline
		\hline
	\end{tabular}
	\caption[signal parameters]{Spin-down parameter ranges for the injected signals analysed in Section \ref{sec:include_exclude_q1q2} in order to compare the results of searching $\nu_0$ only and searching $\nu_0$, $Q_1$ and $Q_2$. A grid of $15\times 17$ values of ${Q_1}_\text{signal}$ and ${Q_2}_\text{signal}$, evenly spaced on a logarithmic scale, are chosen within the ranges. The corresponding ranges of the astrophysical parameters $\epsilon$ and $B$ in equation (\ref{eq:spindownmodel_full}) are also quoted.}
	\label{tab:sig-range-Q1Q2}
\end{table}

\begin{table}
	\centering
	\setlength{\tabcolsep}{8pt}
	\begin{tabular}{lllll}
		\hline
		\hline
		&Search parameter & Range & Resolution & Units \\
		\hline
		Search $\nu_0$ only &$\nu_0$ & $135.45 - 150.15$ & 0.01 & Hz \\
		\hline
		Search $\nu_0$, $Q_1$ and $Q_2$ &$\nu_0$ & $150.095 - 150.105$ & 0.001 & Hz \\
		&$Q_1$& $0.9-1.1$ & 0.02 &${Q_1}_\text{signal}$ \\
		&$Q_2$& $0.9-1.1$ & 0.02 &${Q_2}_\text{signal}$ \\
		\hline
		\hline
	\end{tabular}
	\caption[search parameters]{Search parameter ranges for the synthetic signals with injection parameters quoted in Table \ref{tab:sig-range-Q1Q2}. The upper half of the table refers to searching $\nu_0$ only (i.e. $Q_1=Q_2=0$). The lower half refers to searching $\nu_0$, $Q_1$ and $Q_2$. The ranges encompass the injected signals.}
	\label{tab:search-params-Q1Q2}
\end{table}

\begin{figure}
	\centering
	\scalebox{0.16}{\includegraphics{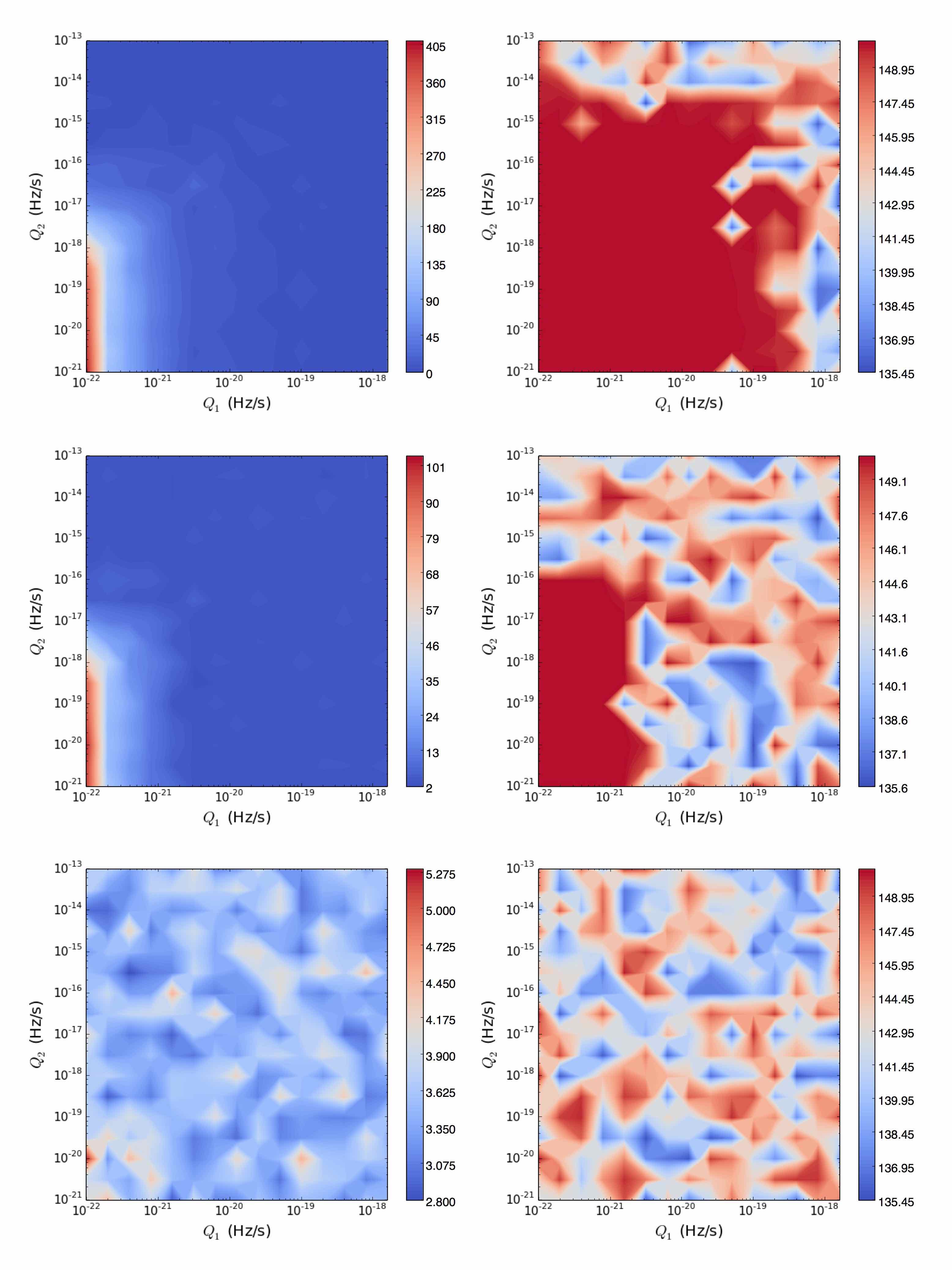}}
	\caption[Contour plots showing the largest $\rho/\sigma_\rho$ returned as functions of $Q_1$ and $Q_2$ values used to create the signal when taking frequency $\nu_0$ as the only search parameter (left panel), and the corresponding frequency at which the largest $\rho/\sigma_\rho$ is found (right panel). ($h_0/\sqrt{S_n(\nu)} = 0.33$\,Hz$^{1/2}$)]{Maximum $\rho/\sigma_\rho$ returned as a function of ${Q_1}_\text{signal}$ and ${Q_2}_\text{signal}$ values when searching over $\nu_0$ only (i.e. ${Q_1}_\text{trial}={Q_2}_\text{trial}=0$) (left), and the corresponding frequency at which $\rho/\sigma_\rho$ peaks (right) for $h_0=1 \times 10^{-23}$ and $h_0/\sqrt{S_n(\nu)} = 0.33$\,Hz$^{1/2}$ (top), $h_0=5 \times 10^{-24}$ and $h_0/\sqrt{S_n(\nu)} = 0.167$\,Hz$^{1/2}$) (middle), $h_0=1 \times 10^{-24}$ and $h_0/\sqrt{S_n(\nu)} = 0.033$\,Hz$^{1/2}$ (bottom).}
	\label{fig:noq1q2}
\end{figure}

Figure \ref{fig:noq1q2} displays the results from the first set of searches, where $\nu_0$ is the only search parameter (i.e. $Q_1=Q_2=0$). The top row displays the results for relatively strong signals ($h_0=1 \times 10^{-23}$, $h_0/\sqrt{S_n(\nu)} = 0.33$\,Hz$^{1/2}$) on the ${Q_1}_\text{signal}-{Q_2}_\text{signal}$ plane. The left panel shows that $\rho/\sigma_\rho$ peaks at $\sim 400$ in the bottom-left corner of the plot and drops dramatically when ${Q_1}_\text{signal}\geq 10^{-19}\,\text{Hz\,s}^{-1}$ and ${Q_2}_\text{signal}\geq 10^{-16}\,\text{Hz\,s}^{-1}$. In the right panel, the frequency at which $\rho/\sigma_\rho$ peaks is lower than ${\nu_0}_\text{signal}=150.1$\,Hz and decreases, as ${Q_1}_\text{signal}$ and ${Q_2}_\text{signal}$ increase. We expect the latter discrepancy; we are searching for a constant-$\nu$ signal, while the injection is spinning down, and the discrepancy grows as ${\dot{\nu}}_\text{signal}$ increases. The middle row of Figure \ref{fig:noq1q2} shows the same thing for weaker signals with $h_0=5 \times 10^{-24}$ and $h_0/\sqrt{S_n(\nu)} = 0.167$\,Hz$^{1/2}$. Here $\rho/\sigma_\rho$ peaks at $\sim 100$, and the frequency where it peaks decreases faster than in the previous case. In the bottom row, with $h_0=1 \times 10^{-24}$ and $h_0/\sqrt{S_n(\nu)} = 0.033$\,Hz$^{1/2}$, the signals are too weak to be detectable. Excluding spin down therefore leads to significant loss in sensitivity, as compared to Section \ref{sec:rhosignal}. 

\begin{figure}
	\centering
	\scalebox{0.16}{\includegraphics{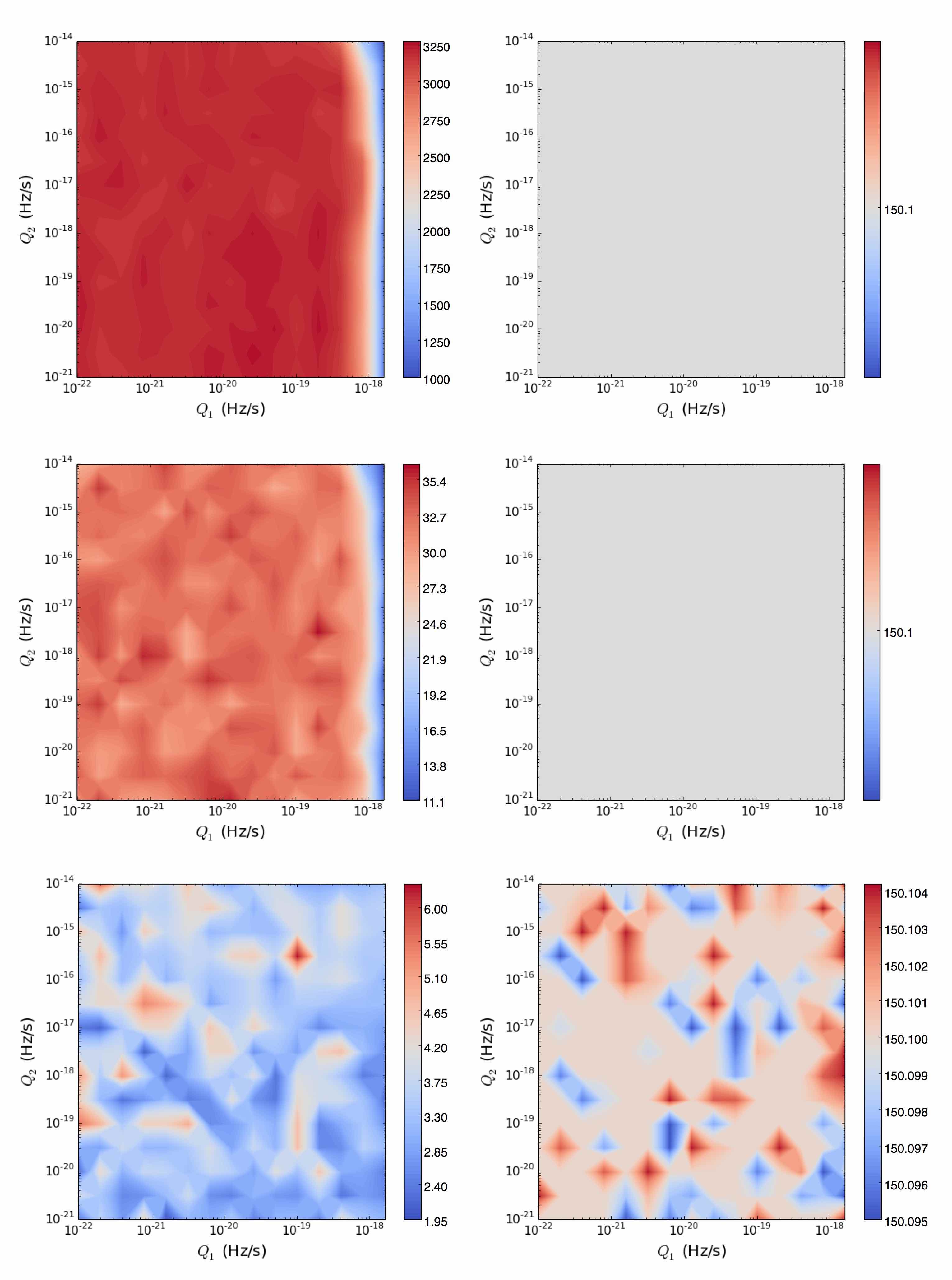}}
	\caption[Contour plots showing the largest $\rho/\sigma_\rho$ returned as functions of $Q_1$ and $Q_2$ values used to create the signal when searching over $\nu_0, Q_1$ and $Q_2$ (left panel), and the corresponding frequency at which the largest $\rho/\sigma_\rho$ is found (right panel). ($h_0=1 \times 10^{-23}$, $h_0/\sqrt{S_n(\nu)} = 0.33$\,Hz$^{1/2}$)]{Maximum $\rho/\sigma_\rho$ returned as a function of ${Q_1}_\text{signal}$ and ${Q_2}_\text{signal}$ values when searching over $\nu_0, Q_1$ and $Q_2$ (left), and the corresponding frequency at which $\rho/\sigma_\rho$ peaks (right) for $h_0=1 \times 10^{-23}$ and $h_0/\sqrt{S_n(\nu)} = 0.33$\,Hz$^{1/2}$ (top), $h_0=1 \times 10^{-24}$ and $h_0/\sqrt{S_n(\nu)} = 0.033$\,Hz$^{1/2}$ (middle), $h_0=3 \times 10^{-25}$ and $h_0/\sqrt{S_n(\nu)} = 0.01$\,Hz$^{1/2}$ (bottom).}
	\label{fig:q1q2}
\end{figure}

Figure \ref{fig:q1q2} displays the results from the second set of searches, where not only $\nu_0$ but also $Q_1$ and $Q_2$ are searched. In the top row ($h_0=1 \times 10^{-23}$, $h_0/\sqrt{S_n(\nu)} = 0.33$\,Hz$^{1/2}$), $\rho/\sigma_\rho$ is larger than in the top row of Figure \ref{fig:noq1q2} (i.e. same $h_0$), reaching as high as $\sim 3.26 \times 10^3$ over a broad range of ${Q_1}_\text{signal}$ and ${Q_2}_\text{signal}$ (${Q_1}_\text{signal} \lesssim 4 \times 10^{-19}\,\text{Hz\,s}^{-1}$ and the whole range of ${Q_2}_\text{signal}$ tested). In the right panel in the top row, the largest $\rho/\sigma_\rho$ always occurs at the injected frequency ${\nu_0}_\text{signal}=150.1$\,Hz. In the middle row ($h_0=1 \times 10^{-24}$, $h_0/\sqrt{S_n(\nu)} = 0.033$\,Hz$^{1/2}$), signals which are undetectable in Figure \ref{fig:noq1q2} remain detectable in Figure \ref{fig:q1q2}. Again $\rho/\sigma_\rho$ peaks at ${\nu_0}_\text{signal}=150.1$\,Hz. In the bottom row ($h_0=3 \times 10^{-25}$, $h_0/\sqrt{S_n(\nu)} = 0.01$\,Hz$^{1/2}$), the signals become lost in the noise in both figures, close to the minimum detectable $h_0$ calculated in Section \ref{sec:rhosignal}.

In summary, we verify that as long as a spinning down signal is strong enough or spins down slowly, it can be detected whether or not $Q_1$ and $Q_2$ are excluded from the search. However, when a signal is weak ($h_0 \lesssim 5 \times 10^{-24}$, $h_0/\sqrt{S_n(\nu)} \lesssim 0.167$\,Hz$^{1/2}$) or the frequency evolves quickly ($|\dot{\nu}| \gtrsim 1 \times 10^{-8}\,\text{Hz\,s}^{-1}$), excluding $Q_1$ and $Q_2$ causes significant loss in sensitivity, enlarging detectable $h_0$ threshold for $\sim 10$ times. 

\subsubsection{Braking index evolution}
\label{sec:brakingindex}
The electromagnetic braking index $n_\text{em}$ for radio pulsars is observed to satisfy $n_\text{em} < 3$ \cite{Livingstone2007}, in contrast with classical magnetic dipole braking ($n_\text{em} = 3$). This raises the possibility that $n_\text{em}$ evolves, as the neutron star spins down, increasing $N_\text{total}$ over and above the already heavy cost of searching over $Q_1$ and $Q_2$. We now quantify how much sensitivity is sacrificed by assuming $n_\text{em}$ to be constant.

Specifically, if we fix $n_\text{em}=3$ in the search, yet the true value is $n_\text{em}=3-\Delta n_\text{em}(t)$, we find that the sensitivity does not change significantly, as long as $T_\text{lag}$ (the maximum interval over which the cross-correlation algorithm requires phase coherence) is smaller than $|\xi|^{-1}T_\text{age}$. Instead, the signal is recovered with similar signal-to-noise ratio but at a modified value of $Q_2$. The result holds if $n_\text{em}$ is constant or evolves slowly on the time-scale $|\xi|^{-1}T_\text{age}$, with the signal location in $Q_2$ evolving on a similar time-scale.

\begin{table}
	\centering
	\setlength{\tabcolsep}{8pt}
	\begin{tabular}{lll}
		\hline
		\hline
		Injection parameter & Value & Units \\
		\hline
		${\nu_0}_\text{signal}$ & 150.1 & Hz \\
		${Q_1}_\text{signal}$ & $3.5 \times 10^{-19}$ & Hz\,s$^{-1}$\\
		${Q_2}_\text{signal}$& $2 \times 10^{-17}$ & Hz\,s$^{-1}$ \\
		$\sqrt{S_n(\nu)}$& $3\times10^{-23}$ & Hz$^{-1/2}$\\
		$h_0$ & $10^{-24}$, $10^{-23}$, $10^{-22}$ & \\
		${n_\text{em}}_\text{signal}$& 2.3, 2.4, 2.5, 2.6, 2.7, 2.8, 2.9, 3.0& \\
		\hline
		\hline
	\end{tabular}
	\caption[signal parameters]{Injection parameters used to create the synthetic data analysed in Section \ref{sec:brakingindex} to study braking index evolution. Three values of $h_0$ and eight values of ${n_\text{em}}_\text{signal}$ are chosen.}
	\label{tab:sig-params-braking-idx}
\end{table}

\begin{table}
	\centering
	\setlength{\tabcolsep}{8pt}
	\begin{tabular}{llll}
		\hline
		\hline
		Search parameter & Range & Resolution & Units \\
		\hline
		$\nu_0$ & $149.6 - 150.6$ & 0.01 & Hz \\
		$Q_1$& $3.0 \times 10^{-19} - 3.9 \times 10^{-19}$ & $0.1 \times 10^{-19}$ &Hz\,s$^{-1}$ \\
		$Q_2$& $1.0 \times 10^{-18} - 2.5 \times 10^{-17}$ & $0.1 \times 10^{-18}$ &Hz\,s$^{-1}$ \\
		\hline
		\hline
	\end{tabular}
	\caption[search parameters]{Search parameter ranges for the targets in Table~\ref{tab:sig-params-braking-idx} with $T_\text{lag}=3600$\,s. The electromagnetic braking index $n_\text{em}=3$ is held fixed in every search.}
	\label{tab:search-params-braking-idx}
\end{table}

\begin{figure}
	\centering
	\scalebox{0.2}{\includegraphics{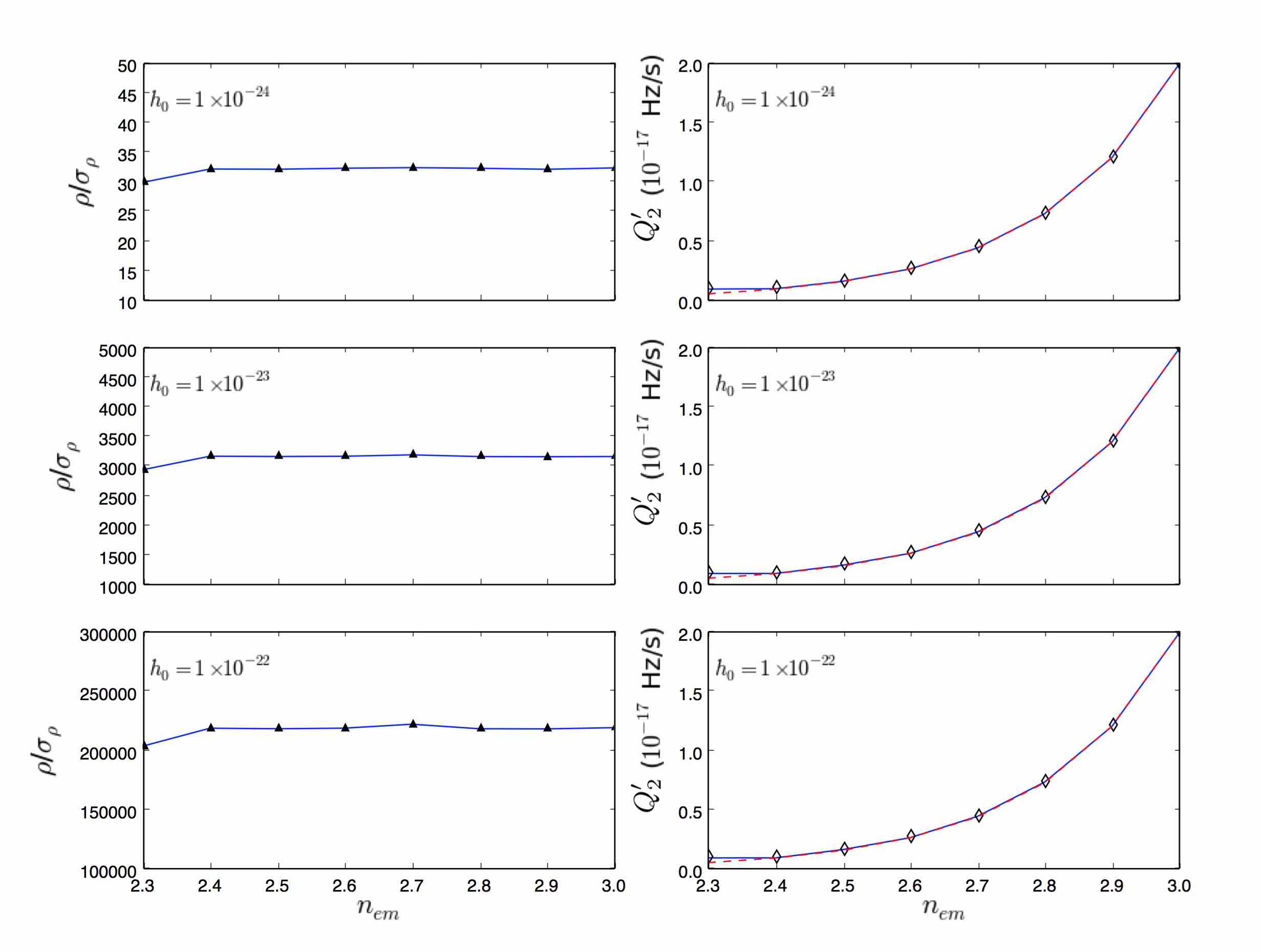}}
	\caption[Detection statistic $\rho/\sigma_\rho$ (left) and equivalent $Q_2'$ (right) as functions of the real electromagnetic braking index $n_\text{em}$]{Maximum $\rho/\sigma_\rho$ (left) and equivalent $Q_2'$ (right) obtained by fixing $n_\text{em}=3$, averaged over 100 trials, as functions of the true, injected electromagnetic braking index ${n_\text{em}}_\text{signal}$ for (top to bottom) $h_0=1 \times 10^{-24}$, $1 \times 10^{-23}$ and $1 \times 10^{-22}$ ($\sqrt{S_n(\nu)}=3\times10^{-23}$\,Hz$^{-1/2}$), with ${\nu_0}_\text{signal}=150.1$\,Hz and ${Q_2}_\text{signal}=2 \times 10^{-17}\,\text{Hz\,s}^{-1}$. The red dashed curves plot the theoretically predicted $Q_2'$ from (\ref{eq:offsetQ2}) as a function of ${n_\text{em}}_\text{signal}$, which mainly overlap the empirical curves.}
	\label{fig:brakingindex}
\end{figure}

Figure \ref{fig:brakingindex} presents results from mock searches demonstrating the behaviour above. We simulate spinning-down signals at three different wave strains whose parameters are quoted in Table \ref{tab:sig-params-braking-idx}, generating one year of 30-min SFTs (from H1 and L1). For each value of $h_0$, we inject signals with eight different values of ${n_\text{em}}_\text{signal}$. The search parameters are quoted in Table \ref{tab:search-params-braking-idx}. The electromagnetic braking index $n_\text{em}=3$ is held fixed in every search. For each value of ${h_0}_\text{signal}$ and ${n_\text{em}}_\text{signal}$, the above test is repeated 100 times. We extract the maximum $\rho/\sigma_\rho$ as well as the corresponding $Q_2'$ value which maximizes $\rho/\sigma_\rho$ from each of the 100 trials, and plot the mean values of $(\rho/\sigma_\rho)_\text{max}$ and $Q_2'$ as functions of ${n_\text{em}}_\text{signal}$ in Figure \ref{fig:brakingindex}. The variation in $(\rho/\sigma_\rho)_\text{max}$ is modest over the full range of ${n_\text{em}}_\text{signal}$, with $(\rho/\sigma_\rho)_\text{max} \simeq 32$, $3.2 \times 10^3$, $2.2 \times 10^5$ for ${h_0}_\text{signal} = 10^{-24}$, $10^{-23}$, $10^{-22}$ respectively. The reason why $\rho/\sigma_\rho$ is always relatively lower for ${n_\text{em}}_\text{signal}\simeq 2.3$ is that $\rho/\sigma_\rho$ peaks at a smaller $Q_2'$ value than the smallest searched. We do not expand the $Q_2$ band to such a small value because that introduces a finer resolution and thus require much larger number of templates. We also find that the $Q_2'$ which maximizes $\rho/\sigma_\rho$ shifts relative to the injected value according to
\begin{equation}
\label{eq:offsetQ2}
Q_2' = \nu_0^{-\Delta n_\text{em}}{Q_2}_\text{signal}, 
\end{equation}
as expected from Taylor expanding (\ref{eq:numodel}) in $|\xi|T_\text{lag}/T_\text{age} \ll 1$. The red dashed curves overplotted in the right panels of Figure \ref{fig:brakingindex} display the theoretically predicted $Q_2'$ values as a function of ${n_\text{em}}_\text{signal}$ from equation (\ref{eq:offsetQ2}) at ${\nu_0}_\text{signal}=150.1$\,Hz and ${Q_2}_\text{signal}=2 \times 10^{-17}\,\text{Hz\,s}^{-1}$. They are consistent with the empirical results. This fact makes it possible to fix $n_\text{em}=3$ in the search, taking only $Q_1$ and $Q_2$ as spin-down variables and reducing $N_\text{total}$ without sacrificing sensitivity.

\section{Sensitivity}
\label{sec:sensitivity}
In this section, we present Monte-Carlo tests to determine the smallest gravitational wave signal detectable by the pipeline in Section \ref{sec:pipeline}. Specifically, we determine empirically the value of $h_0^{\alpha_\text{c}}$, for which a fraction $\alpha_\text{c}$ (normally $\alpha_\text{c}=0.95$) of the Monte-Carlo trials yield $\rho/\sigma_\rho \geq \rho_\text{th}$, where $\rho_\text{th}$ is the agreed detection threshold. We discuss the choice of $\rho_\text{th}$ in Section \ref{sec:sensiA} and estimate $h_0^{\alpha_\text{c}}$ in Sections \ref{sec:sensiB}--\ref{sec:sensiD}. 

\subsection{Threshold $\rho_\text{th}$}
\label{sec:sensiA}
For a given false alarm rate $\alpha_\text{f}$, $\rho_\text{th}$ is estimated as follows. SFTs containing pure noise are generated, and a search is run over signal parameters (i.e. $\nu_0, Q_1, Q_2$, and $n_\text{em}$). The value of $\rho/\sigma_\rho$ which yields a fraction $\alpha_\text{f}$ of positive detections is then $\rho_\text{th}$. In this paper, we consider $\alpha_\text{f} = 1$\%. Specifically, for $10^3$ searches over pure noise, we adjust $\rho_\text{th}$ such that 10 trials have $\rho/\sigma_\rho > \rho_\text{th}$. 

\begin{figure}
\centering
\scalebox{0.16}{\includegraphics{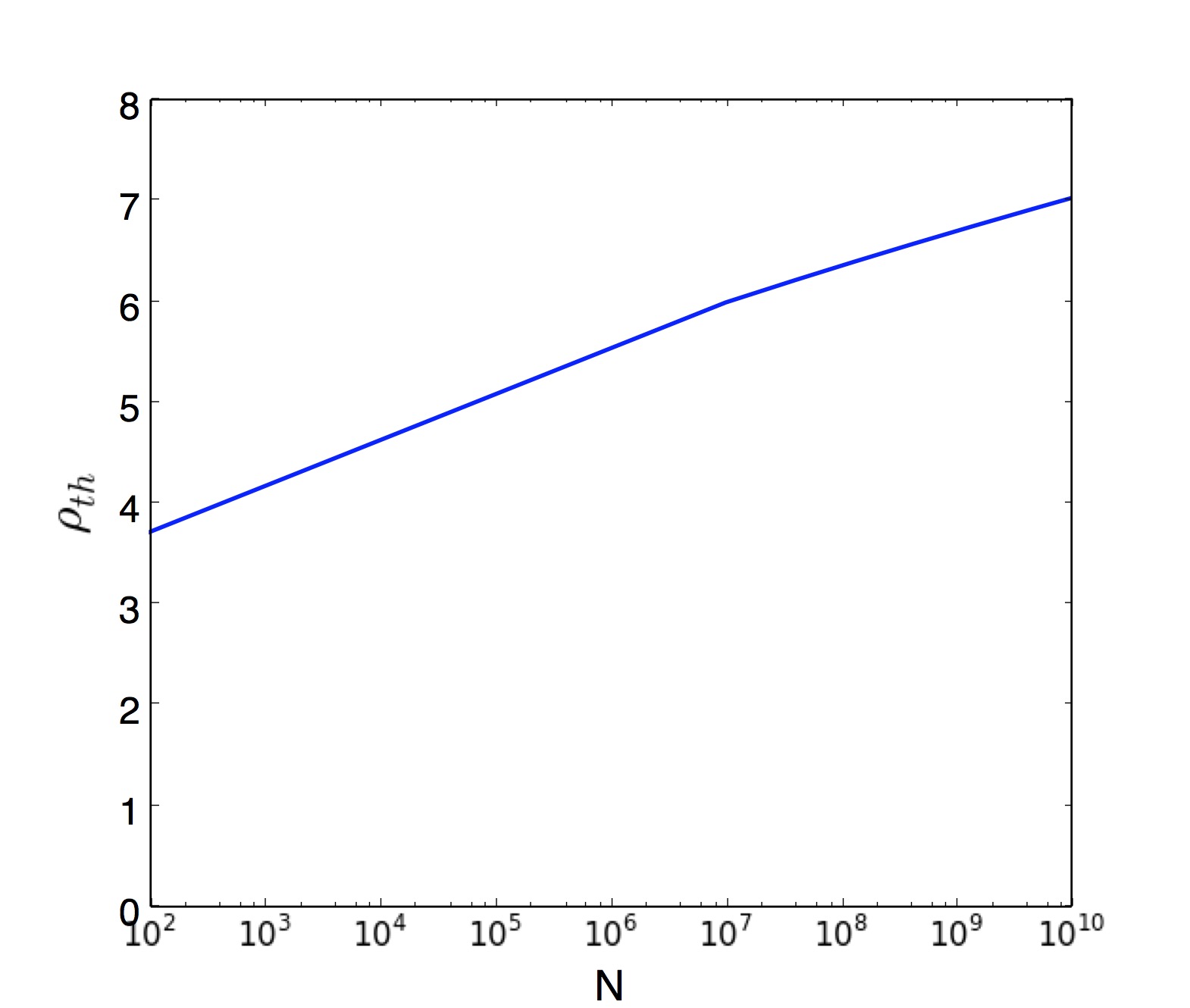}}
\caption[Analytic estimates for $\rho_\text{th}$ as a function of $N$.]{Analytic detection threshold $\rho_\text{th}$ as a function of $N_\text{total}$ from equation (\ref{eq:cdf}) with $\alpha_\text{f}=1$\% and $\sigma_\rho = 1$.}
\label{fig:rhothanalytic}
\end{figure}

An analytic expression for $\rho_\text{th}$ (before normalizing by $\sigma_\rho$) given $\alpha_\text{f}$ and $\sigma_\rho$ is presented by \citet{dhurandhar08}, viz.
\begin{equation}
\label{eq:rhoth}
\rho_\text{th} = \sqrt{2} \sigma_\rho \text{erfc}^{-1}(2 \alpha_\text{f} / N_\text{total}),
\end{equation}
where erfc is the complementary error function, and $N_\text{total}$ is the number of search templates. As shown in Section \ref{sec:noise}, for pure noise, the PDF of $\rho/\sigma_\rho$ is a Gaussian with zero mean and unit variance, assuming that all pairs are independent. As foreshadowed in Section \ref{sec:noise} and \ref{sec:rhosignal}, we do not discuss the non-Gaussian corrections caused by dependent pairs in this paper, because they are negligible for $T_{\rm obs}=1$\,yr; for more details see Ref.~\cite{Coyne2016}. Hence the threshold reduces to
\begin{equation}
\label{eq:cdf}
\rho_\text{th} \approx F^{-1}[1 - (\alpha_\text{f}/N_\text{total})],
\end{equation}
where $F^{-1}(x)$ is the inverse cumulative distribution function (CDF) of $x$. Figure \ref{fig:rhothanalytic} plots $\rho_\text{th}$ as a function of $N_\text{total}$ from equation (\ref{eq:cdf}); it ranges from 3.72 for $N = 10^2$ to 7.03 for $N = 10^{10}$.

Searches without and with spin-down require different numbers of templates.

\subsubsection{Pure noise, zero-spin-down search}
The signal power is concentrated within one frequency bin when searching for a zero spin-down signal. We therefore search 0.1-Hz bands centred on 150.05, 300.05, 600.05\,Hz respectively with a resolution of $10^{-4}$\,Hz using SFTs containing only noise. Each trial consists of $N_\text{total} = 10^3$ search templates. For $\alpha_\text{f} = 1$\% and $\sigma_\rho=1$, equation (\ref{eq:cdf}) yields $\rho_\text{th} = 4.265$. For each band searched, we adjust $\rho_\text{th}$ such that it is exceeded by only 10 out of $10^3$ of the $\rho/\sigma_\rho$ values. Table \ref{tab:thresholdnosd} lists $\rho_\text{th}$ for the three bands. The result in each band agrees with the analytic value to better than $6\%$. 

\begin{table}
\centering
\setlength{\tabcolsep}{8pt}
\begin{tabular}{cc}
\hline
\hline
Band\,(Hz) & $\rho_\text{th}$\\
\hline
150.0--150.1 & 4.440\\
300.0--300.1 &  4.433\\
600.0--600.1 &  4.516\\
\hline
\hline
\end{tabular}
\caption[$\rho_\text{th}$ for each frequency band used in the Monte Carlo searches over pure noise, without spin down.]{Monte-Carlo detection threshold $\rho_\text{th}$ for three 0.1-Hz bands when searching pure noise, assuming zero spin down, $N_\text{total}=10^3$, and $\alpha_\text{f}=1\%$. Equation (\ref{eq:cdf}) yields $\rho_\text{th}=4.265$ analytically.}
\label{tab:thresholdnosd}
\end{table}

\subsubsection{Pure noise, spin-down search}
\label{sec:rhothwithsp}
Searching for a spinning-down signal involves more parameters (i.e. $Q_1,Q_2,n_\text{em}$) and hence larger $N_\text{total}$. Here we consider the parameter space defined in Table \ref{tab:search-params-rho-noise}. Each trial consists of $N_\text{total} = 100 \times 7 \times 11 = 7700$ search templates. The analytic estimate from equation (\ref{eq:cdf}) yields $\rho_\text{th} = 4.700$. Results from the Monte-Carlo tests, adjusting $\rho_\text{th}$ to be exceeded by only 10 out of $10^3$ of the $\rho/\sigma_\rho$ values, yield $\rho_\text{th} = 4.898$, which is 4\% larger than the analytic estimate.

\begin{table}[!tbh]
	\centering
	\setlength{\tabcolsep}{8pt}
	\begin{tabular}{llll}
		\hline
		\hline
		Search parameter & Range & Resolution & Units \\
		\hline
		$\nu_0$ & $150.0 - 150.1$ & $10^{-3}$ & Hz \\
		$Q_1$& $1.0 \times 10^{-19} - 7.0 \times 10^{-19}$ & $1 \times 10^{-19}$ &Hz\,s$^{-1}$ \\
		$Q_2$& $1.0 \times 10^{-18} - 1.1 \times 10^{-17}$ & $1 \times 10^{-18}$ &Hz\,s$^{-1}$ \\
		\hline
		\hline
	\end{tabular}
	\caption[search parameters]{Search parameter ranges used to estimate the threshold for a spin-down search in Section \ref{sec:rhothwithsp} with $n_\text{em}=3$ fixed. The data contain Gaussian noise ($h_0=0$). The ranges of $Q_1$ and $Q_2$ correspond to $7.6 \times 10^{-9}$\,Hz s$^{-1} \leq \lvert \dot{\nu} \rvert \leq 5.3 \times 10^{-8}$\,Hz s$^{-1}$, $2.4 \times 10^{-4} \leq \epsilon \leq 6.4 \times 10^{-4}$, and $1.3 \times 10^{11}$\,G $\leq B \leq 4.2 \times 10^{11}$\,G.}
	\label{tab:search-params-rho-noise}
\end{table}

\subsection{Sensitivity for zero-spinning-down signals}
\label{sec:sensiB}
Without considering the spin down of a signal, we determine $h_0^{95\%}$ using $\rho_\text{th}$ from Table \ref{tab:thresholdnosd}.

\begin{figure*}
	\centering
	\subfigure[]
	{
		\label{fig:sensi-nosp-a}
		\scalebox{0.14}{\includegraphics{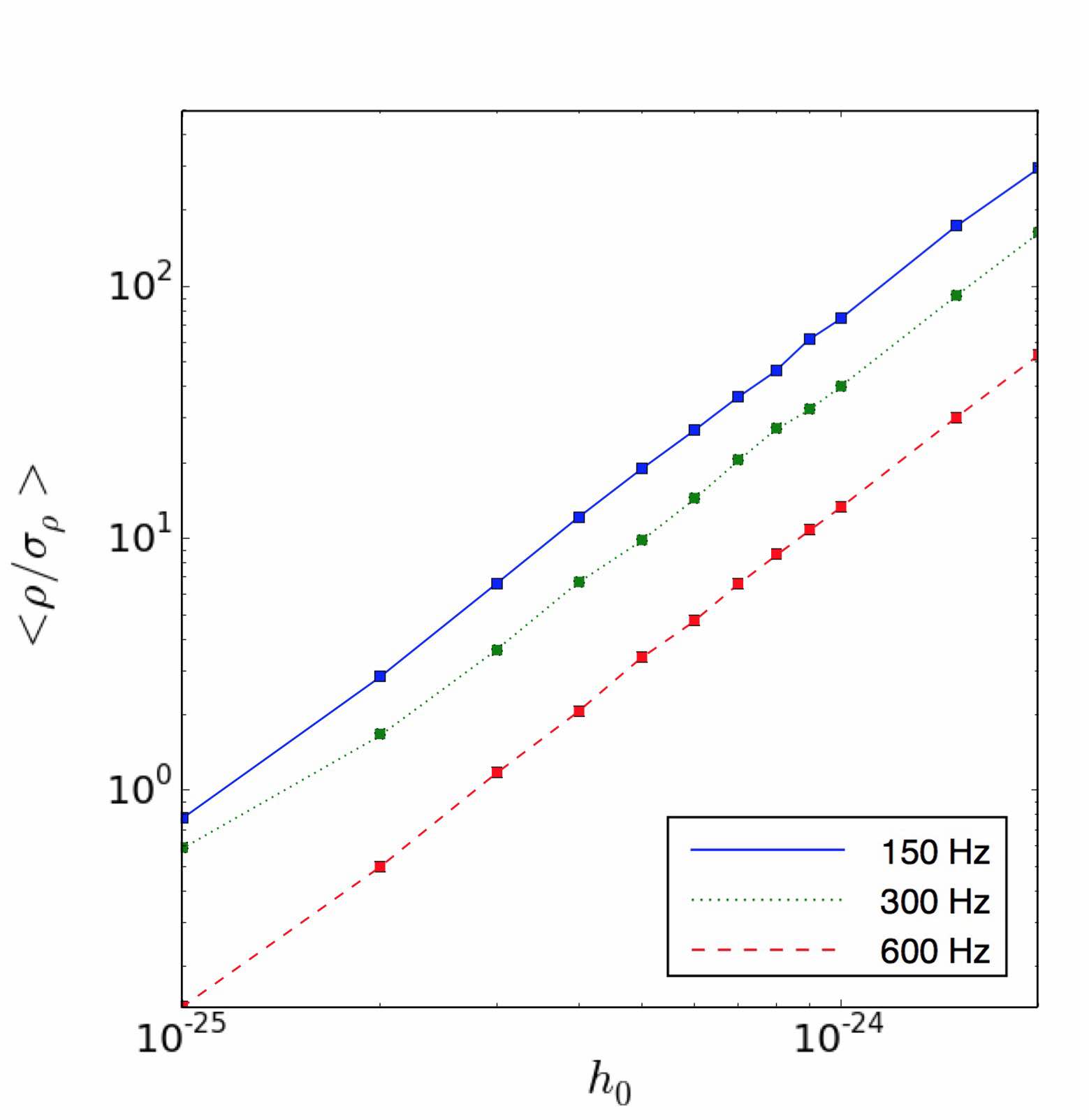}}
	}
	\subfigure[]
	{
		\label{fig:sensi-nosp-b}
		\scalebox{0.14}{\includegraphics{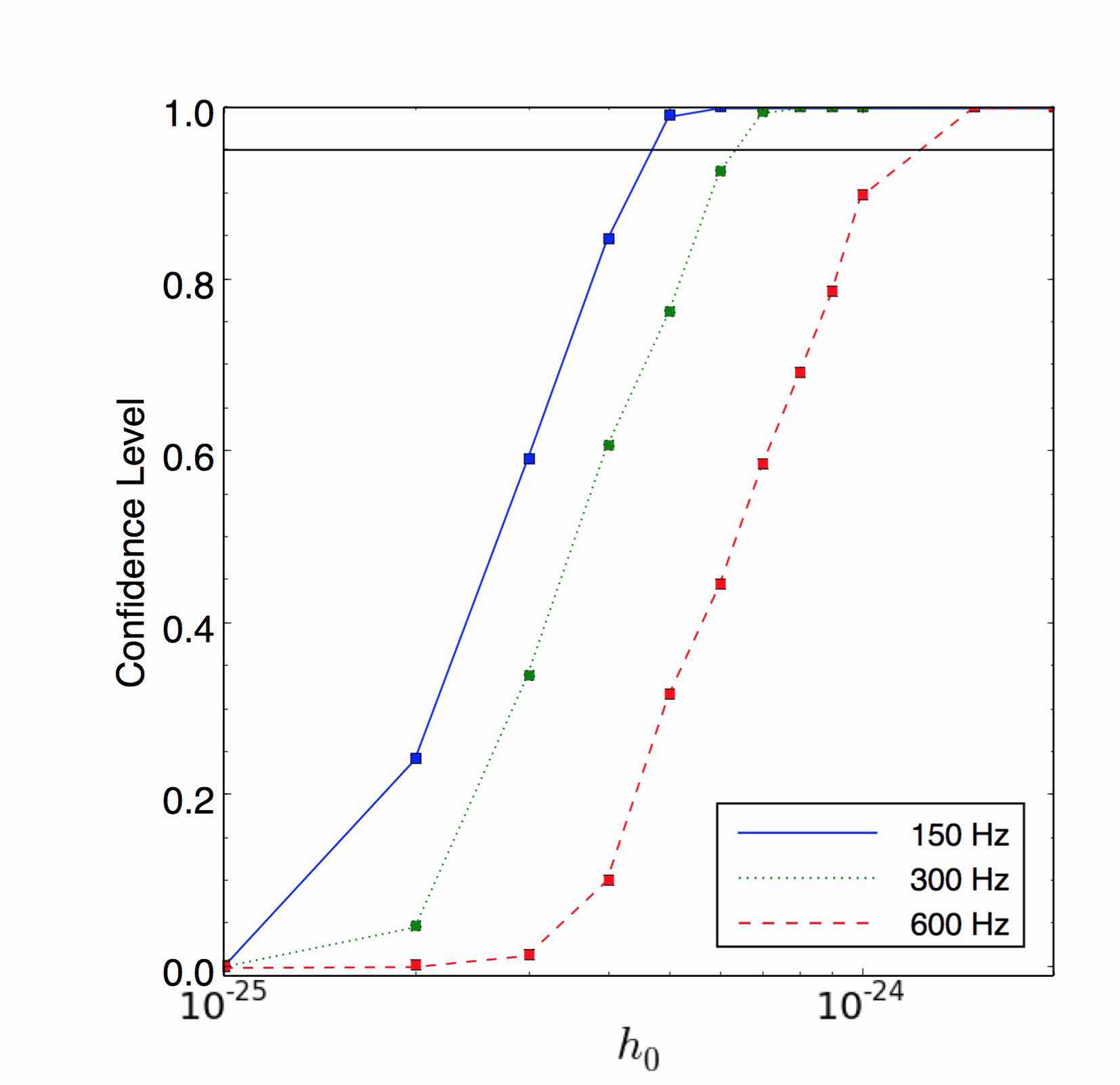}}
	}
	\caption[Normalized detection statistic $\langle \rho/\sigma_\rho \rangle$ averaged over $10^3$ confidence level as functions of $h_0$.]{Sensitivity without spin down. (a) Normalized detection statistic $\langle \rho/\sigma_\rho \rangle$ averaged over $10^3$ trials, and (b) confidence level $C$, as functions of injected gravitational wave strain $h_0$. The injected signals have random ${\cos \iota}_\text{signal}$ and $\psi_\text{signal}$, fixed sky positions ($\alpha, \delta$) = (1.46375\,rad, $-$1.20899\,rad), and ${\nu_0}_\text{signal}=150.05$\,Hz (solid curve), 300.05\,Hz (dotted curve), and 600.05\,Hz (dashed curve). For each injected frequency, $\rho_\text{th}$ is listed in Table \ref{tab:thresholdnosd} for $N_\text{total}=10^3$. The horizontal line in (b) indicates $C = 0.95$.}
	\label{fig:sensi-nosp}
\end{figure*}

We inject signals with constant ${\nu_0}_\text{signal} = 150.05$, 300.05, and 600.05\,Hz and strains ranging between $1 \times 10^{-25} \leq h_0 \leq 2 \times 10^{-24}$ into one year of 30-min SFTs from H1 and L1, with signal parameters $(\alpha, \delta)$ = (1.46375, $-$1.20899) (the coordinates of SNR 1987A), random ${\cos \iota}_\text{signal}$ and $\psi_\text{signal}$, and $S_n(\nu)$ given by equation (\ref{eq:psd}). We then search a 0.1-Hz band centred on ${\nu_0}_\text{signal}$ with a resolution of $10^{-4}$\,Hz, using the exact sky position $(\alpha, \delta)$ and averaging over $\cos \iota$ and $\psi$ ($T_\text{lag}=3600$\,s). The normalized detection statistic $\langle \rho/\sigma_\rho \rangle$ averaged over $10^3$ trials is plotted as a function of $h_0$ in Figure \ref{fig:sensi-nosp-a}. The solid, dotted, and dashed lines correspond to ${\nu_0}_\text{signal} = $ 150.05\,Hz, 300.05\,Hz, and 600.05\,Hz respectively. As expected, $\langle \rho/\sigma_\rho \rangle$ grows $\propto h_0^2$ from equation (\ref{eq:generalnormmu}) for a given ${\nu_0}_\text{signal}$, and it drops when ${\nu_0}_\text{signal}$ and hence $\sqrt{S_n}$ increase. In Figure \ref{fig:sensi-nosp-b}, we plot the confidence level $C$ (i.e. the fraction of $\rho/\sigma_\rho$ values, in each set of $10^3$ trials, which exceed $\rho_\text{th}$) as a function of $h_0$. Linear interpolation in Figure \ref{fig:sensi-nosp-b} implies that $C$ increases to $\geq 95\%$ for $h_0 \geq h_0^{95\%}$ for the $h_0^{95\%}$ values listed in Table \ref{tab:sensi-no-spindown}.

The results interpolated from Figure \ref{fig:sensi-nosp-b} are for a search over $10^3$ templates. The full search involves $\sim 10^9$ templates, corresponding to $\rho_\text{th} \approx 6.71$ from (\ref{eq:cdf}). The estimated strain limits, which are $\sim 20\%$ larger, appear in the lower half of Table \ref{tab:sensi-no-spindown}.

\begin{table}[!tbh]
	\centering
	\setlength{\tabcolsep}{8pt}
	\begin{tabular}{lll}
		\hline
		\hline
		$N_\text{total}$ & ${\nu_0}_\text{signal}$ (Hz)& $h_0^{95\%}$\\
		\hline
		$10^3$ & 150.05 &  $4.72 \times 10^{-25}$\\
		&300.05 &  $6.36 \times 10^{-25}$\\
		&600.05 &  $1.26 \times 10^{-24}$\\
		\hline
		$10^9$ & 150.05 &  $5.64 \times 10^{-25}$\\
		&300.05 &  $7.60 \times 10^{-25}$\\
		&600.05 &  $1.42 \times 10^{-24}$\\
		\hline
		\hline
	\end{tabular}
	\caption[]{Wave strain threshold $h_0^{95\%}$ (i.e. confidence $C\geq 0.95$) estimated for three 0.1-Hz bands containing pure Gaussian noise. For $N_\text{total}=10^3$, the thresholds are estimated from linear interpolation of Monte-Carlo simulation results plotted in Figure \ref{fig:sensi-nosp-b}. For $N_\text{total}=10^9$, the thresholds are based on analytical estimation from equation (\ref{eq:cdf}).}
	\label{tab:sensi-no-spindown}
\end{table}

\subsection{Sensitivity for spinning-down signals}
\label{sec:sensiC}

\begin{figure*}
	\centering
	\subfigure[]
	{
		\label{fig:sensi-sp-vary-h0-a}
		\scalebox{0.36}{\includegraphics{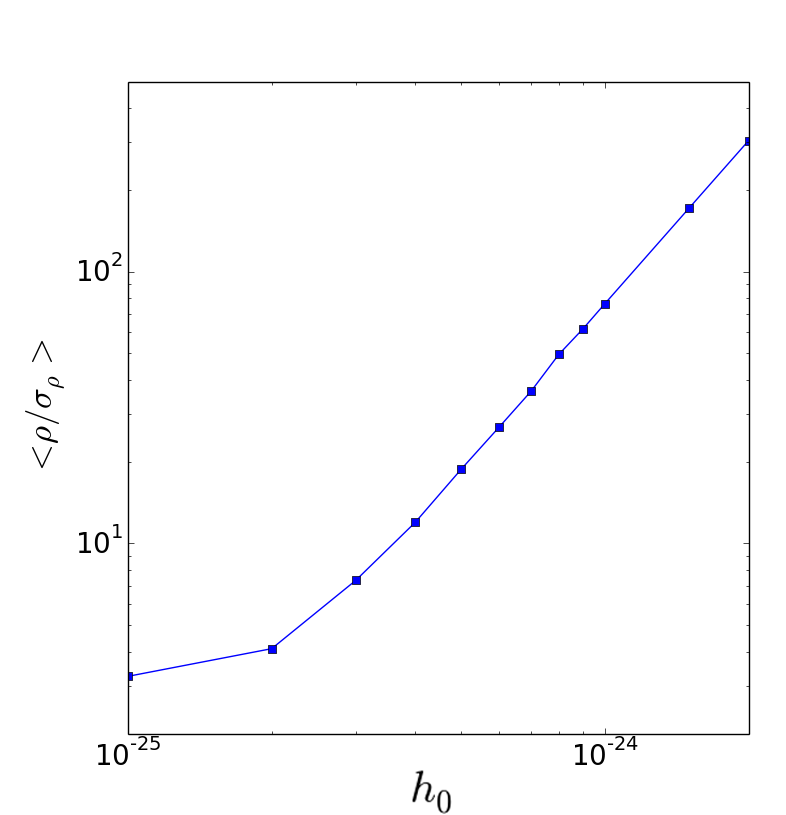}}
	}
	\subfigure[]
	{
		\label{fig:sensi-sp-vary-h0-b}
		\scalebox{0.36}{\includegraphics{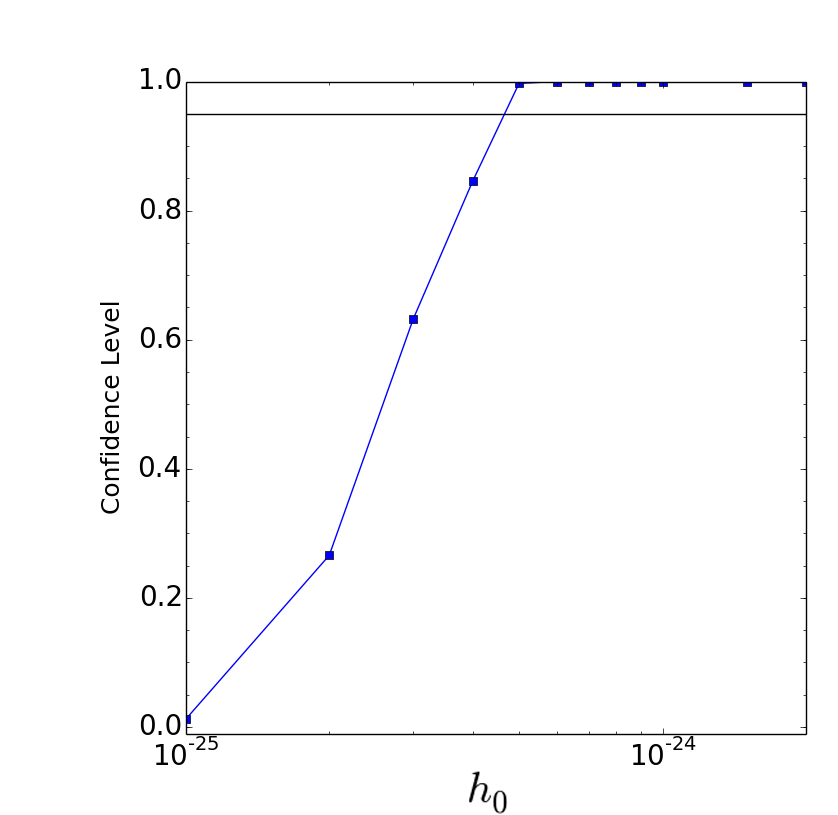}}
	}
	\caption[with spin down]{Sensitivity with spin down (${\nu_0}_\text{signal} = 150.05$\,Hz, ${Q_1}_\text{signal} = 5 \times 10^{-19}\,\text{Hz\,s}^{-1}$,${Q_2}_\text{signal} = 1 \times 10^{-17}\,\text{Hz\,s}^{-1}$). (a) Normalized detection statistic $\langle \rho/\sigma_\rho \rangle$ averaged over $10^3$ trials, and (b) confidence level $C$, as functions of injected gravitational wave strain $h_0$. The injected signals have random ${\cos \iota}_\text{signal}$ and $\psi_\text{signal}$ and fixed sky positions ($\alpha, \delta$) = (1.46375\,rad, $-$1.20899\,rad). The horizontal line in (b) indicates $C = 0.95$.}
	\label{fig:spin-down-no-limit-h0}
\end{figure*}

We inject spin-down signals with the parameters quoted in Table \ref{tab:sensi-sig-paras}. The wave strain is still ranging between $1 \times 10^{-25} \leq h_0 \leq 2 \times 10^{-24}$, and all other parameters remain the same as those in Section \ref{sec:sensiB}. The same searches with $10^3$ templates are carried out and the normalized detection statistics averaged over $10^3$ trials as well as the confidence levels are shown in Figure \ref{fig:spin-down-no-limit-h0}. Despite the signal spin down, the values of $\langle \rho/\sigma_\rho \rangle$ (Figure \ref{fig:sensi-sp-vary-h0-a}) and $C$ (Figure \ref{fig:sensi-sp-vary-h0-b}) are close to those plotted in Figure \ref{fig:sensi-nosp} at the same $h_0$. And hence we find a similar $h_0^{95\%} \sim 4.72 \times 10^{-25}$.

\begin{table}
	\centering
	\setlength{\tabcolsep}{8pt}
	\begin{tabular}{lll}
		\hline
		\hline
		Injection parameter & Value & Astrophysical parameter \\
		\hline
		${\nu_0}_\text{signal}$ & 150.05\,Hz& \\
		${Q_1}_\text{signal}$ & $5 \times 10^{-19}$\,Hz\,s$^{-1}$ & $\epsilon = 5.4 \times10^{-4}$ \\
		${Q_2}_\text{signal}$& $1 \times 10^{-17}$\,Hz\,s$^{-1}$ & $B=4.05 \times 10^{11}$\,G\\
		\hline
		\hline
	\end{tabular}
	\caption[signal parameters]{Injection parameters used to create the synthetic data analysed in Section \ref{sec:sensiC} containing spinning-down signals. The injection parameters ${Q_1}_\text{signal}$ and ${Q_1}_\text{signal}$ are computed from the astrophysical parameters using equations (\ref{eq:spindownmodel_full}) and (\ref{eq:numodel}). The corresponding initial spin-down rate $|\dot{\nu}(0)|$ is $3.81 \times 10^{-8}$\,Hz s$^{-1}$.}
	\label{tab:sensi-sig-paras}
\end{table}

\subsection{Sensitivity for spinning-down signals with limitation on $h_0$}
\label{sec:sensiD}
The search sensitivity depends on two factors, (1) spin-down rate (i.e. combination of spin-down parameters $\nu_0$, $Q_1$, $Q_2$ and $n_\text{em}$) and (2) wave strain $h_0$. Firstly, the cross-correlation pipeline tracks up to $\nu^{(6)}(t)$ terms using the astrophysical model so that the search is most sensitive for the regime with spin-down rate $|\dot{\nu}(0)|\lesssim 10^{-7}\,\text{Hz\,s}^{-1}$, above which the sensitivity starts to drop quickly because the error in tracked signal phase increases to $\gtrsim \pi/2$ after one year's observation. Second, given $\epsilon$ and $\nu$, the gravitational wave strain at Earth is \cite{jara98}
 \begin{equation}
 \label{eqn:h0_epsilon}
 h_0=\frac{4 \pi^2 G}{c^4}\frac{I \epsilon \nu^2}{D}.
 \end{equation}
A stronger signal indicates larger $\epsilon$ and $\nu_0$ and hence higher spin-down rate which inversely decreases the sensitivity.
 
\begin{figure*}
	\centering
	\subfigure[]
	{
		\label{fig:Q1-sp-a}
		\scalebox{0.15}{\includegraphics{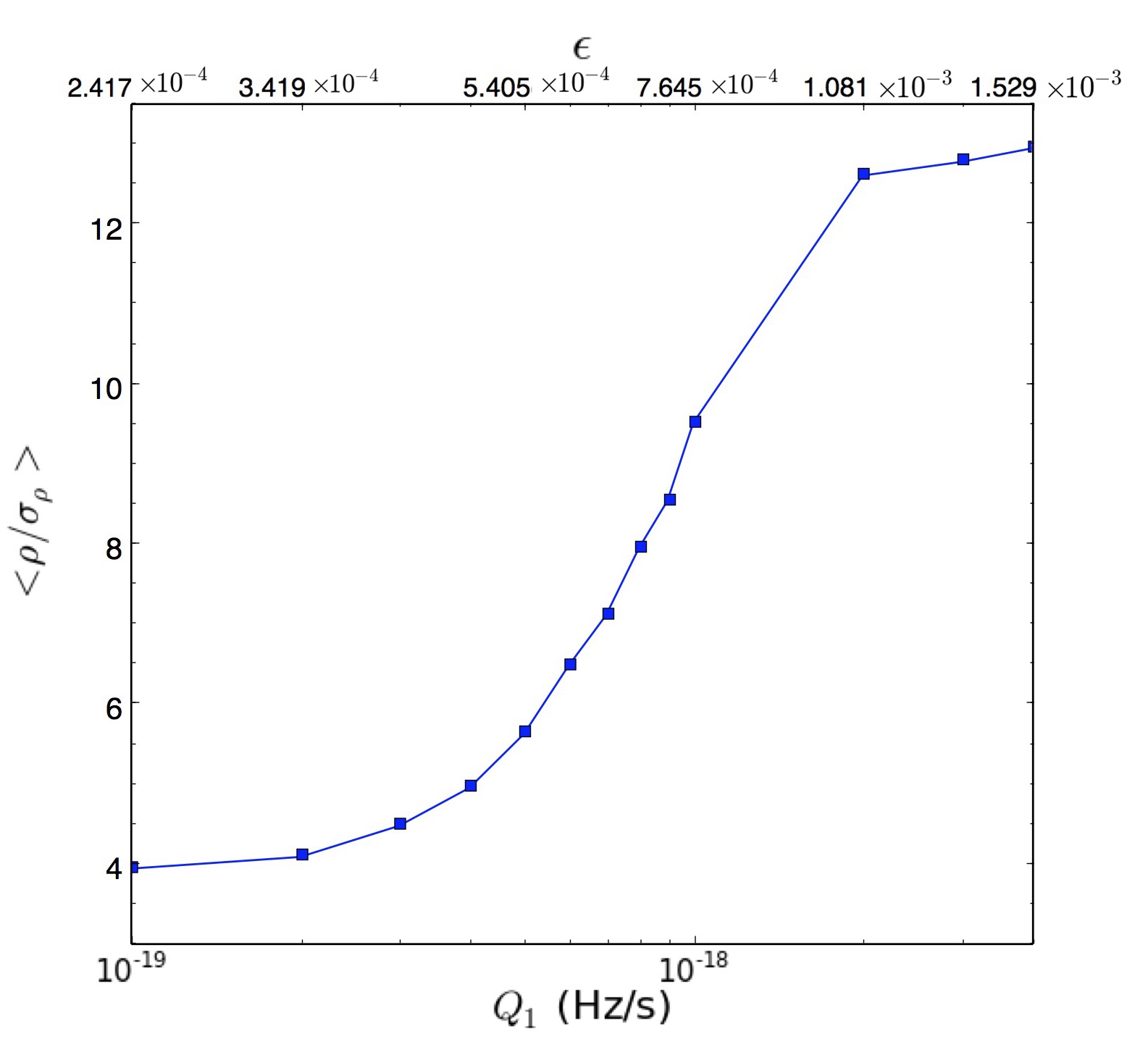}}
	}
	\subfigure[]
	{
		\label{fig:Q1-sp-b}
		\scalebox{0.15}{\includegraphics{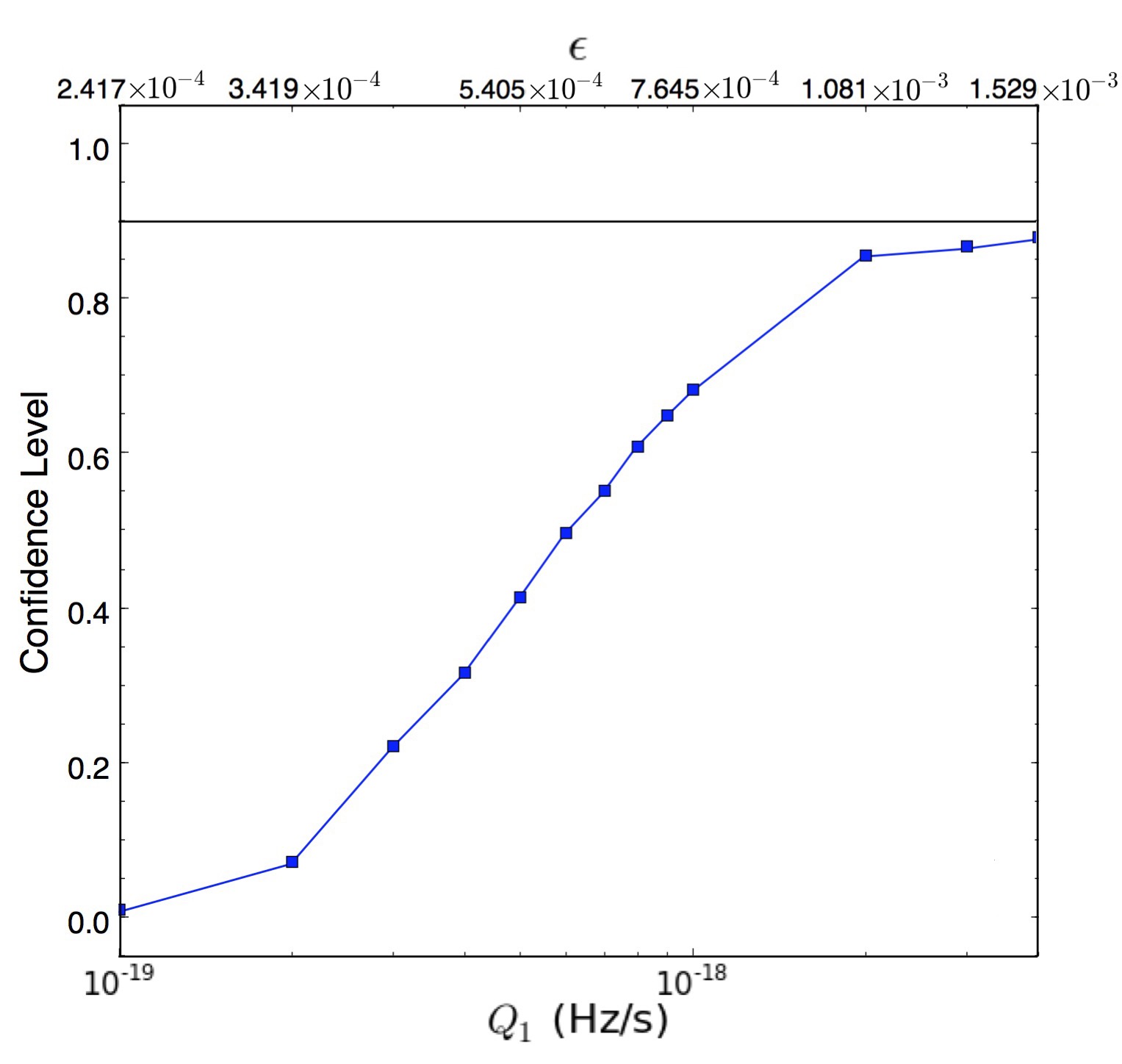}}
	}
	\caption[Normalized detection statistic $\langle \rho/\sigma_\rho \rangle$ averaged over $10^3$ trials and confidence level $C$ as functions of ${Q_1}_\text{signal}$ and $\epsilon$.]{Sensitivity with spin down. (a) Normalized detection statistic $\langle \rho/\sigma_\rho \rangle$ averaged over $10^3$ trials and (b) confidence level $C$ as functions of ${Q_1}_\text{signal}$ (bottom axis) and $\epsilon$ (top axis). The injected signals have random ${\cos \iota}_\text{signal}$ and $\psi_\text{signal}$, fixed sky positions ($\alpha, \delta$) = (1.46375\,rad, $-$1.20899\,rad), ${\nu_0}_\text{signal}=150.05$\,Hz and ${Q_2}_\text{signal}=1 \times 10^{-17}\,\text{Hz\,s}^{-1}$. From Section \ref{sec:rhothwithsp}, we set $\rho_\text{th}=4.898$ for $N_\text{total}=7700$. The horizontal line in (b) indicates $C = 0.90$.}
	\label{fig:Q1-sp}
\end{figure*}

We first inject spin-down signals with the parameters quoted in Table \ref{tab:sensi-sig-paras-D1} into one year of 30-min SFTs from H1 and L1. Wave strain ranges $1 \times 10^{-25} \lesssim h_0 \lesssim 7 \times 10^{-25}$ by using equation (\ref{eqn:h0_epsilon})\footnote{For comparison, we have $h_0=2.5 \times 10^{-25}$ computed from equation (\ref{eqn:h0_epsilon}) using parameters quoted in Table \ref{tab:sensi-sig-paras} in Section \ref{sec:sensiC}, which is relatively low compared to the range $1 \times 10^{-25} \lesssim h_0 \lesssim 7 \times 10^{-25}$ we test in Section \ref{sec:sensiD}.}. We search the parameter ranges in Table \ref{tab:sensi-search-paras-D}. We have $\rho_\text{th}=4.898$ for a signal with ${\nu_0}_\text{signal}=150.05$\,Hz in Section \ref{sec:rhothwithsp}, given $N_\text{total}=7700$. Figure \ref{fig:Q1-sp-a} plots the normalized detection statistic $\langle \rho/\sigma_\rho \rangle$ averaged over $10^3$ trials as a function of ${Q_1}_\text{signal}$ (bottom axis) and corresponding $\epsilon$ (top axis). We plot the confidence level $C$ as a function of ${Q_1}_\text{signal}$ and $\epsilon$ in Figure \ref{fig:Q1-sp-b}. The confidence level $C$ increases with ${Q_1}_\text{signal}$ and $\epsilon$ but saturates at $\sim 0.9$ for ${Q_1}_\text{signal} \gtrsim 2 \times 10^{-18}\,\text{Hz\,s}^{-1}$ ($|\dot{\nu}(0)| \sim 1.5 \times 10^{-7}\text{Hz\,s}^{-1}$,). This result is consistent with our expectation that, as $h_0$ increases, larger $\epsilon$ and $\nu_0$ (i.e. higher spin-down rate) lead to difficulty in phase tracking and prevent achieving better sensitivity.

\begin{table}
	\centering
	\setlength{\tabcolsep}{8pt}
	\begin{tabular}{lll}
		\hline
		\hline
		Injection parameter & Value & Astrophysical parameter \\
		\hline
		${\nu_0}_\text{signal}$ & 150.05\,Hz& \\
		${Q_1}_\text{signal}$ & $1 \times 10^{-19} - 4 \times 10^{-18}$\,Hz\,s$^{-1}$ & $2.4 \times 10^{-4} \leq \epsilon \leq 1.5 \times 10^{-3}$\\
		${Q_2}_\text{signal}$& $1 \times 10^{-17}$\,Hz\,s$^{-1}$ & $B=4.05 \times 10^{11}$\,G\\
		${n_\text{em}}_\text{signal}$&3&\\
		\hline
		\hline
	\end{tabular}
	\caption[signal parameters]{Injection parameters used to create the first set of synthetic data analysed in Section \ref{sec:sensiD} containing spinning-down signals, in which ${Q_2}_\text{signal}$ is fixed and a group of ${Q_1}_\text{signal}$ values are tested. The corresponding astrophysical parameters $\epsilon$ and $B$ in equation (\ref{eq:spindownmodel_full}) are also quoted in the last column. The wave strain covers the range $1 \times 10^{-25} \lesssim h_0 \lesssim 7 \times 10^{-25}$ as calculated from equation (\ref{eqn:h0_epsilon}).}
	\label{tab:sensi-sig-paras-D1}
\end{table}

\begin{table}
	\centering
	\setlength{\tabcolsep}{8pt}
	\begin{tabular}{llll}
		\hline
		\hline
		Search parameter & Range width & Resolution & Units \\
		\hline
		$\nu_0$ & 0.1 & $10^{-3}$ & Hz \\
		$Q_1$& $7 \times 10^{-19}$ & $1 \times 10^{-19}$ &Hz\,s$^{-1}$ \\
		$Q_2$& $1.1 \times 10^{-17}$ & $1 \times 10^{-18}$ &Hz\,s$^{-1}$ \\
		\hline
		\hline
	\end{tabular}
	\caption[search parameters]{Search parameter ranges for injected spin-down signals in Table \ref{tab:sensi-sig-paras-D1} and \ref{tab:sensi-sig-paras-D2}. The ranges are centred on the injected signals. The range width column defines the domain of the search parameter assuming that it is centred on the injection.}
	\label{tab:sensi-search-paras-D}
\end{table}

\begin{figure}[!htb]
	\centering
	\scalebox{0.17}{\includegraphics{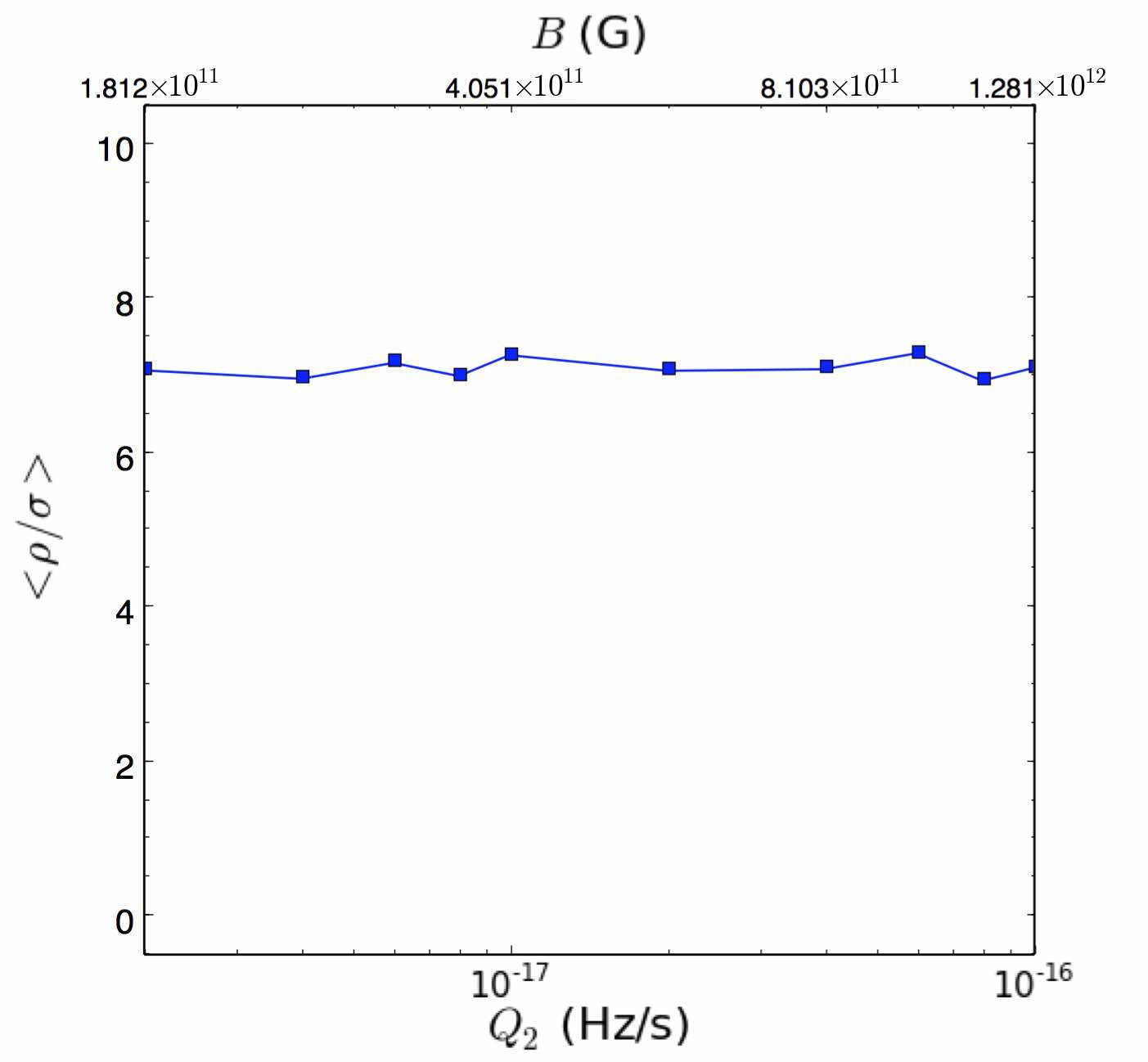}}
	\caption[Normalized detection statistic $\langle \rho/\sigma_\rho \rangle$ averaged over $10^3$ trials as a function of $Q_2$ and $B$.]{Normalized detection statistic $\langle \rho/\sigma_\rho \rangle$ averaged over $10^3$ trials as a function of $Q_1$ (bottom axis) and magnetic field strength $B$ (G) (top axis). The injected signals have random ${\cos \iota}_\text{signal}$ and $\psi_\text{signal}$, fixed sky positions ($\alpha, \delta$) = (1.46375\,rad, $-$1.20899\,rad), ${\nu_0}_\text{signal}=150.05$\,Hz and ${Q_1}_\text{signal}=7 \times 10^{-19}\,\text{Hz\,s}^{-1}$.}
	\label{fig:Q2-sp}
\end{figure}

Next we inject signals with the parameters quoted in Table \ref{tab:sensi-sig-paras-D2}. This time we fix ${Q_1}_\text{signal}$ and test a group of ${Q_2}_\text{signal}$ values for the same ranges as in Table \ref{tab:sensi-search-paras-D}. Figure \ref{fig:Q2-sp} plots the normalized detection statistic $\langle \rho/\sigma_\rho \rangle$ averaged over $10^3$ trials as a function of ${Q_2}_\text{signal}$ (bottom axis) and corresponding $B$ (top axis). As expected, varying ${Q_2}_\text{signal}$ within a reasonable range of magnetic field strength does not impact the sensitivity much for given ${Q_1}_\text{signal}$, because the wave strain depends more on $\epsilon$ than $B$.

\begin{table}[!tbh]
	\centering
	\setlength{\tabcolsep}{8pt}
	\begin{tabular}{lll}
		\hline
		\hline
		Injection parameter & Value & Astrophysical parameter \\
		\hline
		${\nu_0}_\text{signal}$ & 150.05\,Hz& \\
		${Q_1}_\text{signal}$ & $7 \times 10^{-19}$\,Hz\,s$^{-1}$ & $\epsilon=6.4 \times 10^{-4}$\\
		${Q_2}_\text{signal}$&$2 \times 10^{-18} - 1 \times 10^{-16}$ \,Hz\,s$^{-1}$ & $1.8 \times 10^{11}$\,G$\leq B \leq 1.3 \times 10^{12}$\,G\\
		${n_\text{em}}_\text{signal}$&3&\\
		\hline
		\hline
	\end{tabular}
	\caption[signal parameters]{Injection parameters used to create the second set of synthetic data analysed in Section \ref{sec:sensiD} containing spinning-down signals, in which ${Q_1}_\text{signal}$ is fixed and a group of ${Q_2}_\text{signal}$ values are tested. The corresponding astrophysical parameters $\epsilon$ and $B$ in equation (\ref{eq:spindownmodel_full}) are also quoted in the last column. The wave strain $h_0 = 3 \times 10^{-25}$ is calculated from equation (\ref{eqn:h0_epsilon}).}
	\label{tab:sensi-sig-paras-D2}
\end{table}

\section{LIGO S5 Search}
\label{sec:search}
\subsection{Data and templates}
\label{sec:data_templates}

The S5 data contain two years of short Fourier transforms (SFTs), collected from Nov 2005 to Oct 2007. A search of the band 75--450\,Hz is conducted, using SFTs from the H1 and L1 interferometers from 01 Nov 2006 to 30 Oct 2007 UTC. The second year of S5 is chosen, because we are limited computationally to $T_\text{obs} \leq 1$\, year, and the noise power spectral density is lower during the second year than the first. We analyse 23223 30-min SFTs in total, with 12590 from H1 and 10633 from L1.

In view of the substantial computational cost, we select the template grid with an eye towards efficiency. In Section 6 of Ref. \cite{chung11}, a semi-coherent phase metric was developed to calculate the mismatch $m$ as a function of the template spacing along each of the four axes of the parameter space $(\nu_0,Q_1,Q_2,n_\text{em})$. For the search in this paper, we elect to tolerate a maximum mismatch $m\leq0.2$ for the template closest to the true source parameters. Drawing on the analysis in Section 6 of Ref. \cite{chung11}, specifically equations (39)--(41), we construct a set of templates $\{\nu_0,Q_1,Q_2\}$ across the astrophysically relevant parameter range quoted in Table \ref{tab:search_para}. The largest values of $Q_1$ and $Q_2$ are limited by the maximum number of templates we can afford computationally ($N_\text{total} \sim 10^9$). Only two values of $Q_2$ are needed to sample the relevant range at the resolution required for $m\leq0.2$. We fix $n_\text{em}=3$ (see Section \ref{sec:brakingindex}), the sky position $(\alpha, \delta)$ = (1.46375, $-$1.20899), and $T_\text{lag}=3600$\,s, and average over $\cos \iota$ and $\psi$. As the pipeline loses track of the signal phase quickly with a spin-down rate $|\dot{\nu}(0)| \gtrsim 10^{-7}\text{Hz\,s}^{-1}$ (see Section \ref{sec:sensiD}), a narrower band of $Q_1$ is searched for $350 \leq \nu_0/\text{Hz} \leq 450$. The total number of templates is $N_\text{total}= 2,373,875,000$, implying $\rho_\text{th} \approx 6.50$ ($\alpha_\text{f}=10\%$). 

\begin{table}[!htb]
	\centering
	\setlength{\tabcolsep}{5pt}
	\begin{tabular}{lllll}
		\hline
		\hline
		$\nu_0$ (Hz)  & Resolution (Hz) & $Q_1,Q_2$ (Hz/s)& Resolution (Hz/s) & $\epsilon,B$(G) \\
		\hline
		$75\leq \nu_0 < 350$ & $4 \times 10^{-4}$ & $2 \times 10^{-22}  \leq Q_1 \leq 6.17 \times 10^{-19}$ & $3.68 \times 10^{-22}$ & $1.08 \times 10^{-5}\leq \epsilon \leq 6.0 \times10^{-4}$\\
		&&$1 \times 10^{-21} \leq Q_2 \leq 2 \times 10^{-16}$ & $2 \times 10^{-16}$ & $4.05 \times 10^9\leq B$$\leq 1.81 \times10^{12} $\\
		\hline
		$350\leq \nu_0 \leq 450$ & $4 \times 10^{-4}$ & $2 \times 10^{-22}  \leq Q_1 \leq 4.99 \times 10^{-20}$ & $3.68 \times 10^{-22}$ & $1.08 \times 10^{-5}\leq \epsilon \leq 1.71 \times10^{-4}$\\
		&&$1 \times 10^{-21} \leq Q_2 \leq 2 \times 10^{-16}$ & $2 \times 10^{-16}$ & $4.05 \times 10^9\leq B$$\leq 1.81 \times10^{12} $\\
		\hline
		\hline
	\end{tabular}
	\caption[search parameters]{Ranges and resolutions of template parameters $\nu_0$, $Q_1$ and $Q_2$ ($n_\text{em}$ fixed). Only two $Q_2$ values are searched. Corresponding ranges of $\epsilon$ and $B$ are listed in the right column.}
	\label{tab:search_para}
\end{table}

\subsection{Candidates and line vetoes}
Templates $\{\nu_0,Q_1,Q_2\}$ with $\rho/\sigma_\rho >\rho_\text{th}$ are found to cluster at 19 narrow $\nu_0$ bands, each spanning $\sim 0.5$\,Hz and extending over the entire ranges of $Q_1$ and $Q_2$. We list the peak $\rho$ and corresponding $\nu_0$, $Q_1$ and $Q_2$ values in Table \ref{tab:S5-candidate-peaks} for each cluster.

\begin{table}
	\centering
	\setlength{\tabcolsep}{8pt}
	\begin{tabular}{llll}
		\hline
		\hline
		$\rho/\sigma_\rho$ & $\nu_0$ (Hz) & $Q_1$ ($\text{Hz\,s}^{-1}$) & $Q_2$ ($\text{Hz\,s}^{-1}$) \\
		\hline
		23.81 & 75.0240 & $1.0872\times 10^{-20}$ & $1 \times 10^{-21}$ \\
		3774.10 & 91.1360 & $1.80888\times 10^{-19}$ & $1 \times 10^{-21}$ \\
		7.77 & 93.2896 & $2.20264\times 10^{-19}$ & $2 \times 10^{-16}$ \\
		11.07 & 96.4980 & $2.31304\times 10^{-19}$ & $2 \times 10^{-16}$ \\
		35.73 & 100.0008 & $1.0136\times 10^{-20}$ & $1 \times 10^{-21}$ \\
		90.11& 108.8632 & $5.0984\times 10^{-20}$ & $1 \times 10^{-21}$ \\
		10.90 & 112.0000 & $2.408\times 10^{-21}$ & $1 \times 10^{-21}$ \\
		47.07 & 119.8792 & $4.984\times 10^{-21}$ & $1 \times 10^{-21}$ \\
		27.73 & 128.0012 & $4.616\times 10^{-21}$ & $1 \times 10^{-21}$ \\
		49.61 & 139.5112 & $2.776\times 10^{-21}$ & $1 \times 10^{-21}$ \\
		7.72 & 144.8112 & $1.31944\times 10^{-19}$ & $2 \times 10^{-16}$ \\
		7.26 & 145.3072 & $5.64344\times 10^{-19}$ & $1 \times 10^{-21}$ \\
		21.89 & 179.8132 & $9.36\times 10^{-22}$ & $1 \times 10^{-21}$ \\
		23.93 & 193.5700 & $4.3256\times 10^{-20}$ & $1 \times 10^{-21}$ \\
		8.22& 200.0304 & $9.032\times 10^{-21}$ & $2 \times 10^{-16}$ \\
		7.79 & 329.7820 & $2.00\times 10^{-22}$ & $1 \times 10^{-21}$ \\
		2891.52 & 381.9036 & $2.040\times 10^{-21}$ & $1 \times 10^{-21}$ \\
		1093.79 & 393.1372 & $9.36\times 10^{-22}$ & $1 \times 10^{-21}$ \\
		6243.65 & 396.9736 & $9.36\times 10^{-22}$ & $1 \times 10^{-21}$ \\
		\hline
		\hline
	\end{tabular}
	\caption[S5 search candidates]{First-pass candidates from the LIGO S5 search for SNR 1987A, listing the maximum $\rho/\sigma_\rho$ in each cluster with $\rho/\sigma_\rho >\rho_\text{th} = 6.50$ and the corresponding $\nu_0$, $Q_1$ and $Q_2$, sorted according to $\nu_0$.}
	\label{tab:S5-candidate-peaks}
\end{table}

Continuous waves emitted by non-spherical spinning neutron stars appear as narrow spectral lines. The instrumental power line at 60\,Hz with wings extending $\pm2$\,Hz, its harmonics, and noise lines from electronics, wire, calibration, etc., impact the search by obscuring astrophysical signals in that band. Within the frequency range we are searching, the most prominent known peaks lie at low frequencies $\sim 90$--100\,Hz (electronic lines), and at $\sim 329$--350\,Hz (mirror suspensions). An instrumental line catalogue can be found in Appendix B of Ref. \cite{Aasi2013} and at the LIGO Open Science Center\footnote{https://losc.ligo.org/speclines/}. We notch out bands contaminated by known noise lines. For each candidate cluster with peak at frequency $\nu_0$ and $\dot{\nu}_0$ (calculated from $Q_1$ and $Q_2$), we veto the cluster if the band $\nu_0 - \Delta \nu \leq \nu \leq \nu_0 + \Delta \nu$, with $\Delta \nu \approx \nu_0 \times 10^{-4} + |\dot{\nu} _0|\times 3.14 \times 10^7$\,s, overlaps with a known noise line. This criterion takes into account the maximum possible Doppler shift due to the Earth's orbit and the maximum frequency shift due to the spin down of the source \cite{Aasi2013}. The surviving candidates are listed in Table \ref{tab:S5-candidate-remove-knownlines}.

\begin{table}
	\centering
	\setlength{\tabcolsep}{8pt}
	\begin{tabular}{llll}
		\hline
		\hline
		$\rho/\sigma_\rho$ & $\nu_0$ (Hz) & $Q_1$ ($\text{Hz\,s}^{-1}$) & $Q_2$ ($\text{Hz\,s}^{-1}$) \\
		\hline
		3774.10 & 91.1360 & $1.80888\times 10^{-19}$ & $1 \times 10^{-21}$ \\
		35.73 & 100.0008 & $1.0136\times 10^{-20}$ & $1 \times 10^{-21}$ \\
		10.90 & 112.0000 & $2.408\times 10^{-21}$ & $1 \times 10^{-21}$ \\
		27.73 & 128.0012 & $4.616\times 10^{-21}$ & $1 \times 10^{-21}$\\
		8.22 & 200.0304 & $9.032\times 10^{-21}$ & $2 \times 10^{-16}$ \\
		2891.52 & 381.9036 & $2.040\times 10^{-21}$ & $1 \times 10^{-21}$\\
		\hline
		\hline
	\end{tabular}
	\caption[S5 search candidates]{Second-pass candidates from the LIGO S5 search for SNR 1987A after instrumental line veto, listing the maximum $\rho/\sigma_\rho$ in each cluster with $\rho/\sigma_\rho >\rho_\text{th} = 6.50$ and the corresponding $\nu_0$, $Q_1$ and $Q_2$, sorted according to $\nu_0$.}
	\label{tab:S5-candidate-remove-knownlines}
\end{table}

\subsection{Manual vetoes}
We now examine the survivors in Table \ref{tab:S5-candidate-remove-knownlines} manually to check if they are false alarms. We do this in two ways. First, we search the second year of S5 (01 Nov 2006 -- 30 Oct 2007 UTC) from H1 and L1 separately to test if the signal appears in both interferometers. The sensitivities of the two interferometers during S5 are comparable to one another, implying that a signal is expected to meet the same detection criterion in both detectors. Second,  we search the first year of S5 (04 Nov 2005 -- 30 Oct 2006 UTC) from H1 and L1 to test if the candidate persists in both years. As with the first detection criterion, the strain sensitivities of the detectors in the first and second years of observation are comparable, implying that a gravitational-wave signal present in one year of data should also be present in both.

\begin{figure*}[!htb]
	\centering
	\subfigure[]
	{
		\label{100hz}
		\scalebox{0.185}{\includegraphics{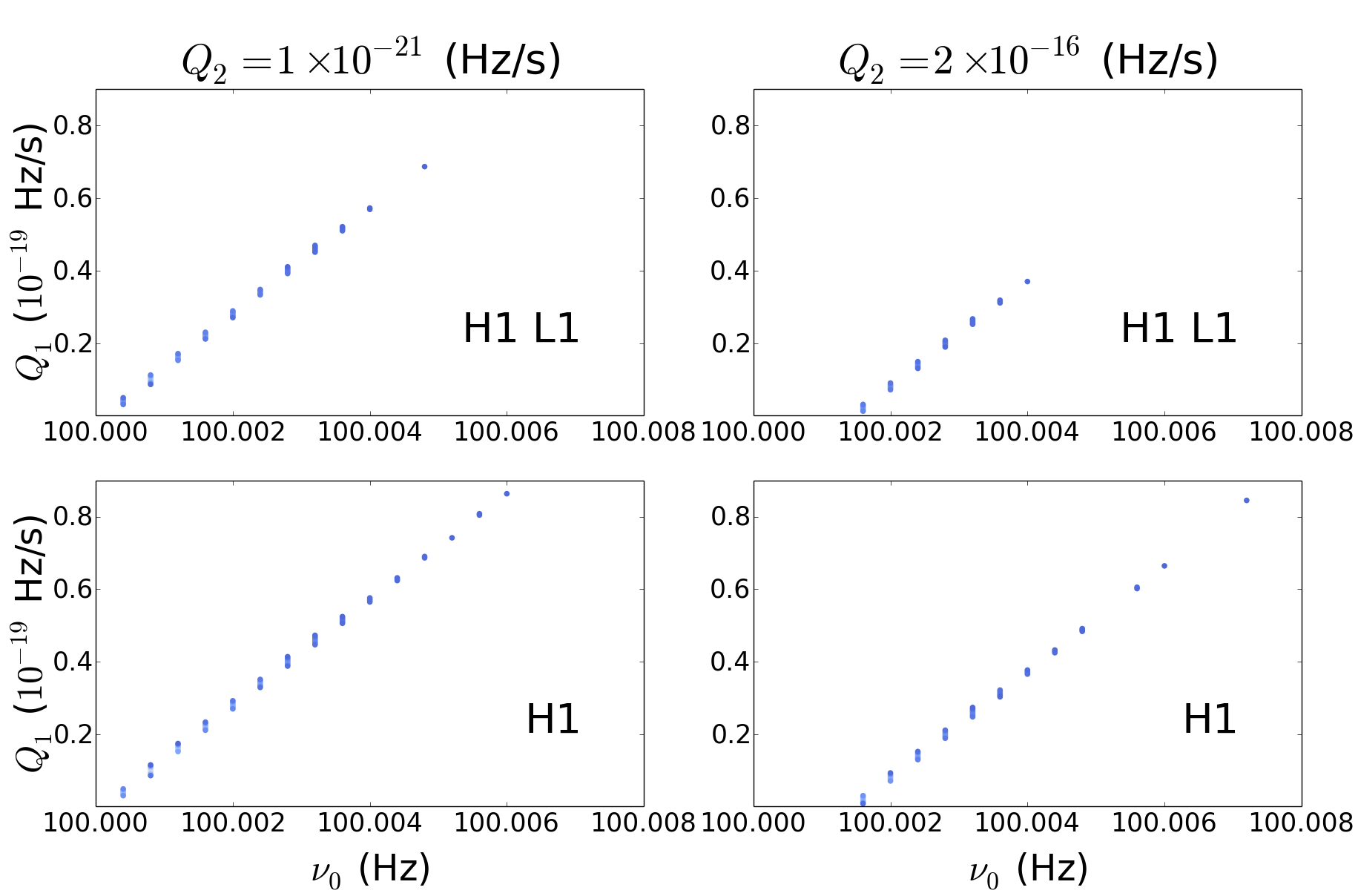}}
	}
	\subfigure[]
	{
		\label{200hz}
		\scalebox{0.185}{\includegraphics{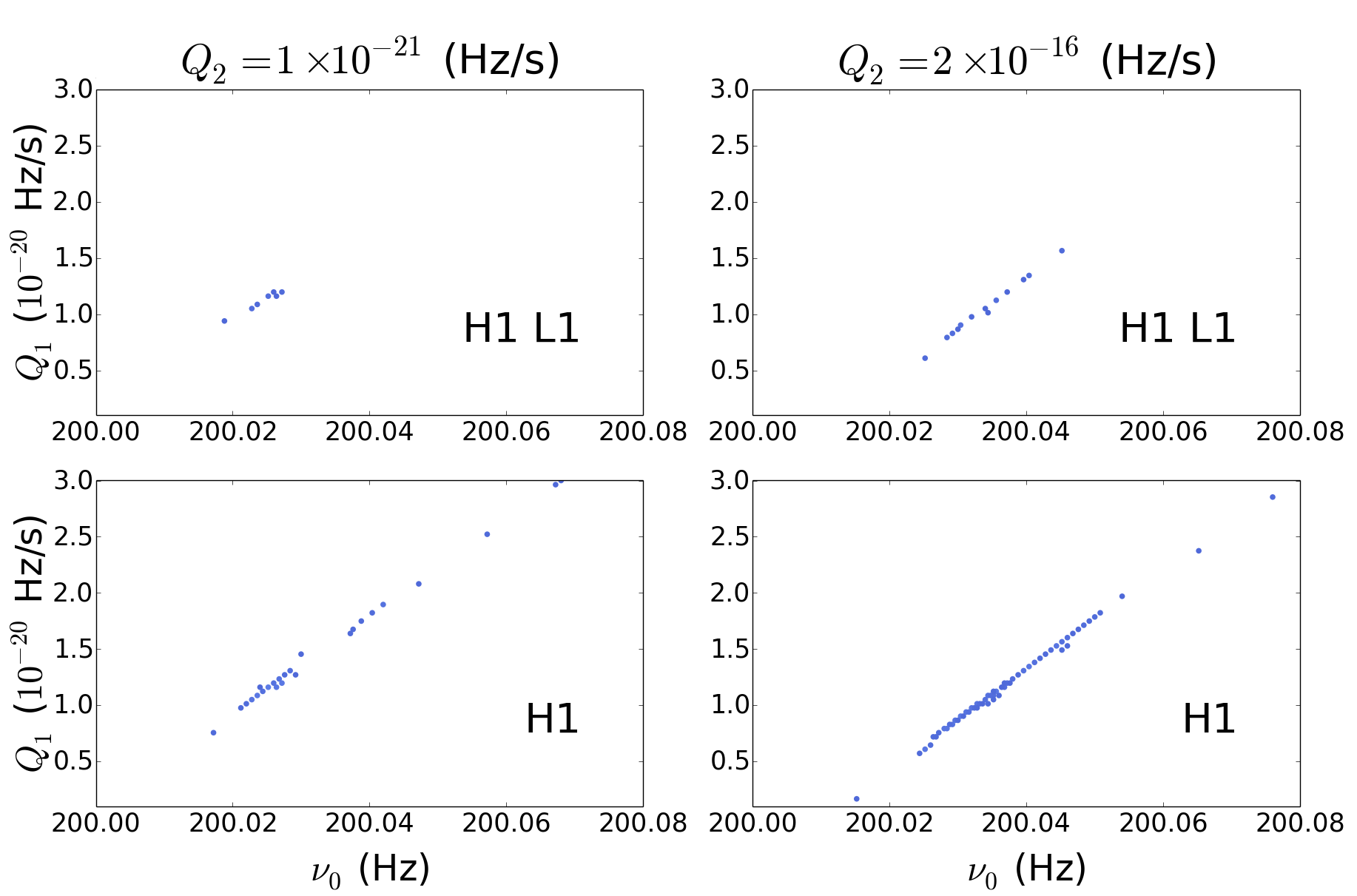}}
	}
	\subfigure[]
	{
		\label{91hz}
		\scalebox{0.185}{\includegraphics{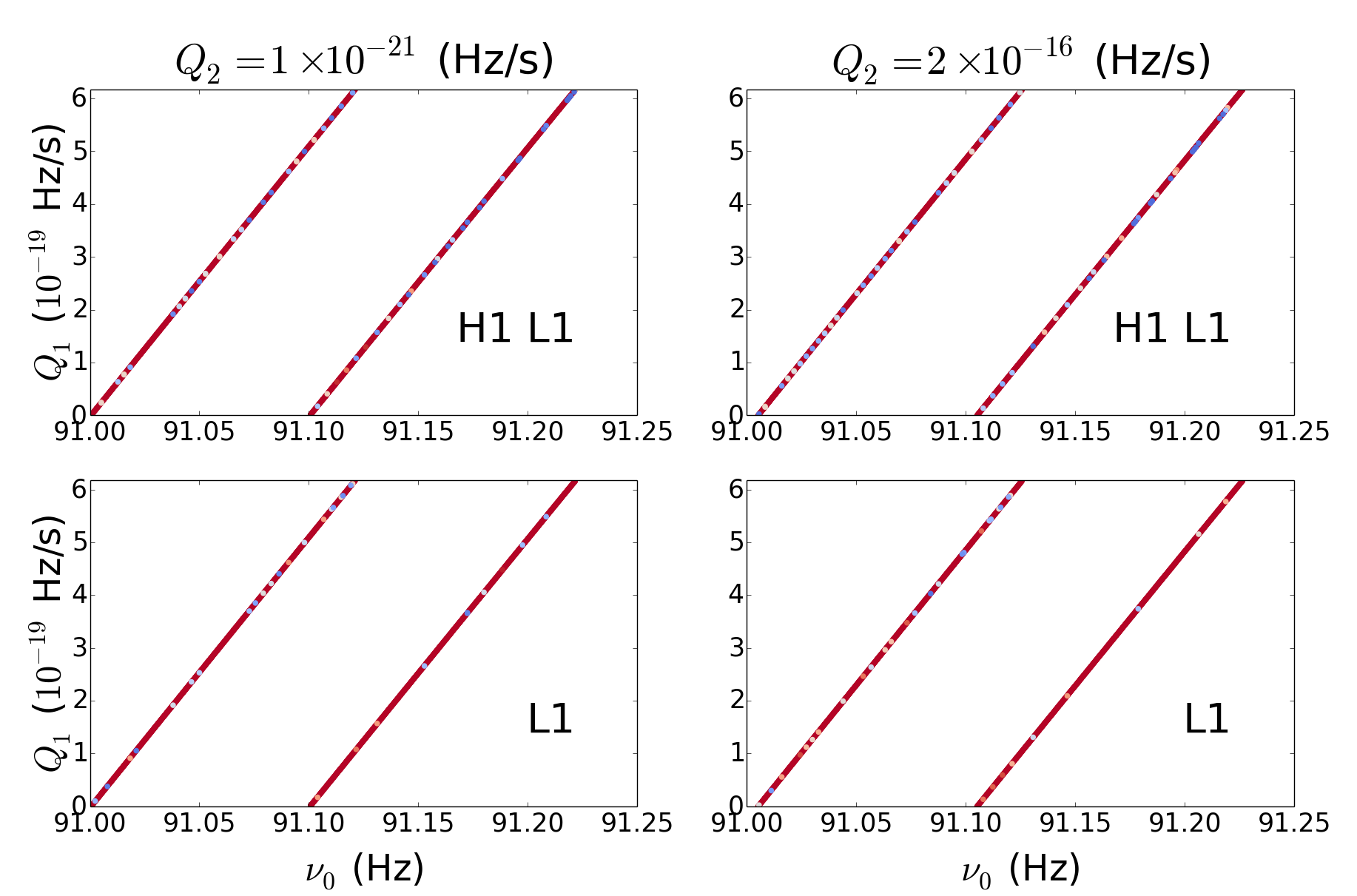}}
	}
	\subfigure[]
	{
		\label{128hz}
		\scalebox{0.185}{\includegraphics{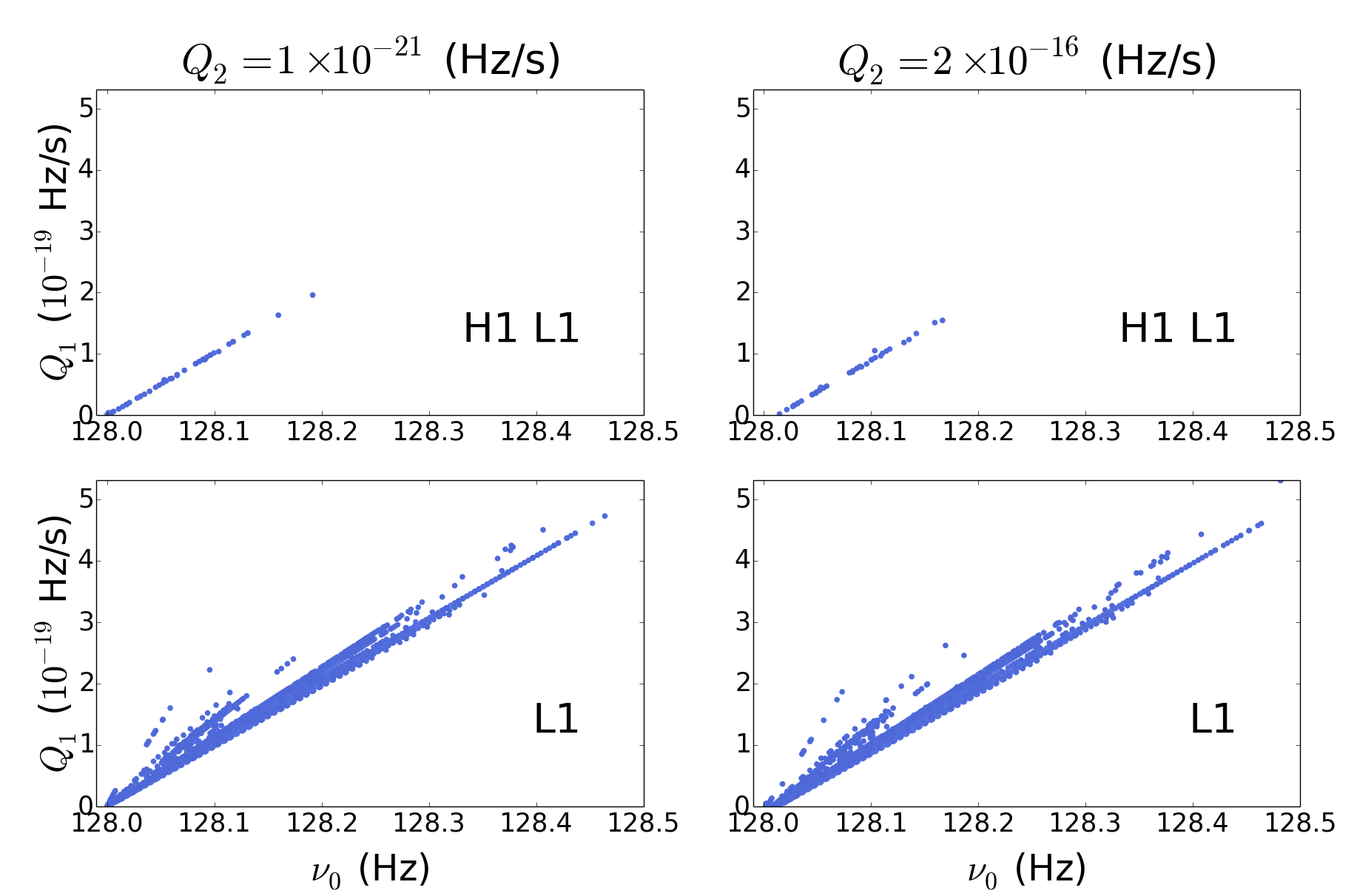}}
	}
	\subfigure[]
	{
		\label{381hz}
		\scalebox{0.185}{\includegraphics{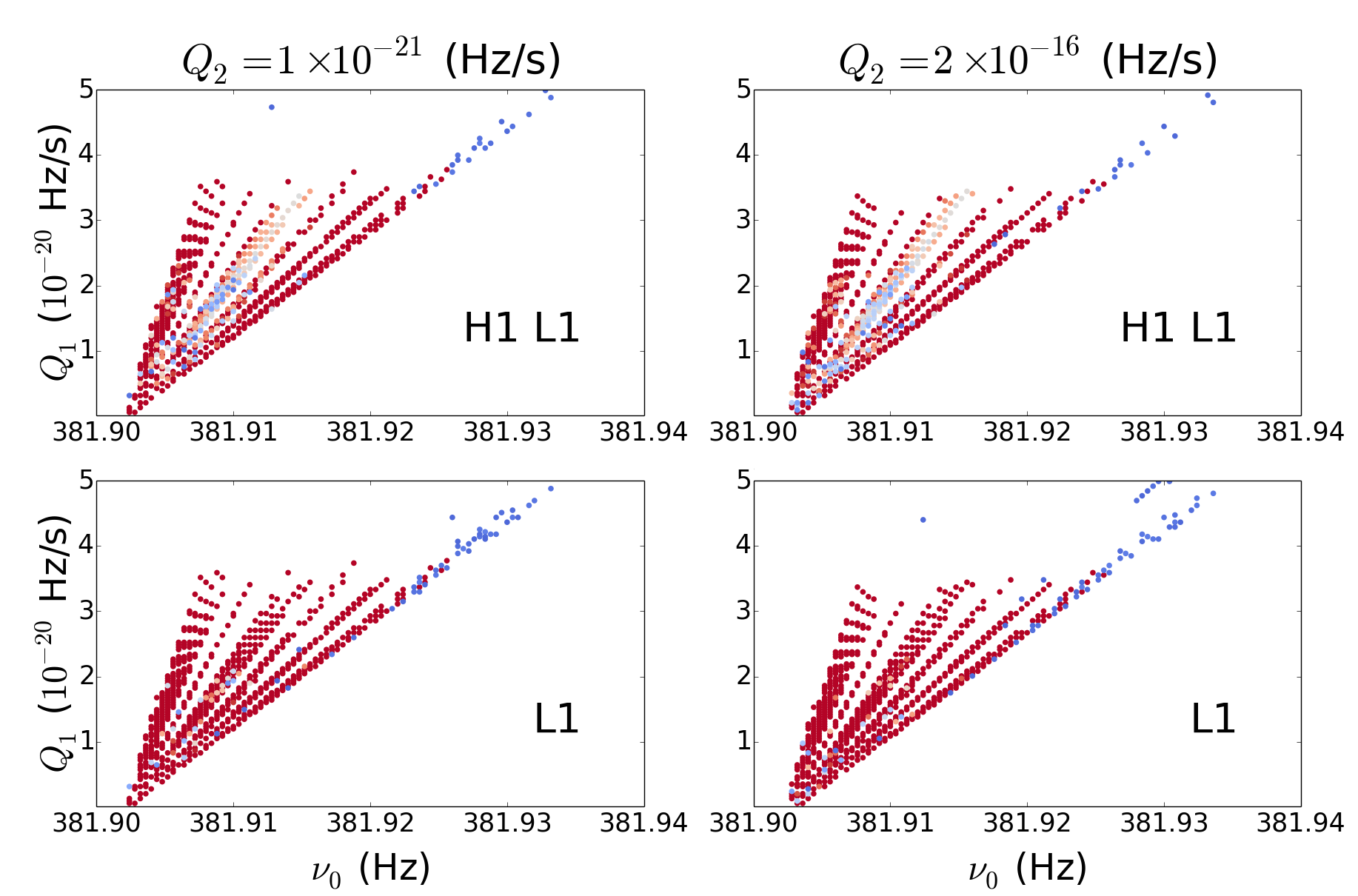}}
	}
	\caption[candidates]{Surviving candidates after instrumental line veto. Normalized detection statistic $\rho/\sigma_\rho$ as a function of $Q_1$ and $\nu_0$ for five clusters (group of four panels). Each dot on the plots stands for one template $\{\nu_0, Q_1, Q_2\}$ with $\rho/\sigma_\rho > \rho_\text{th}=6.5$. The colour of the dots indicates values of $\rho/\sigma_\rho$ (larger $\rho/\sigma_\rho$ in warmer colour). The dots merge into continuous lines or thick bars because they are closely spaced. For each cluster, two values of $Q_2$ are searched: $Q_2 = 1 \times 10^{-21}\text{Hz\,s}^{-1}$ (left two panels in each group of four) and $Q_2 = 2 \times 10^{-16}\text{Hz\,s}^{-1}$ (right two panels in each group of four). The top and bottom panels in each group correspond to two detectors (H1 and L1) and one detector (H1 or L1) respectively. For clusters (a) and (b), no detection is found in L1; for (c), (d) and (e), no detection is found in H1.}
	\label{fig:candidates}
\end{figure*}

Figure \ref{fig:candidates} compares the output from both detectors (H1 and L1; top two panels in each group of four) and from one detector (H1 or L1; bottom two panels in each group of four) for each candidate cluster. For cluster (a) and (b), the bottom two panels are from H1 and no detection is found in L1; for (c), (d) and (e), the bottom two panels are from L1 and no detection is found in H1. The left panels in each group of four are for $Q_2=1 \times 10^{-21}\,\text{Hz\,s}^{-1}$ and the right panels are for $Q_2=2 \times 10^{-16}\,\text{Hz\,s}^{-1}$. Red and blue dots stand for $\rho$ higher and lower than $50$ respectively. In some cases, the cluster spreads wider across parameter space {$\nu_0, Q_1$}, and $\rho/\sigma_\rho$ is higher, from the output of one detector than both detectors, because the loud noise line causing these candidates in one detector is weakened by the noise in the other detector. 

\begin{figure}[!htb]
	\centering
	\scalebox{0.32}{\includegraphics{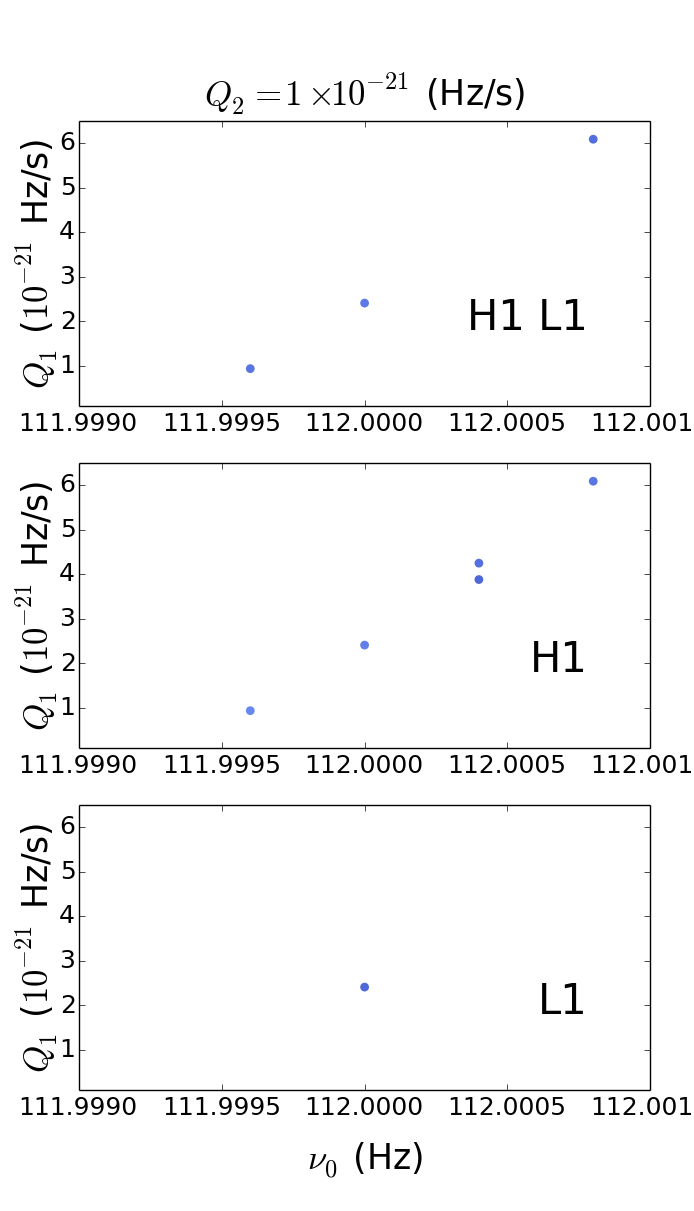}}
	\caption[candidate 112hz]{Surviving candidate cluster around 112\,Hz seen in both detectors. Normalized detection statistic $\rho/\sigma_\rho$ as a function of $Q_1$ and $\nu_0$. Each dot on the plots stands for one template $\{\nu_0, Q_1, Q_2\}$ with $\rho/\sigma_\rho > \rho_\text{th}=6.5$. All templates are obtained at $Q_2=1 \times 10^{-21}\,\text{Hz\,s}^{-1}$, and no hits are at $Q_2=2 \times 10^{-16}\,\text{Hz\,s}^{-1}$. The top, middle, and bottom panels correspond to two detectors (H1 and L1), H1, and L1 respectively.}
	\label{fig:candidate_112}
\end{figure}

The only candidate seen in both detectors lies around 112\,Hz. The detection statistic $\rho/\sigma_\rho$ for this candidate is plotted as a dot for each template $\{\nu_0, Q_1, Q_2\}$ with $\rho/\sigma_\rho > \rho_\text{th}=6.5$ in Figure \ref{fig:candidate_112}. At $Q_2=1 \times 10^{-21}\,\text{Hz\,s}^{-1}$, three templates exceed $\rho_\text{th}$ obtained for using both detectors, five templates exceed $\rho_\text{th}$ for H1, and one template exceeds $\rho_\text{th}$ for L1. No hits are at $Q_2=2 \times 10^{-16}\,\text{Hz\,s}^{-1}$. Following up further, we take the first year of S5 data from H1 and L1, and run the same search around 112\,Hz. If the candidate is astrophysical in origin, we expect a detection with similar statistical significance at a slightly higher $\nu_0$ consistent with the astrophysical spin-down model. However, nothing is detected in the first year of S5 data for $111.5 \leq \nu_0/\text{Hz} \leq 112.5$, $2 \times 10^{-22}  \leq Q_1/\text{Hz\,s}^{-1} \leq 6.17 \times 10^{-19}$ and $1 \times 10^{-21} \leq Q_2/\text{Hz\,s}^{-1} \leq 2 \times 10^{-16}$. As the candidate comprises relatively few templates, and $\rho/\sigma_\rho$ stands just above the threshold, a false alarm is strongly implied. 

In summary, no candidate survives the manual vetoes. The false alarm rate selected for the whole search is $10\%$, so a single false alarm candidate cluster is consistent with our expectation.

\begin{figure*}[!htb]
	\centering
	\subfigure[]
	{
		\label{half_year_91hz}
		\scalebox{0.3}{\includegraphics{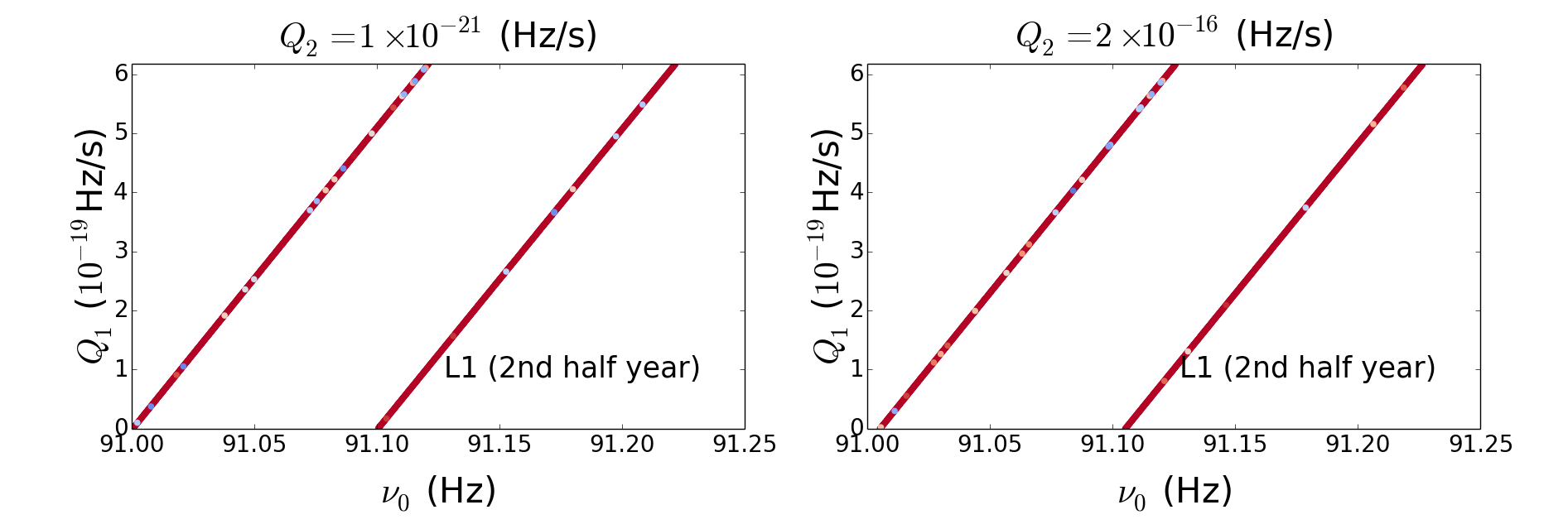}}
	}
	\subfigure[]
	{
		\label{half_year_381hz}
		\scalebox{0.3}{\includegraphics{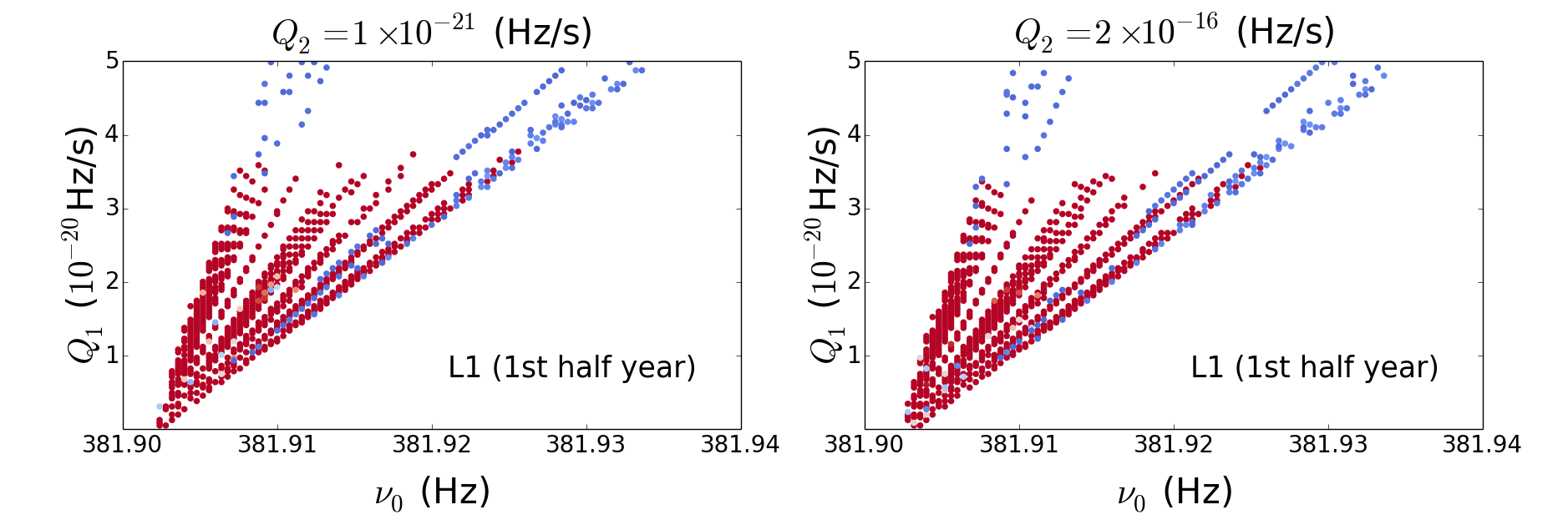}}
	}
	\caption[half year]{Two strongest vetoed candidate clusters around (a) 91\,Hz and (b) 381\,Hz (both found in L1) seen in half of the second year of S5 data. The normalized detection statistic $\rho/\sigma_\rho$ is plotted as a function of $Q_1$ and $\nu_0$ for the two clusters. Each dot on the plots stands for one template $\{\nu_0, Q_1, Q_2\}$ with $\rho/\sigma_\rho > \rho_\text{th}=6.5$. The colour of the dots indicates values of $\rho/\sigma_\rho$ (larger $\rho/\sigma_\rho$ in warmer colour). The cluster around 91\,Hz only exists in the second half year (01 May 2007 -- 30 Oct 2007 UTC). The cluster around 381\,Hz only exists in the first half year (01 Nov 2006 -- 30 Apr 2007 UTC). For each cluster, two values of $Q_2$ are searched: $Q_2 = 1 \times 10^{-21}\text{Hz\,s}^{-1}$ (left panels) and $Q_2 = 2 \times 10^{-16}\text{Hz\,s}^{-1}$ (right panels).}
	\label{fig:half_year}
\end{figure*}

To better understand the cause of the strongest vetoed candidates (i.e. clusters around 91\,Hz and 381\,Hz; both found in L1), we divide the second year of S5 data from the L1 detector into two halves (01 Nov 2006 -- 30 Apr 2007 UTC and 01 May 2007 -- 30 Oct 2007 UTC), search them separately, and compare the two outputs. The cluster around 91\,Hz only exists in the second half year. The cluster around 381\,Hz only exists in the first half year. The normalized detection statistic for each template $\{\nu_0, Q_1, Q_2\}$ with $\rho/\sigma_\rho > \rho_\text{th}=6.5$ from L1 is plotted in Figure \ref{fig:half_year} [(a) for cluster around 91\,Hz in the second half year, and (b) for cluster around 381\,Hz in the first half year]. The patterns of dots in the ($Q_1, \nu_0$) plane from the second half year [around 91\,Hz; Figure \ref{half_year_91hz}] and the first half year [around 381\,Hz; Figure \ref{half_year_381hz}] are exactly the same as those from the whole year [see Figure \ref{91hz} and \ref{381hz}]. Hence, instead of being some persistent noise line throughout the whole observation period, the candidate is probably a short-term glitch. 

\begin{figure*}[!htb]
	\centering
	\subfigure[]
	{
		\label{known_line_108}
		\scalebox{0.3}{\includegraphics{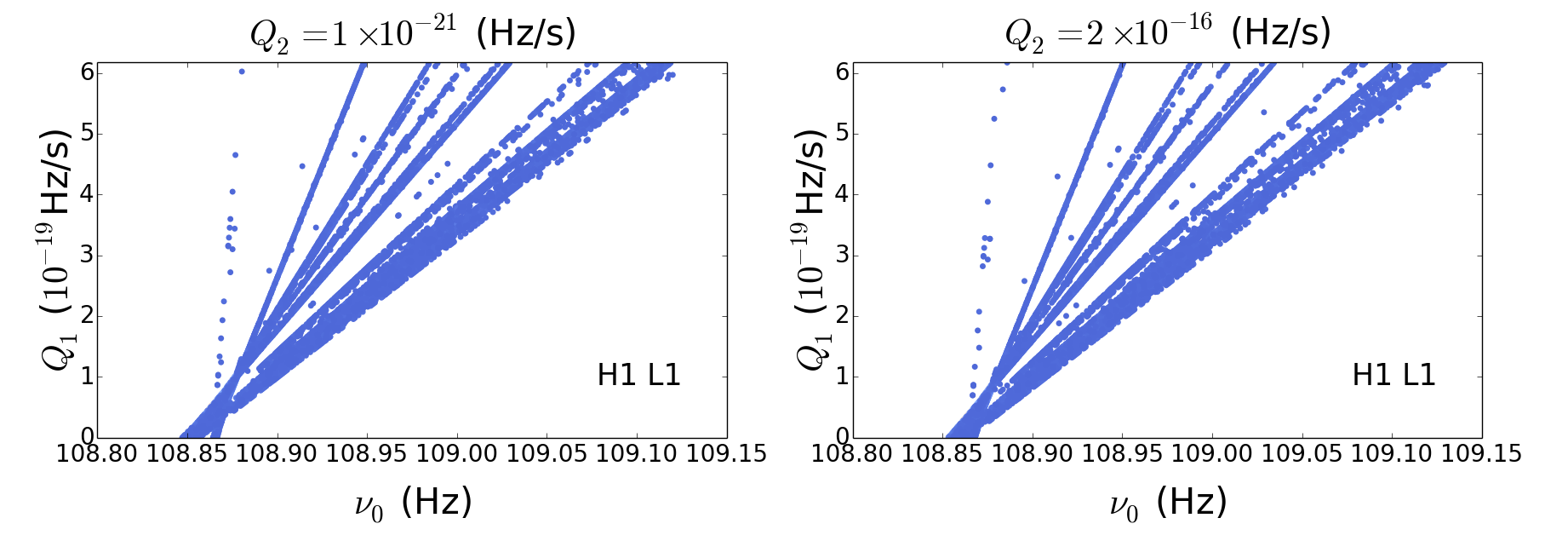}}
	}
	\subfigure[]
	{
		\label{know_line_193}
		\scalebox{0.3}{\includegraphics{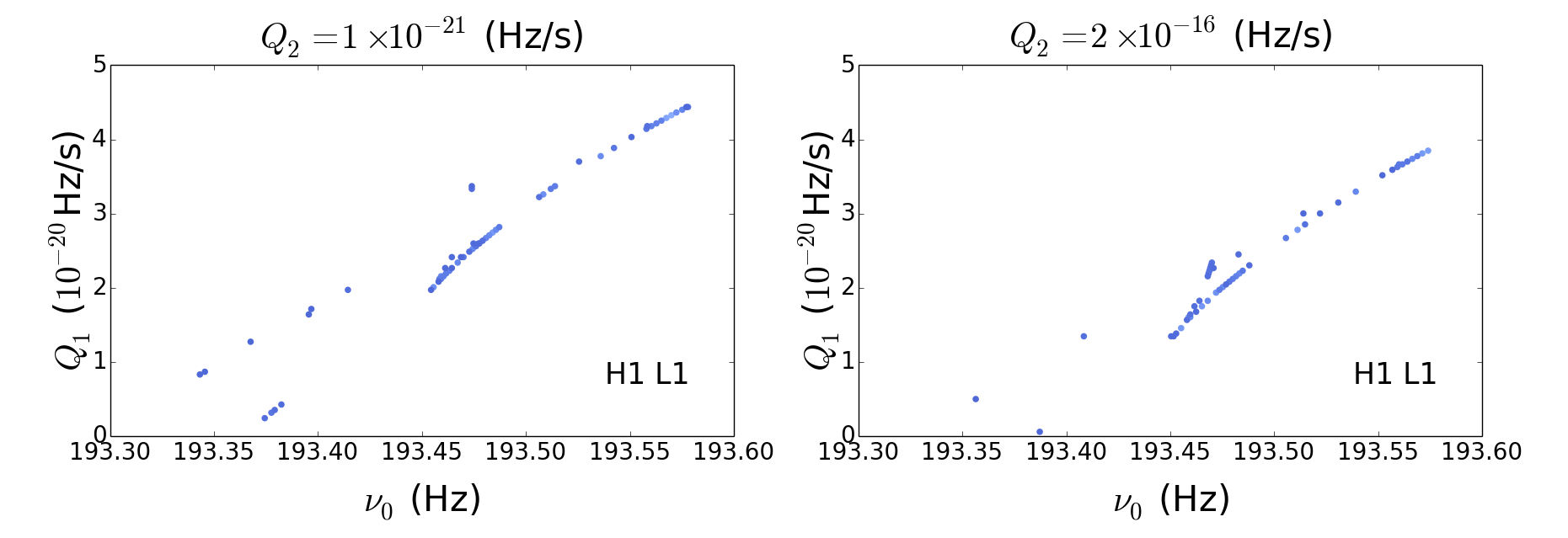}}
	}
	\caption[known line sample]{Two examples of the clusters caused by instrumental lines at (a) 108.8\,Hz and (b) 193.4\,Hz. The normalized detection statistic $\rho/\sigma_\rho$ is plotted as a function of $Q_1$ and $\nu_0$. Each dot on the plots stands for one template $\{\nu_0, Q_1, Q_2\}$ with $\rho/\sigma_\rho > \rho_\text{th}=6.5$. For each cluster, two values of $Q_2$ are searched: $Q_2 = 1 \times 10^{-21}\text{Hz\,s}^{-1}$ (left panels) and $Q_2 = 2 \times 10^{-16}\text{Hz\,s}^{-1}$ (right panels). Both clusters are obtained from the second year of S5 data from two detectors (H1 and L1).}
	\label{fig:known_line_sample}
\end{figure*}

We also check how the pattern of dots caused by a glitch differs from that of a known instrumental spectral line. We plot two examples of the clusters caused by instrumental lines at 108.8\,Hz and 193.4\,Hz in Figure \ref{fig:known_line_sample}. We find that the pattern of dots are similar to a glitch, with dots spreading $\sim 0.5$\,Hz in frequency across the whole $Q_1$ and $Q_2$ band searched. Interestingly, therefore, we cannot differentiate reliably between a persistent line and a transient glitch from the super-threshold template distribution in the ($Q_1, \nu_0$) plane.  

\subsection{Wave strain upper limit}
Without a detection, we are able to place an upper limit on $h_0$ as a function of $\nu_0$. 

Given the one-sided power spectral density $S_n^{(1)}(\nu)$ and $S_n^{(2)}(\nu)$ for each interferometer, and assuming that $\rho$ is normally distributed, the lowest detectable gravitational wave strain $h_\text{th}(\nu)$ calculated by \citet{dhurandhar08} is
\begin{equation}
\label{eqn:theoretical_sensitivity}
h_\text{th} (\nu) = \frac{\mathcal{S}^{1/2}}{\sqrt{2}\langle|\tilde{\mathcal{G}}_{IJ}|^2\rangle^{1/4}N_\text{pairs}^{1/4}}\left[\frac{\left(S_n^{(1)}(\nu)S_n^{(2)}(\nu)\right)^{1/2}}{\Delta T}\right]^{1/2},
\end{equation}
with $\mathcal{S}=\text{erfc}^{-1}(2\alpha_\text{f}) + \text{erfc}^{-1}(2\alpha_\text{d})$, where $\alpha_\text{f}$ is the false alarm rate, $\alpha_\text{d}$ is the false dismissal rate, $\langle|\tilde{\mathcal{G}}_{IJ}|^2\rangle$ is the cross-correlation function defined in (\ref{eqngij}) averaged over $\cos \iota$ and $\psi$, and $N_\text{pairs}$ is the number of SFT pairs. The theoretical sensitivity is analysed as a function of $\nu_0$ in Section 4.1 by \citet{chung11}, who found $h_0 \leq 1.6 \times 10^{-25}$ at the most sensitive frequency around 150\,Hz, with $\alpha_\text{f}=\alpha_\text{d}=0.1$. This estimate in Ref. \cite{chung11} is also based on the S5 noise curve, and hence it is approximately the theoretical sensitivity we expect.

The upper limit we are able to place is more conservative than $h_\text{th}$ in equation (\ref{eqn:theoretical_sensitivity}), because the sensitivity drops significantly for $|\dot{\nu}(0)| \gtrsim 10^{-7}\,\text{Hz\,s}^{-1}$ (i.e. large $Q_1$, $Q_2$ and $\nu_0$), where the pipeline loses track of the signal phase ($\gtrsim \pi/2$) after a year's observation (see Section \ref{sec:sensiD}). The observation period during which the phase tracking remains accurate is shorter than one year for $|\dot{\nu}(0)| \gtrsim 10^{-7}\,\text{Hz\,s}^{-1}$, reducing $N_\text{pairs}$ and hence the sensitivity. At a given $\nu_0$, when the largest $Q_1$ and $Q_2$ in our parameter space are set in the template, the search is least sensitive because of the largest $|\dot{\nu}(0)|$ leading to a quickest loss in phase tracking. Hence the upper limit on $h_0$ at this $\nu_0$ is most conservative with the largest $Q_1$ and $Q_2$. We analyse the upper limit on $h_0$ as function of $\nu_0$ with both largest and smallest $Q_1$ and $Q_2$.

\begin{figure*}[!htb]
	\centering
	\subfigure[]
	{
		\label{75-300-UL}
		\scalebox{0.36}{\includegraphics{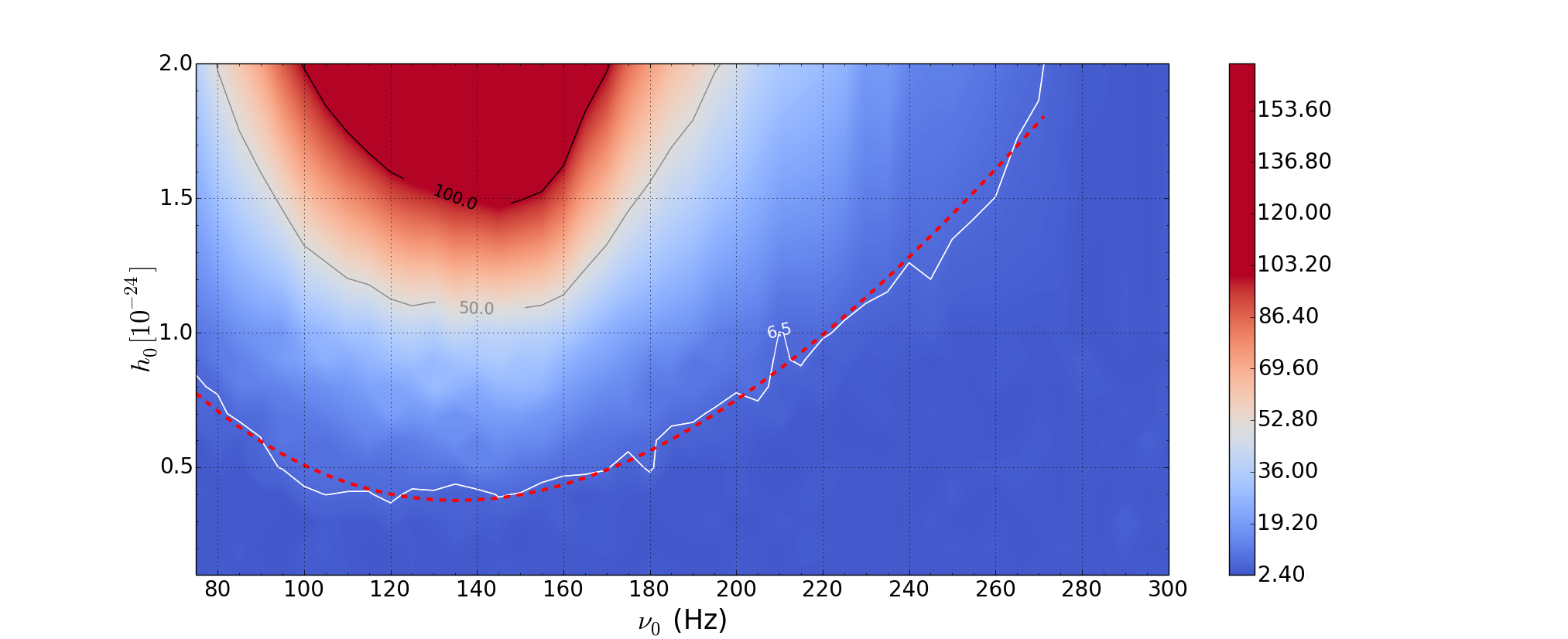}}
	}
	\subfigure[]
	{
		\label{255-450-UL}
		\scalebox{0.36}{\includegraphics{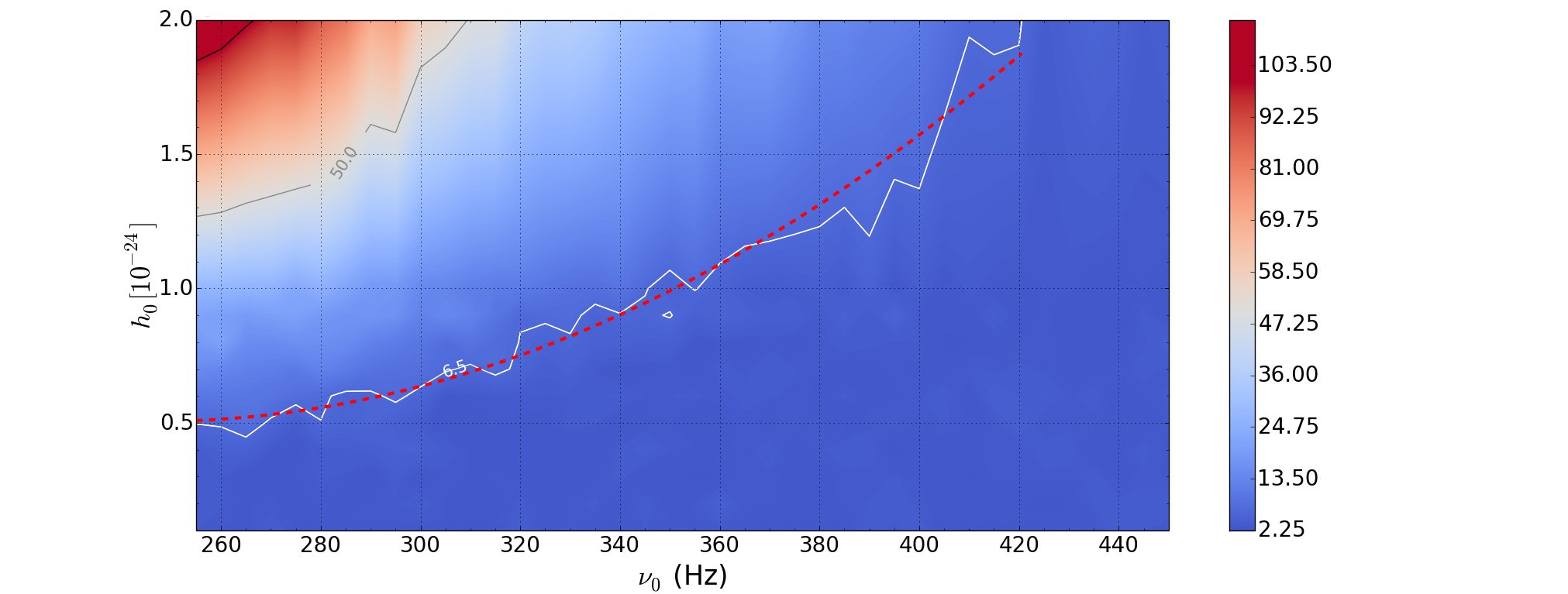}}
	}
	\caption[h0 sensitivity and UL]{Contours of normalized detection statistic $\rho/\sigma_\rho$ from searching the synthetic data with different values of $h_0$ and $\nu_0$ injected. The signals are generated with $1 \times 10^{-25} \leq h_0 \leq 2 \times 10^{-24}$ and (a) $75 \leq \nu_0/\text{Hz} \leq 300$, $Q_1 = 6.17 \times 10^{-19}\,\text{Hz\,s}^{-1}$, $Q_2 = 2 \times 10^{-16}\,\text{Hz\,s}^{-1}$, and (b) $255 \leq \nu_0/\text{Hz} \leq 450$, $Q_1 = 5 \times 10^{-20}\,\text{Hz\,s}^{-1}$, $Q_2 = 2 \times 10^{-16}\,\text{Hz\,s}^{-1}$. The red dashed curves are the contours $\rho/\sigma_\rho = \rho_\text{th} =6.5$, implying the 90\% confidence level upper limit on $h_0$. The best upper limit $h_0 \leq 3.8 \times 10^{-25}$ is obtained near 150\,Hz with $Q_1 \leq 6.17 \times 10^{-19}\,\text{Hz\,s}^{-1}$ and $Q_2 \leq 2 \times 10^{-16}\,\text{Hz\,s}^{-1}$}
	\label{fig:h0_upper_limit}
\end{figure*} 

We first evaluate the upper limit on $h_0$ with the largest $Q_1$ and $Q_2$ values in the two frequency bands searched separately. The largest $Q_1$ and $Q_2$ values are listed in Table \ref{tab:h0-UL-space}. At each given $\nu_0$, we find out the smallest $h_0$, above which we have $\rho/\sigma_\rho > \rho_\text{th}=6.5$ (i.e. a detection with 90\% confidence level). Hence this $h_0$ is the 90\% confidence level upper limit without a detection.

\begin{table}
	\centering
	\setlength{\tabcolsep}{8pt}
	\begin{tabular}{lll}
		\hline
		\hline
		$\nu_0$ range (Hz) & ${Q_1}_\text{max}$ (Hz\,s$^{-1}$) & ${Q_2}_\text{max}$ (Hz\,s$^{-1}$)\\
		\hline
		75--300 & $6.17 \times 10^{-19}$& $2\times 10^{-16}$\\
		255--450 & $5 \times 10^{-20}$& $2\times 10^{-16}$\\
		\hline
		\hline
	\end{tabular}
	\caption[]{Maximum $Q_1$ and $Q_2$ values in the two frequency bands of the search.}
	\label{tab:h0-UL-space}
\end{table}

The analysis is described in three steps. First, we inject synthetic signals for wave strains in the range $1 \times 10^{-25} \leq h_0 \leq 2 \times 10^{-24}$ spinning down with ${Q_1}_\text{max}$ and ${Q_2}_\text{max}$ in Table \ref{tab:h0-UL-space}. Second, we search these synthetic data sets with the same templates as we use searching the LIGO S5 data in Section \ref{sec:data_templates}, and plot the normalized detection statistic $\rho/\sigma_\rho$ as contours on the ($h_0, \nu_0$) planes in Figure \ref{fig:h0_upper_limit} for two frequency ranges respectively. Third, we draw the contour $\rho/\sigma_\rho = \rho_\text{th} = 6.5$ as a red dashed curve. For given $\nu_0$, any $h_0$ above the curve leads to $\rho/\sigma_\rho > \rho_\text{th} = 6.5$, which stands for a detection with $\alpha_\text{f}=\alpha_\text{d}=0.1$. Hence the red dashed curve is the 90\% confidence level upper limits on $h_0$ given no detection is found.

As the injected $\nu_0$ gets larger, the sensitivity decreases and the upper limit on $h_0$ increases. Comparing Figure \ref{75-300-UL} and \ref{255-450-UL} in the frequency range 255--300\,Hz, the upper limit on $h_0$ is larger in panel (a) than in panel (b) by a factor of $\sim 3$. The lower upper limit on $h_0$ in panel (b) does not indicate better sensitivity because we sacrifice $\sim 90\%$ of the $Q_1$ parameter space compared to (a). Generally speaking, the pipeline is most sensitive for the parameter domain defined in Table \ref{tab:most-sensi-range}, reaching $h_0 \leq 8 \times 10^{-25}$. The best upper limit $h_0 \leq 3.8 \times 10^{-25}$ is obtained near 150\,Hz with $Q_1 \leq 6.17 \times 10^{-19}\,\text{Hz\,s}^{-1}$ and $Q_2 \leq 2 \times 10^{-16}\,\text{Hz\,s}^{-1}$.

\begin{table}[!tbh]
	\centering
	\setlength{\tabcolsep}{8pt}
	\begin{tabular}{lll}
		\hline
		\hline
		Search Parameter & Range &  Astrophysical parameter\\
		\hline
		${\nu_0}$ & 75--200 Hz& \\
		$Q_1$ & $2 \times 10^{-22}-6.17 \times 10^{-19}$\,Hz\,s$^{-1}$ & $1.08 \times 10^{-5}\leq \epsilon \leq 6.0 \times10^{-4}$\\
		$Q_2$&$1 \times 10^{-21} - 2 \times 10^{-16}$ \,Hz\,s$^{-1}$ &  $4.05 \times 10^9\leq B/\text{G} \leq 1.81 \times10^{12}$\\
		\hline
		\hline
	\end{tabular}
	\caption[]{Parameter domain with sensitivity $h_0 \leq 8 \times 10^{-25}$ for the search in Section \ref{sec:search}. The corresponding astrophysical parameters $\epsilon$ and $B$ in equation (\ref{eq:spindownmodel_full}) are quoted in the last column.}
	\label{tab:most-sensi-range}
\end{table}

\begin{figure}[!htb]
	\centering
	\scalebox{0.35}{\includegraphics{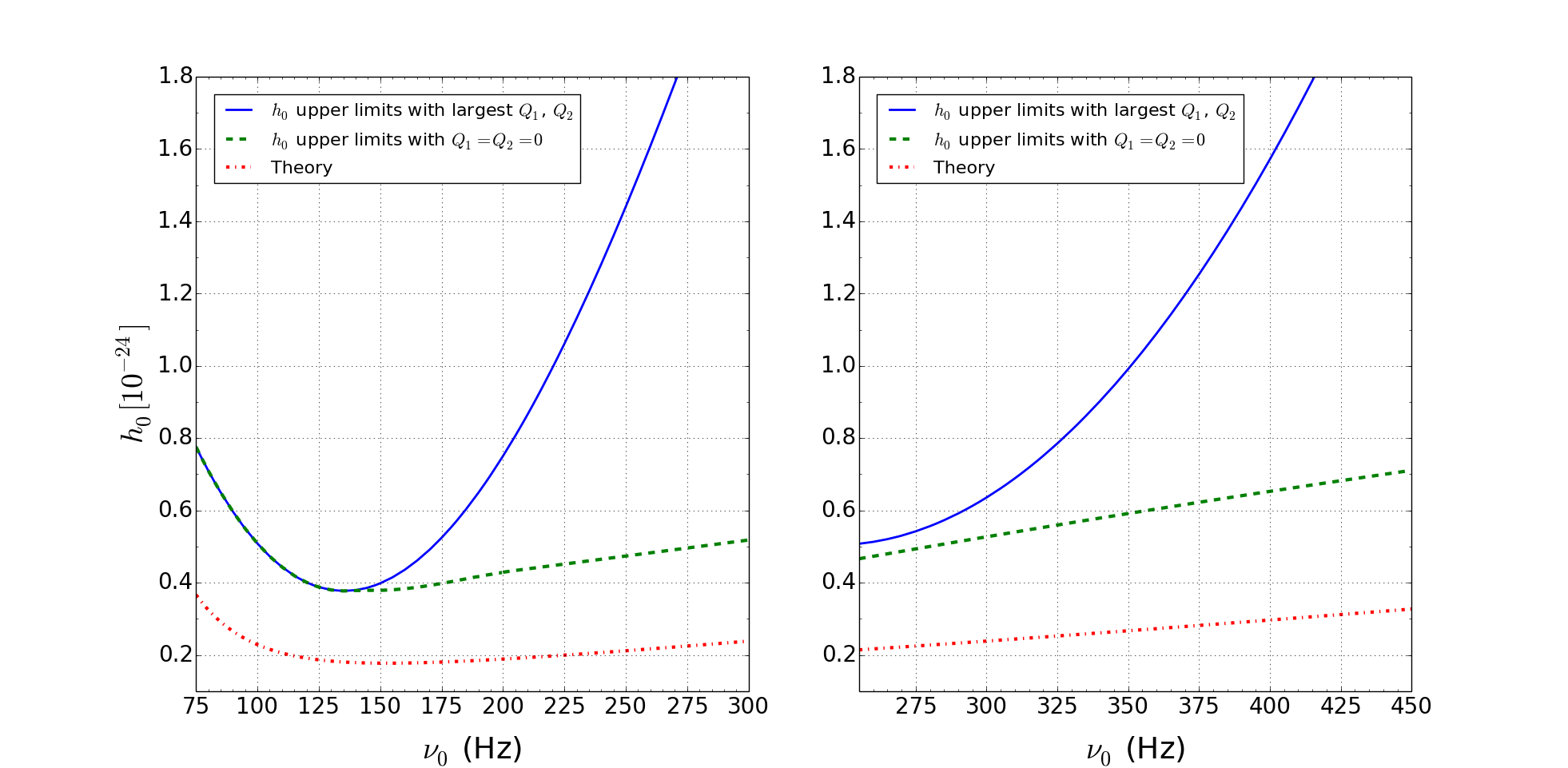}}
	\caption[compare-UL]{Comparison among the $h_0$ upper limits with largest $Q_1$ and $Q_2$ (blue solid curves), the $h_0$ upper limits with $Q_1 = Q_2 =0$ (green dash-dot curves), and the theoretical sensitivity from equation (\ref{eqn:theoretical_sensitivity}) (red dashed curves). The left panel shows the band $75 \leq \nu_0/\text{Hz} \leq 300$ with $Q_1 = 6.17 \times 10^{-19}\,\text{Hz\,s}^{-1}$ and $Q_2 = 2 \times 10^{-16}\,\text{Hz\,s}^{-1}$ injected for the blue curve. The right panel shows the band $255 \leq \nu_0/\text{Hz} \leq 450$ with $Q_1 = 5 \times 10^{-20}\,\text{Hz\,s}^{-1}$ and $Q_2 = 2 \times 10^{-16}\,\text{Hz\,s}^{-1}$ injected for the blue curve. For $\nu_0 \leq 150$\,Hz, the blue curve and green curve almost overlap, because we have $|\dot{\nu}(0)| \lesssim 10^{-7}$\,Hz\,s$^{-1}$ for all ($Q_1$, $Q_2$), and the pipeline tracks signal phase accurately with an error $\lesssim 10^{-8}$ over a year. The best upper limit $h_0 \leq 3.8 \times 10^{-25}$ is obtained near 150\,Hz with $Q_1 \leq 6.17 \times 10^{-19}\,\text{Hz\,s}^{-1}$ and $Q_2 \leq 2 \times 10^{-16}\,\text{Hz\,s}^{-1}$.}
	\label{fig:compare-UL}
\end{figure}

\begin{figure}[!htb]
	\centering
	\scalebox{0.36}{\includegraphics{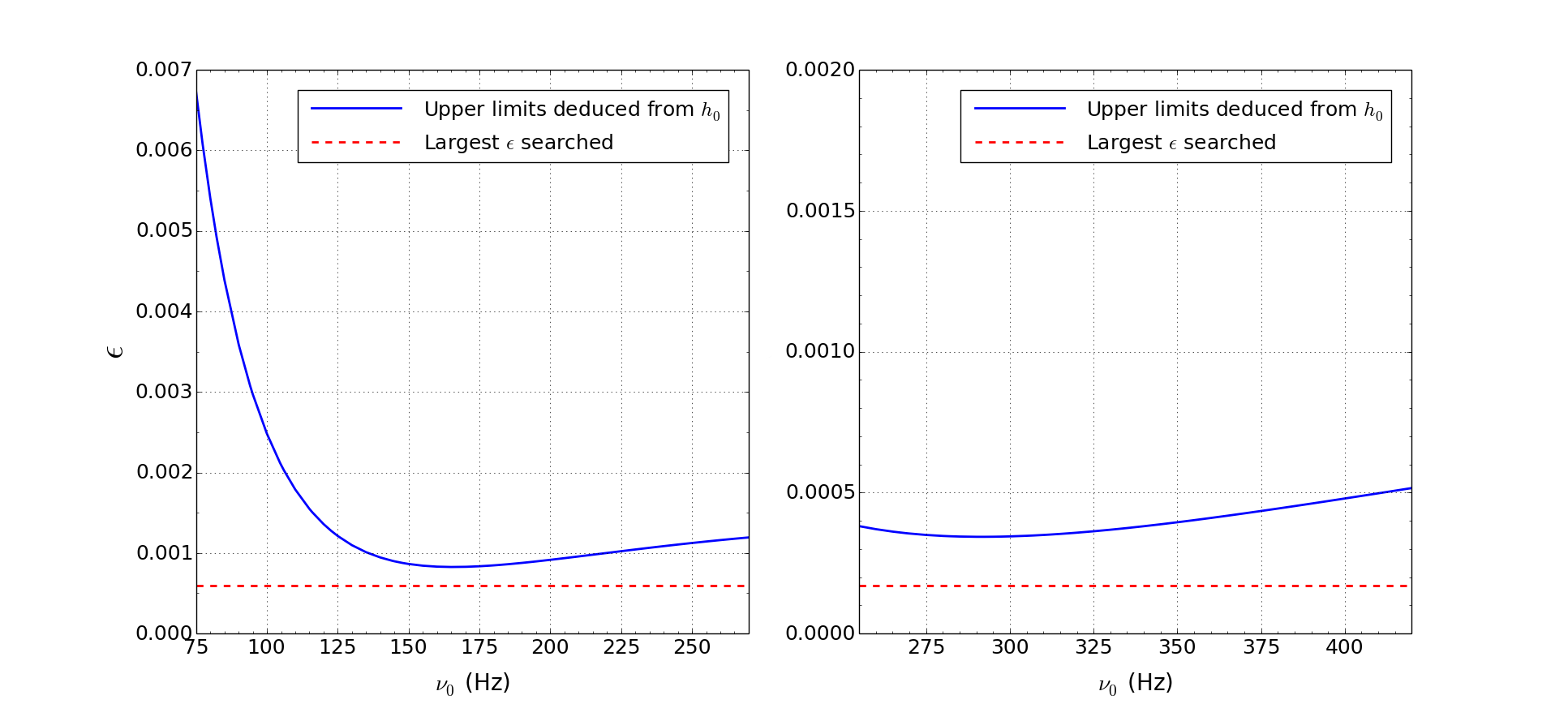}}
	\caption[epsilon-UL]{Upper limits on ellipticity $\epsilon$ deduced from the $h_0$ upper limits (with largest $Q_1$ and $Q_2$) using equation (\ref{eqn:h0_epsilon}). The left and right panels are for the bands $75 \leq \nu_0/\text{Hz} \leq 300$ and $255 \leq \nu_0/\text{Hz} \leq 450$ respectively. The dashed horizontal lines indicate the largest $\epsilon$ searched in each panel, with $\epsilon = 6.0 \times 10^{-4}$ (left) and $1.71 \times 10^{-4}$ (right). The best upper limit $\epsilon \leq 8.2 \times 10^{-4}$ is obtained near 150\,Hz, $\sim 35\%$ higher than the largest ellipticity being searched.}
	\label{fig:epsilon-UL}
\end{figure}

Similarly, we also evaluate the upper limit on $h_0$ with $Q_1 = Q_2 =0$ (i.e. $\dot{\nu}=0$). We inject synthetic signals with $1 \times 10^{-25} \leq h_0 \leq 2 \times 10^{-24}$, $75 \leq \nu_0/\text{Hz} \leq 450$ and $\dot{\nu}=0$, plot the $\rho/\sigma_\rho$ as contours on the ($h_0, \nu_0$) plane, and draw the $\rho/\sigma_\rho = \rho_\text{th} = 6.5$ curve, which represents the best upper limit achievable by the cross-correlation pipeline.

Figure \ref{fig:compare-UL} displays the comparison among the $h_0$ upper limits with largest $Q_1$ and $Q_2$ (blue solid curves; same as red dashed curves in Figure \ref{fig:h0_upper_limit}), the $h_0$ upper limits with $Q_1 = Q_2 =0$ (green dash-dot curves), and the theoretical sensitivity from equation (\ref{eqn:theoretical_sensitivity}) (red dashed curves). The band is separated into the same two segments as in Figure \ref{fig:h0_upper_limit}. For $\nu_0 \leq 150$\,Hz, the blue curve and green curve almost overlap, because we have $|\dot{\nu}(0)| \lesssim 10^{-7}$\,Hz\,s$^{-1}$ for all ($Q_1$, $Q_2$), and the pipeline tracks signal phase accurately with an error $\lesssim 10^{-8}$ over a year. For $\nu_0 \gtrsim 150$\,Hz, the difference between upper limits with largest $Q_1$ and $Q_2$ and upper limits with $Q_1 = Q_2 =0$ increases with $\nu_0$. If we diminish the $Q_1$ and $Q_2$ parameter space being searched, the corresponding upper limits with largest $Q_1$ and $Q_2$ (blue curves) get closer to the green curves. Hence the real $h_0$ upper limits always lie between the blue curves and green curves for the parameter space listed in Table \ref{tab:search_para}. As a reference, the sensitivity in theory from equation (\ref{eqn:theoretical_sensitivity}) is plotted as red dashed curves in Figure \ref{fig:compare-UL}. It is $\sim 2 \times 10^{-25}$ to $4 \times 10^{-25}$ lower than the best upper limits from the green curves. The discrepancy arises in at least two ways. First, the theoretical calculation pertains to the special case where $N_{\rm pairs}=10^5$ and the noise floor in all SFTs is the same (see Section IV in Ref.~\cite{dhurandhar08} and Section 4.1 in Ref.~\cite{chung11}). Second, equation (\ref{eqn:theoretical_sensitivity}) is valid under the assumption that $\rho$ is normally distributed (i.e. all SFT pairs are independent), which is not true in reality. From the analysis in Section \ref{sec:noise}, the moments of the noise-only PDF of $\rho/\sigma_{\rho}$ agree with those of a Gaussian distribution to an accuracy over 95\% for $T_\text{obs}= 1$\,yr. Hence we do not expect the latter cause to contribute more than $\sim 5\%$ to the overall discrepancy, consistent with the discrepancy between the theoretical and empirical values of $\rho_{\rm th}$ in Section \ref{sec:sensiA}. A more accurate statistical investigation lies outside the scope of this paper.

Upper limits on ellipticity $\epsilon$ can be deduced from the $h_0$ upper limits (with largest $Q_1$ and $Q_2$) in Figure \ref{fig:h0_upper_limit} and \ref{fig:compare-UL}, using the relationship between wave strain at the Earth and the ellipticity of the star described in equation (\ref{eqn:h0_epsilon}), and are plotted in Figure \ref{fig:epsilon-UL} as blue curves. The dashed horizontal lines indicate the largest $\epsilon$ (see Table \ref{tab:search_para}) searched in each panel, with $\epsilon = 6.0 \times 10^{-4}$ (left) and $1.71 \times 10^{-4}$ (right). The upper limits on $\epsilon$ derived from $h_0$ is larger than the maximum values being searched, which indicates that the upper limits are still above the largest spin-down rate we are sensitive to. The best upper limit $\epsilon \leq 8.2 \times 10^{-4}$ is obtained near 150\,Hz, $\sim 35\%$ higher than the largest ellipticity being searched.

\section{Conclusion}
\label{sec:codeconclusion}

In this paper, we perform a cross-correlation search for SNR 1987A using the second year of LIGO Science Run 5 data. The frequency band 75--450\,Hz is searched. Six out of the total 19 first-pass candidates survive line vetoes. One out of the six second-pass candidates remains after the first stage of manual veto (search two interferometers separately), but does not survive the second stage (search first year of S5). With zero survivors, a 90\% confidence level upper limit is placed on the wave strain given by $h_0 \approx 3.8 \times 10^{-25}$ at 150\,Hz, the most sensitive frequency, corresponding to $\epsilon \approx 8.2 \times 10^{-4}$. The previous most sensitive search for SNR 1987A conducted with the radiometer pipeline yielded a 90 \% confidence level upper limit on the wave strain of $h_0 \approx 1.57 \times 10^{-24}$ (converted from the original value by the correction factor \cite{MessengerNote}) at the most sensitive frequency range \cite{Eric2011}. Hence the strain upper limit yielded from our search improves on previously published result by a factor $\approx4$.

To verify the algorithm, we conduct a battery of tests on synthetic data and verify that the cross-correlation data analysis pipeline is functioning correctly for gravitational wave signals from a continuous-wave source obeying the spin-down law described by equation (\ref{eq:spindownmodel_full}). It is demonstrated that averaging over $\cos \iota$ and $\psi$ sacrifices typically $\lesssim 10\%$ and at worst $\lesssim 50\%$ sensitivity while delivering computational savings. It is also shown that the electromagnetic braking index $n_\text{em}$ can be excluded from the search parameters (by setting $n_\text{em}=3$) without sacrificing sensitivity, alleviating concerns expressed in previous work \cite{chung11}. We estimate the detection threshold and sensitivity with a group of Monte-Carlo tests. Without spin down, the estimated strain limits are $h_0^{95\%} \approx 5.64 \times 10^{-25}$, $7.60 \times 10^{-25}$, and $1.42 \times 10^{-24}$ for 150, 300, and 600\,Hz respectively. 

The next step in this investigation is to search Advanced LIGO data, as they become available. Despite a shorter observation period of 4 months for the first Advanced LIGO Science Run O1 (i.e. a threefold reduction in $N_\text{pairs}$), the noise power spectral density of Advanced LIGO is $\sim 4$ times better than Initial LIGO. Hence referring to equation (\ref{eqn:theoretical_sensitivity}), we can expect improvement in the theoretical sensitivity $h_\text{th}$. On the other hand, the remnant has aged since S5, so the expected signal amplitude is lower in O1. 

\section{Acknowledgements}
We are grateful to Keith Riles, John Whelan, Badri Krishnan, Karl Wette, Benjamin Owen, and the LIGO Scientific Collaboration Continuous Wave Working Group for informative discussions. We are also grateful to Eric Thrane, Alberto Vecchio and Teviet Creighton for their comprehensive formal review of the computer code and validation results of the SN1987A cross-correlation pipeline. L. Sun is supported by an Australian Postgraduate Award. The research was supported by Australian Research Council (ARC) Discovery Project DP110103347. P. D. Lasky is also supported by ARC DP140102578. Computations were done on the ATLAS Hannover cluster operated by the Albert Einstein Institute in support of research conducted by the LIGO and Virgo
Scientific Collaborations, and the Pawsey Supercomputing Centre supported by the Australian Government and the Government of Western Australia.

\end{document}